%% file: report.tex
\documentclass[11pt,twoside,a4paper]{cernrep} 
\usepackage{rep_common}
\usepackage{cancel}
\usepackage{placeins}
\usepackage{booktabs}
\usepackage{slashbox}
\usepackage{subfig}
\usepackage{amsmath,amssymb,amsthm}
\usepackage{rotating}
\usepackage{multicol}
\usepackage{tabulary}
\usepackage[colorlinks, urlcolor=blue, linkcolor=blue, citecolor=blue, plainpages=false]{hyperref}
\usepackage{float}
\floatstyle{plaintop}
\restylefloat{table}
\usepackage{xargs}
\usepackage{xstring,ragged2e}
\usepackage{lineno}

\usepackage{graphicx,multirow,colortbl}

\newcommand{\mc}{\textrm{MuC}}

\begin{document}
\newcounter{ncontacts}
\newcommand{\fcontact}[1]{\StrCount{#1}{,}[\tmp]\setcounter{ncontacts}{\tmp}
  Contact\ifthenelse{\value{ncontacts} > 0}{s}{}: #1}
\newcommand{\feditor}[1]{\StrCount{#1}{,}[\tmp]\setcounter{ncontacts}{\tmp}
  Contact Editor\ifthenelse{\value{ncontacts} > 0}{s}{}: #1}
\newcommandx{\asection}[2][1=NONE]{
  \ifthenelse{\equal{#1}{NONE}}
  {\section{#2}}{\section[#2]{#2\footnote{\feditor{#1}}}}}
\newcommandx{\asubsection}[2][1=NONE]{
  \ifthenelse{\equal{#1}{NONE}}
             {\subsection{#2}}{
               \subsection[#2]{#2\footnote{\fcontact{#1}}}}}
\newcommandx{\asubsubsection}[2][1=NONE]{
  \ifthenelse{\equal{#1}{NONE}}
             {\subsubsection{#2}}{
               \subsubsection[#2]{#2\footnote{\fcontact{#1}}}}}
\newcommandx{\asubsubsubsection}[2][1=NONE]{
  \ifthenelse{\equal{#1}{NONE}}
             {\subsubsubsection{#2}}{
               \subsubsubsection[#2]{#2\footnote{\fcontact{#1}}}}}

\newcommand\snowmass{\begin{center}\rule[-0.2in]{\hsize}{0.01in}\\\rule{\hsize}{0.01in}\\
\vskip 0.1in Submitted to the  Proceedings of the US Community Study\\ 
on the Future of Particle Physics (Snowmass 2021)\\ 
\rule{\hsize}{0.01in}\\\rule[+0.2in]{\hsize}{0.01in} \end{center}}

\title{{\normalfont\bfseries\boldmath\huge
\begin{center}
The physics case of a 3~TeV muon collider stage
\end{center}
\vspace*{-35pt}
}
{\textnormal{\normalsize \snowmass
\vspace*{-40pt}
}}
{\textnormal{\normalsize
\abstract{
In the path towards a muon collider with center of mass energy of 10~TeV or more, a stage at 3~TeV emerges as an appealing option. Reviewing the physics potential of such 
collider is the main purpose of this document. In order to outline the progression of the physics performances across the stages, a few sensitivity projections for higher energy are also presented. 
\ \\[2pt]
There are many opportunities for probing new physics at a 3~TeV muon collider. Some of them are in common with the extensively documented physics case of the CLIC 3~TeV energy stage, and include measuring the Higgs trilinear coupling and testing the possible composite nature of the Higgs boson and of the top quark at the 20~TeV scale. 
\ \\[2pt]
Other opportunities are unique of a 3~TeV muon collider, and stem from the fact that muons are collided rather than electrons. This is exemplified by studying the potential to explore the microscopic origin of the current $g$-2 and $B$-physics anomalies, which are both related with muons.
}\\[30pt]}}
{\textnormal{\normalsize\justifying
This is one of the five reports submitted to Snowmass by the muon colliders community at large. The reports preparation effort has been coordinated by the International Muon Collider Collaboration. Authors and Signatories have been collected with a 
\href{https://indico.cern.ch/event/1130036/}{subscription page}, and are defined as follows:
\begin{itemize}
 \item An ``Author'' contributed to the results documented in the report in any form, including e.g.~by participating to the discussions of the community meetings and sending comments on the draft, or plans to contribute to the future work.
\item
A ``Signatory'' expresses support to the efforts described in the report and endorses the Collaboration plans.
\end{itemize}
}}
}
\input{authors}

\begin{titlepage}

\vspace*{-1.8cm}

\noindent
\begin{tabular*}{\linewidth}{lc@{\extracolsep{\fill}}r@{\extracolsep{0pt}}}
\vspace*{-1.2cm}\mbox{\!\!\!\includegraphics[width=.14\textwidth]{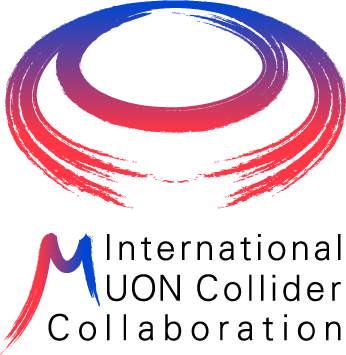}} & &  \\
 & & \today \\  
 & & \href{https://muoncollider.web.cern.ch}{https://muoncollider.web.cern.ch} \\ 
 & & \\
\hline
\end{tabular*}
\vspace*{0.3cm}
%
%
\maketitle
\vspace{\fill}

\end{titlepage}

\setcounter{tocdepth}{3}
\tableofcontents

\newpage

\subfile{\main/Section_Intro/section}

\
\newpage

\subfile{\main/Section_Higgs_EFT/section}

\ 
\newpage

\subfile{\main/Section_BSM/section}

\ 
\newpage
\subfile{\main/Section_Muon_Specific/section}

\ 
\newpage

\subfile{\main/Section_Conclusions/section}


\addcontentsline{toc}{chapter}{References}
\bibliographystyle{report}
\bibliography{report}
\end{document}

%% file: authors.tex
\newcounter{instituteref}
\newcommand{\iinstitute}[2]{\refstepcounter{instituteref}\label{#1}$^{\ref{#1}}$\href{http://inspirehep.net/record/#1}{#2}}
\newcommand{\iauthor}[3]{\href{http://inspirehep.net/record/#1}{#2}$^{#3}$}
\author{Editors: \\
    \iauthor{1021004}{J.~de~Blas}{\ref{903836}},
    \iauthor{1077579}{D.~Buttazzo}{\ref{902886}},
    \iauthor{1275234}{R.~Capdevilla}{\ref{908474},\ref{903282}},
    \iauthor{1024481}{D.~Curtin}{\ref{903282}},
    \iauthor{1052115}{R.~Franceschini}{\ref{906528},\ref{907692}},
    \iauthor{999053}{F.~Maltoni}{\ref{910783},\ref{902674}},
    \iauthor{1025277}{P.~Meade}{\ref{910429}},
    \iauthor{1074063}{F.~Meloni}{\ref{902770}},
    \iauthor{987285}{S.~Su}{\ref{902647}},
    \iauthor{1077733}{E.~Vryonidou}{\ref{902984}},
    \iauthor{1037622}{A.~Wulzer}{\ref{903113}}
    \\ \vspace*{4mm}Authors: \\
    \iauthor{1757334}{C.~Aim\`e}{\ref{943385},\ref{902885}},
    \iauthor{1060487}{A.~Apyan}{\ref{902682}},
    \iauthor{1491320}{P.~Asadi}{\ref{1237813}},
    \iauthor{1067349}{M.A.~Mahmoud.}{\ref{912409}},
    \iauthor{1041900}{A.~Azatov}{\ref{904416},\ref{902888}},
    \iauthor{1073143}{N.~Bartosik}{\ref{902889}},
    \iauthor{1029828}{A.~Bertolin}{\ref{902884}},
    \iauthor{1794682}{S.~Bottaro}{\ref{903128},\ref{902886}},
    \iauthor{1894439}{L.~Buonincontri}{\ref{902884},\ref{903113}},
    \iauthor{1057458}{M.~Casarsa}{\ref{902888}},
    \iauthor{}{L.~Castelli}{\ref{903113}},
    \iauthor{1014281}{M.G.~Catanesi}{\ref{902877}},
    \iauthor{1418744}{F.G.~Celiberto}{\ref{906718},\ref{912328}},
    \iauthor{1021757}{A.~Cerri}{\ref{1241166}},
    \iauthor{1793525}{C.~Cesarotti}{\ref{902835}},
    \iauthor{1037833}{G.~Chachamis}{\ref{905303}},
    \iauthor{2037614}{S.~Chen}{\ref{1471035}},
    \iauthor{1069708}{Y.-T.~Chien}{\ref{1275736}},
    \iauthor{1272180}{M.~Chiesa}{\ref{943385},\ref{902885}},
    \iauthor{1862239}{M.~Costa}{\ref{903128},\ref{902886}},
    \iauthor{2049478}{G.~Da~Molin}{\ref{903113}},
    \iauthor{1012395}{S.~Dasu}{\ref{903349}},
    \iauthor{1012030}{D.~Denisov}{\ref{902689}},
    \iauthor{1012025}{H.~Denizli}{\ref{908452}},
    \iauthor{1011983}{R.~Dermisek}{\ref{902874}},
    \iauthor{1063643}{L.~Luzio}{\ref{903113},\ref{902884}},
    \iauthor{1031269}{B.~Di~Micco}{\ref{906528},\ref{907692}},
    \iauthor{1011804}{K.~R.~Dienes}{\ref{902647},\ref{902990}},
    \iauthor{1011508}{T.~Dorigo}{\ref{902884}},
    \iauthor{1010482}{M.~Fabbrichesi}{\ref{902888}},
    \iauthor{1719039}{D.~Fiorina}{\ref{902885}},
    \iauthor{1894571}{M.~Forslund}{\ref{910429}},
    \iauthor{1009120}{E.~Gabrielli}{\ref{903287},\ref{902888}},
    \iauthor{1946817}{F.~Garosi}{\ref{904416}},
    \iauthor{1706734}{A.~Glioti}{\ref{1471035}},
    \iauthor{}{M.~Greco}{\ref{906528}},
    \iauthor{1198373}{A.~Greljo}{\ref{902668}},
    \iauthor{1258537}{R.~Gr\"ober}{\ref{1513358},\ref{902884}},
    \iauthor{1007486}{C.~Grojean}{\ref{902770},\ref{902858}},
    \iauthor{1274618}{J.~Gu}{\ref{903628}},
    \iauthor{1259916}{C.~Han}{\ref{903702}},
    \iauthor{1006825}{T.~Han}{\ref{903130}},
    \iauthor{1912097}{K.~Hermanek}{\ref{902874}},
    \iauthor{1006149}{M.~Herndon}{\ref{903349}},
    \iauthor{1067690}{T.R.~Holmes}{\ref{1623978}},
    \iauthor{1515880}{S.~Homiller}{\ref{902835}},
    \iauthor{1649884}{G.-y.~Huang}{\ref{902841}},
    \iauthor{1475406}{S.~Jana}{\ref{902841}},
    \iauthor{1028687}{S.~Jindariani}{\ref{902796}},
    \iauthor{1051663}{Y.~Kahn}{\ref{902867}},
    \iauthor{1002991}{W.~Kilian}{\ref{903203}},
    \iauthor{1019544}{P.~Koppenburg}{\ref{903832}},
    \iauthor{1904040}{N.~Kreher}{\ref{903203}},
    \iauthor{1252769}{K.~Krizka}{\ref{902953}},
    \iauthor{1077491}{G.~Krnjaic}{\ref{902796}},
    \iauthor{1854911}{N.~Kumar}{\ref{902767}},
    \iauthor{1071846}{L.~Lee}{\ref{1623978}},
    \iauthor{1074984}{Q.~Li}{\ref{903603}},
    \iauthor{1256188}{Z.~Liu}{\ref{903010}},
    \iauthor{999862}{K.R.~Long}{\ref{902868},\ref{903174}},
    \iauthor{999724}{I.~Low}{\ref{902645},\ref{903083}},
    \iauthor{1700371}{Q.~Lu}{\ref{902835}},
    \iauthor{999654}{D.~Lucchesi}{\ref{903113},\ref{902884}},
    \iauthor{1355155}{L.~Ma}{\ref{904187}},
    \iauthor{1514492}{Y.~Ma}{\ref{903130}},
    \iauthor{1668860}{L.~Mantani}{\ref{907623}},
    \iauthor{1078065}{D.~Marzocca}{\ref{902888}},
    \iauthor{1751811}{N.~McGinnis}{\ref{903290}},
    \iauthor{997877}{B.~Mele}{\ref{902887}},
    \iauthor{1461119}{C.~Merlassino}{\ref{903112}},
    \iauthor{2049482}{A.~Montella}{\ref{902888}},
    \iauthor{1069385}{M.~Nardecchia}{\ref{903168},\ref{902887}},
    \iauthor{}{F.~Nardi}{\ref{903113},\ref{902884}},
    \iauthor{1077958}{P.~Panci}{\ref{903129},\ref{902886}},
    \iauthor{1048820}{S.~Pagan~Griso}{\ref{902953}},
    \iauthor{1037621}{G.~Panico}{\ref{902801},\ref{902880}},
    \iauthor{1023838}{P.~Paradisi}{\ref{1513358},\ref{902884}},
    \iauthor{994095}{N.~Pastrone}{\ref{902889}},
    \iauthor{993440}{F.~Piccinini}{\ref{902885}},
    \iauthor{1050691}{K.~Potamianos}{\ref{903112}},
    \iauthor{992463}{E.~Radicioni}{\ref{902877}},
    \iauthor{992222}{R.~Rattazzi}{\ref{1471035}},
    \iauthor{1214912}{D.~Redigolo}{\ref{902880}},
    \iauthor{992031}{L.~Reina}{\ref{902803}},
    \iauthor{1021811}{J.~Reuter}{\ref{902770}},
    \iauthor{1020819}{C.~Riccardi}{\ref{943385},\ref{902885}},
    \iauthor{1885919}{L.~Ricci}{\ref{1471035}},
    \iauthor{1028713}{L.~Ristori}{\ref{902796}},
    \iauthor{1040385}{T.~Robens}{\ref{902678}},
    \iauthor{991509}{W.~Rodejohann}{\ref{902841}},
    \iauthor{1054727}{R.~Ruiz}{\ref{902756}},
    \iauthor{1072232}{F.~Sala}{\ref{908583}},
    \iauthor{1885424}{J.~Salko}{\ref{902668}},
    \iauthor{990505}{P.~Salvini}{\ref{902885}},
    \iauthor{990367}{J.~Santiago}{\ref{909079},\ref{903836}},
    \iauthor{}{I.~Sarra}{},
    \iauthor{989645}{D.~Schulte}{\ref{902725}},
    \iauthor{1039590}{M.~Selvaggi}{\ref{902725}},
    \iauthor{1019799}{A.~Senol}{\ref{908452}},
    \iauthor{1342183}{L.~Sestini}{\ref{902884}},
    \iauthor{1071696}{V.~Sharma}{\ref{903349}},
    \iauthor{1074094}{R.~Simoniello}{\ref{902725}},
    \iauthor{1319078}{G.~Stark}{\ref{1218068}},
    \iauthor{1023964}{D.~Stolarski}{\ref{906105}},
    \iauthor{1027860}{W.~Su}{\ref{907284}},
    \iauthor{1375692}{O.~Sumensari}{\ref{1776405}},
    \iauthor{1071725}{X.~Sun}{\ref{1210798}},
    \iauthor{986860}{T.~Tait}{\ref{903302}},
    \iauthor{1268914}{J.~Tang}{\ref{903702},\ref{903123}},
    \iauthor{1257546}{A.~Tesi}{\ref{902880}},
    \iauthor{1023463}{B.~Thomas}{\ref{904505}},
    \iauthor{1878399}{E.~A.~Thompson}{\ref{902770}},
    \iauthor{1064514}{R.~Torre}{\ref{902881}},
    \iauthor{1778841}{S.~Trifinopoulos}{\ref{902888}},
    \iauthor{1265350}{I.~Vai}{\ref{902885}},
    \iauthor{1880878}{A.~Valenti}{\ref{1513358},\ref{902884}},
    \iauthor{1863232}{L.~Vittorio}{\ref{903128},\ref{902886}},
    \iauthor{984146}{L.-T.~Wang}{\ref{902729}},
    \iauthor{1511975}{Y.~Wu}{\ref{903094}},
    \iauthor{1618109}{K.~Xie}{\ref{903130}},
    \iauthor{1656814}{X.~Zhao}{\ref{906528},\ref{907692}},
    \iauthor{1037623}{J.~Zurita}{\ref{907907}}
    \\ \vspace*{4mm} Signatories: \\
    \iauthor{1018987}{D.~Acosta}{\ref{903156}},
    \iauthor{1018902}{K.~Agashe}{\ref{902990}},
    \iauthor{1018633}{B.C.~Allanach}{\ref{907623}},
    \iauthor{1018264}{F.~Anulli}{\ref{902887}},
    \iauthor{1049113}{A.~Apresyan}{\ref{902796}},
    \iauthor{}{D.~Athanasakos}{\ref{910429}},
    \iauthor{1028433}{J.J.~Back}{\ref{903734}},
    \iauthor{1424044}{L.~Bandiera}{\ref{905268}},
    \iauthor{1017330}{R.~J.~Barlow}{\ref{911708}},
    \iauthor{1037853}{E.~Barzi}{\ref{902796},\ref{903092}},
    \iauthor{2031609}{F.~Batsch}{\ref{902725}},
    \iauthor{1068289}{M.~Bauce}{\ref{902887},\ref{903168}},
    \iauthor{1016672}{J.~S.~Berg}{\ref{902689}},
    \iauthor{1016557}{J.~Berryhill}{\ref{902796}},
    \iauthor{1049763}{A.~Bersani}{\ref{902881}},
    \iauthor{1020223}{K.M.~Black}{\ref{903349}},
    \iauthor{1015872}{M.~Bonesini}{\ref{902882},\ref{907960}},
    \iauthor{1015812}{C.~Booth}{\ref{903196}},
    \iauthor{}{L.~Bottura}{},
    \iauthor{1031261}{D.~Bowring}{\ref{902796}},
    \iauthor{1015478}{A.~Braghieri}{\ref{902885}},
    \iauthor{1015228}{G.~Brooijmans}{\ref{902749}},
    \iauthor{1015214}{A.~Bross}{\ref{902796}},
    \iauthor{1114205}{E.~Brost}{\ref{902689}},
    \iauthor{1670912}{B.~Caiffi}{\ref{902881}},
    \iauthor{1014742}{G.~Calderini}{\ref{926589},\ref{903119}},
    \iauthor{1707397}{S.~Calzaferri}{\ref{902885}},
    \iauthor{}{P.~Cameron}{\ref{902689}},
    \iauthor{1024602}{A.~Canepa}{\ref{902796}},
    \iauthor{}{F.~Casaburo}{},
    \iauthor{1020772}{G.~Cavoto}{\ref{903168},\ref{902887}},
    \iauthor{1075318}{L.~Celona}{\ref{902879}},
    \iauthor{}{G.~Cesarini}{},
    \iauthor{1014143}{Z.~Chacko}{\ref{902990}},
    \iauthor{2023221}{A.~Chanc\'e}{\ref{912490}},
    \iauthor{1380376}{R.~T.~Co}{\ref{908012}},
    \iauthor{1013275}{A.~Colaleo}{\ref{902660},\ref{902877}},
    \iauthor{1013241}{G.~Collazuol}{\ref{902884},\ref{903113}},
    \iauthor{}{D.~J.~Colling}{\ref{902868}},
    \iauthor{1013050}{G.~Corcella}{\ref{902807}},
    \iauthor{1046385}{N.~Craig}{\ref{903307}},
    \iauthor{}{L.~M.~Cremaldi}{},
    \iauthor{1060042}{A.~Crivellin}{\ref{903370},\ref{905405}},
    \iauthor{1035631}{Y.~Cui}{\ref{903304}},
    \iauthor{1937290}{C.~Curatolo}{\ref{902882}},
    \iauthor{1067364}{R.~T.~D'Agnolo}{\ref{1087875}},
    \iauthor{1047966}{F.~D'Eramo}{\ref{903113},\ref{902884}},
    \iauthor{2052018}{M.~Dam}{\ref{902882}},
    \iauthor{1076225}{H.~Damerau}{\ref{902725}},
    \iauthor{1651018}{E.~De~Matteis}{\ref{907142}},
    \iauthor{1012237}{A.~Deandrea}{\ref{1743848}},
    \iauthor{1012143}{J.~Delahaye}{\ref{902725}},
    \iauthor{1019723}{A.~Delgado}{\ref{903085}},
    \iauthor{1039487}{C.~Densham}{\ref{903174}},
    \iauthor{1395010}{K.~F.~Di~Petrillo}{\ref{902796}},
    \iauthor{1246709}{J.~Dickinson}{\ref{902796}},
    \iauthor{1054778}{J.~Duarte}{\ref{903305}},
    \iauthor{1404358}{F.~Errico}{\ref{902660},\ref{902877}},
    \iauthor{1048347}{R.~Essig}{\ref{910429}},
    \iauthor{1064320}{P.~Everaerts}{\ref{903349}},
    \iauthor{1010523}{L.~Everett}{\ref{903349}},
    \iauthor{1045844}{J.~Fan}{\ref{902692}},
    \iauthor{1069878}{S.~Farinon}{\ref{902881}},
    \iauthor{1010105}{A.~Ferrari}{\ref{1276460}},
    \iauthor{1648215}{J.~F.~Somoza}{\ref{902725}},
    \iauthor{1010065}{G.~Ferretti}{\ref{902825}},
    \iauthor{1009979}{F.~Filthaut}{\ref{903075}},
    \iauthor{1046463}{P.~Franchini}{\ref{902948},\ref{903170}},
    \iauthor{1019509}{M.~Frigerio}{\ref{1508424}},
    \iauthor{1009009}{M.~Gallinaro}{\ref{905303}},
    \iauthor{1068164}{I.~Garcia~Garcia}{\ref{903889}},
    \iauthor{1894454}{L.~Giambastiani}{\ref{903113},\ref{902884}},
    \iauthor{}{A.S.~Giannakopoulou}{\ref{903237}},
    \iauthor{1075917}{D.~Giove}{\ref{907142}},
    \iauthor{1971617}{C.~Giraldin}{\ref{903113}},
    \iauthor{1029806}{L.~Gladilin}{},
    \iauthor{1008112}{S.~Goldfarb}{\ref{902999}},
    \iauthor{1037882}{H.M.~Gray}{\ref{903299},\ref{902953}},
    \iauthor{1059457}{L.~Gray}{\ref{902796}},
    \iauthor{1007092}{H.E.~Haber}{\ref{1218068}},
    \iauthor{1055424}{J.~Haley}{\ref{903094}},
    \iauthor{2044726}{J.~Hauptman}{\ref{902893}},
    \iauthor{1383268}{B.~Henning}{\ref{1471035}},
    \iauthor{2049476}{H.~Jia}{\ref{903349}},
    \iauthor{}{C.~Jolly}{\ref{903174}},
    \iauthor{1003695}{D.~M.~Kaplan}{\ref{902865}},
    \iauthor{1653554}{I.~Karpov}{\ref{902725}},
    \iauthor{1067609}{D.~Kelliher}{\ref{903174}},
    \iauthor{1020007}{K.~Kong}{\ref{902912}},
    \iauthor{1345391}{G.K.~Krintiras}{\ref{902912}},
    \iauthor{1001375}{P.~Kyberd}{\ref{903940}},
    \iauthor{999784}{R.~LOSITO}{\ref{902725}},
    \iauthor{1292423}{J.-B.~Lagrange}{\ref{903174}},
    \iauthor{}{S.~Levorato}{},
    \iauthor{1064657}{W.~Li}{\ref{903156}},
    \iauthor{1996476}{R.~L.~Voti}{\ref{902887}},
    \iauthor{1000076}{R.~Lipton}{\ref{902796}},
    \iauthor{1074693}{M.~Liu}{\ref{903142}},
    \iauthor{}{S.~Lomte}{\ref{903349}},
    \iauthor{1041997}{R.~Mahbubani}{\ref{902678}},
    \iauthor{998923}{B.~Mansouli\'e}{\ref{912490}},
    \iauthor{1056868}{A.~Mariotti}{\ref{907933}},
    \iauthor{1670119}{S.~Mariotto}{\ref{903009},\ref{907142}},
    \iauthor{1971307}{P.~Mastrapasqua}{\ref{910783}},
    \iauthor{998430}{K.~Matchev}{\ref{902804}},
    \iauthor{1054925}{A.~Mazzacane}{\ref{902796}},
    \iauthor{1022138}{P.~Merkel}{\ref{902796}},
    \iauthor{1032624}{F.~Mescia}{\ref{905190},\ref{911212}},
    \iauthor{1066275}{R.~K.~Mishra}{\ref{902835}},
    \iauthor{1070072}{A.~Mohammadi}{\ref{903349}},
    \iauthor{}{R.~Mohapatra}{},
    \iauthor{997065}{N.~Mokhov}{\ref{902796}},
    \iauthor{996989}{P.~Montagna}{\ref{943385},\ref{902885}},
    \iauthor{1064125}{R.~Musenich}{\ref{902881}},
    \iauthor{995835}{M.S.~Neubauer}{\ref{902867}},
    \iauthor{995826}{D.~Neuffer}{\ref{902796}},
    \iauthor{995794}{H.~Newman}{\ref{902711}},
    \iauthor{995460}{Y.~Nomura}{\ref{903299}},
    \iauthor{1070110}{I.~Ojalvo}{\ref{16750}},
    \iauthor{1465792}{J.L.~Oliver}{\ref{903302}},
    \iauthor{1274353}{D.~Pagani}{\ref{902878}},
    \iauthor{994435}{M.~Palmer}{\ref{902689}},
    \iauthor{1067971}{R.~Paparella}{\ref{907142}},
    \iauthor{1772198}{A.~Pellecchia}{\ref{902660}},
    \iauthor{1067962}{A.~Perloff}{\ref{902748}},
    \iauthor{1021028}{M.~Pierini}{\ref{902725}},
    \iauthor{1651162}{M.~Prioli}{\ref{907142}},
    \iauthor{1024769}{M.~Procura}{\ref{903326}},
    \iauthor{1217056}{R.~Radogna}{\ref{902660},\ref{902877}},
    \iauthor{991702}{R.A.~Rimmer}{\ref{904961}},
    \iauthor{1056642}{F.~Riva}{\ref{902813}},
    \iauthor{1054170}{C.~Rogers}{\ref{903174}},
    \iauthor{991185}{L.~Rossi}{\ref{903009},\ref{907142}},
    \iauthor{1912150}{R.~Ryne}{\ref{1189711}},
    \iauthor{1077871}{E.~Salvioni}{\ref{1513358},\ref{902884}},
    \iauthor{1042797}{E.~Santopinto}{\ref{902881}},
    \iauthor{989950}{J.~Schieck}{\ref{903324},\ref{904536}},
    \iauthor{1022121}{R.~Schwiehorst}{\ref{903006}},
    \iauthor{1055618}{D.~Sertore}{\ref{907142}},
    \iauthor{988978}{V.~Shiltsev}{\ref{902796}},
    \iauthor{1019902}{J.~Shu}{\ref{903895}},
    \iauthor{1622677}{F.~M.~Simone}{\ref{902660},\ref{902877}},
    \iauthor{1889775}{K.~Skoufaris}{\ref{902725}},
    \iauthor{1046340}{P.~Snopok}{\ref{902865}},
    \iauthor{988143}{F.J.P.~Soler}{\ref{902823}},
    \iauthor{1066476}{M.~Sorbi}{\ref{903009},\ref{907142}},
    \iauthor{}{A.~Stamerra}{\ref{902660},\ref{902877}},
    \iauthor{1057643}{M.~Statera}{\ref{907142}},
    \iauthor{1058749}{D.~Stratakis}{\ref{902796}},
    \iauthor{1077402}{N.~Strobbe}{\ref{903010}},
    \iauthor{1071880}{J.~Stupak}{\ref{1273509}},
    \iauthor{987128}{R.~Sundrum}{\ref{912511}},
    \iauthor{1078570}{M.~Swiatlowski}{\ref{903290}},
    \iauthor{1454316}{A.~Sytov}{\ref{905268}},
    \iauthor{1020371}{A.~Taffard}{\ref{903302}},
    \iauthor{2057899}{J.~Tang}{\ref{903123}},
    \iauthor{1055960}{M.~Taoso}{\ref{902889}},
    \iauthor{1054400}{J.~Thaler}{\ref{1237813}},
    \iauthor{985810}{L.~Tortora}{\ref{907692}},
    \iauthor{1031757}{Y.~Torun}{\ref{902865}},
    \iauthor{2025179}{R.~U.~Valente}{\ref{907142}},
    \iauthor{1613622}{M.~Valente}{\ref{903290}},
    \iauthor{1643523}{N.~Valle}{\ref{943385},\ref{902885}},
    \iauthor{1071756}{R.~Venditti}{\ref{902660},\ref{902877}},
    \iauthor{1063935}{P.~Verwilligen}{\ref{902877}},
    \iauthor{1077738}{N.~Vignaroli}{\ref{902883}},
    \iauthor{984555}{P.~Vitulo}{\ref{943385},\ref{902885}},
    \iauthor{1077733}{E.~Vryonidou}{\ref{902984}},
    \iauthor{1054127}{C.~Vuosalo}{\ref{903349}},
    \iauthor{1073818}{H.~Weber}{\ref{902858}},
    \iauthor{1049718}{M.~Wendt}{\ref{902725}},
    \iauthor{1260509}{C.G.~Whyte}{\ref{904214}},
    \iauthor{982905}{A.~Yamamoto}{\ref{902916}},
    \iauthor{1421892}{W.~Yin}{\ref{903268}},
    \iauthor{1019845}{K.~Yonehara}{\ref{902796}},
    \iauthor{1024759}{H.-B.~Yu}{\ref{903304}},
    \iauthor{1064691}{M.~Zanetti}{\ref{903113}},
    \iauthor{1971310}{A.~Zaza}{\ref{902660},\ref{902877}},
    \iauthor{}{J.~Zhang}{},
    \iauthor{1066114}{Y.~J.~Zheng}{\ref{902912}},
    \iauthor{981974}{A.~Zlobin}{\ref{902796}},
    \iauthor{1863481}{D.~Zuliani}{\ref{903113},\ref{902884}}
    \vspace*{1cm}} \institute{\small
    \iinstitute{903836}{CAFPE and Departamento de F\'isica Te\'orica y del Cosmos, Universidad de Granada, Spain};
    \iinstitute{902886}{INFN Sezione di Pisa, Italy};
    \iinstitute{908474}{Perimeter Institute, Canada};
    \iinstitute{903282}{Department of Physics, University of Toronto, Canada};
    \iinstitute{906528}{Dipartimento di Matematica e Fisica, Universit\`a Roma Tre, Italy};
    \iinstitute{910783}{Center for Cosmology, Particle Physics and Phenomenology, Universit\'e catholique de Louvain, Belgium};
    \iinstitute{910429}{C. N. Yang Institute for Theoretical Physics, Stony Brook University, United States};
    \iinstitute{902770}{Deutsches Elektronen-Synchrotron DESY, Germany};
    \iinstitute{902647}{University of Arizona, United States};
    \iinstitute{902984}{University of Manchester, United Kingdom};
    \iinstitute{903113}{Dipartimento di Fisica e Astronomia, Universit'a di Padova, Italy};
    \iinstitute{907692}{INFN Sezione di Roma Tre, Italy};
    \iinstitute{902674}{Dipartimento di Fisica e Astronomia, Universit\`a di Bologna, Italy};
    \iinstitute{943385}{Universit{\`a} di Pavia, Italy};
    \iinstitute{902682}{Department of Physics, Brandeis University, United States};
    \iinstitute{1237813}{Center for Theoretical Physics, Massachusetts Institute of Technology, United States};
    \iinstitute{912409}{{Center for High Energy Physics (CHEP-FU), Fayoum University, Egypt}};
    \iinstitute{904416}{SISSA International School for Advanced Studies, Italy};
    \iinstitute{902889}{{INFN Sezione di Torino, Italy}};
    \iinstitute{902884}{INFN Sezione di Padova, Italy};
    \iinstitute{903128}{{Scuola Normale Superiore, Italy}};
    \iinstitute{902888}{INFN Sezione di Trieste, Italy};
    \iinstitute{902877}{INFN Sezione di Bari, Italy};
    \iinstitute{906718}{European Centre for Theoretical Studies in Nuclear Physics and Related Areas (ECT*), Italy};
    \iinstitute{1241166}{MPS School, University of Sussex, United Kingdom};
    \iinstitute{902835}{Department of Physics, Harvard University, United States};
    \iinstitute{905303}{Laborat{\' o}rio de Instrumenta\c{c}{\~ a}o e F{\' \i}sica Experimental de Part{\' \i}culas (LIP), Portugal};
    \iinstitute{1471035}{{Theoretical Particle Physics Laboratory (LPTP), Institute of Physics, EPFL, Switzerland}};
    \iinstitute{1275736}{Physics and Astronomy Department, Georgia State University, United States};
    \iinstitute{903349}{University of Wisconsin, United States};
    \iinstitute{902689}{Brookhaven National Laboratory, United States};
    \iinstitute{908452}{Department of Physics, Bolu Abant Izzet Baysal University, Turkey};
    \iinstitute{902874}{Physics Department, Indiana University, United States};
    \iinstitute{902885}{INFN Sezione di Pavia, Italy};
    \iinstitute{903287}{Physics Department, University of Trieste, Italy};
    \iinstitute{902668}{Albert Einstein Center for Fundamental Physics, Institute for Theoretical Physics, University of Bern, Switzerland};
    \iinstitute{1513358}{Universit\`a di Padova, Italy};
    \iinstitute{903628}{Department of Physics, Fudan University, China};
    \iinstitute{903702}{School of Physics, Sun Yat-Sen University, China};
    \iinstitute{903130}{University of Pittsburgh, United States};
    \iinstitute{1623978}{University of Tennessee, United States};
    \iinstitute{902841}{Max-Planck-Institut f{\"u}r Kernphysik, Germany};
    \iinstitute{902796}{Fermi National Accelerator Laboratory, United States};
    \iinstitute{902867}{Department of Physics, University of Illinois at Urbana-Champaign, United States};
    \iinstitute{903203}{Department of Physics, University of Siegen, Germany};
    \iinstitute{903832}{Nikhef National Institute for Subatomic Physics, The Netherlands};
    \iinstitute{902953}{{Physics Division, Lawrence Berkeley National Laboratory, United States}};
    \iinstitute{902767}{Delhi University, India};
    \iinstitute{903603}{Peking University, China};
    \iinstitute{903010}{School of Physics and Astronomy, University of Minnesota, United States};
    \iinstitute{902868}{Imperial College London, United Kingdom};
    \iinstitute{902645}{High Energy Physics Division, Argonne National Laboratory, United States};
    \iinstitute{904187}{Shandong University, China};
    \iinstitute{907623}{{DAMTP, University of Cambridge, United Kingdom}};
    \iinstitute{903290}{TRIUMF, Canada};
    \iinstitute{902887}{{INFN Sezione di Roma, Italy}};
    \iinstitute{903112}{Particle Physics Department, University of Oxford, United Kingdom};
    \iinstitute{903168}{Sapienza University of Rome, Italy};
    \iinstitute{903129}{Pisa University, Italy};
    \iinstitute{902801}{{Dipartimento di Fisica e Astronomia, Universit{\`a} degli Studi di Firenze, Italy}};
    \iinstitute{902880}{INFN Sezione di Firenze, Italy};
    \iinstitute{902803}{Florida State University, United States};
    \iinstitute{902678}{Rudjer Boskovic Institute, Croatia};
    \iinstitute{902756}{Institute of Nuclear Physics -- Polish Academy of Sciences {\rm (IFJ PAN)}, Poland};
    \iinstitute{411233}{International Institute of Physics, Universidade Federal do Rio Grande do Norte, Brazil};
    \iinstitute{908583}{Laboratoire de Physique Th\'eorique et Hautes \'Energies, Sorbonne Universit\'e, CNRS, France};
    \iinstitute{909079}{{CAFPE}, Spain};
    \iinstitute{902725}{CERN, Switzerland};
    \iinstitute{1218068}{SCIPP, UC Santa Cruz, United States};
    \iinstitute{906105}{Ottawa-Carleton Institute for Physics, Carleton University, Canada};
    \iinstitute{907284}{Korea Institute for Advanced Study, South Korea};
    \iinstitute{1776405}{IJCLab, P\^ole Th\'eorie (B\^at.~210), CNRS/IN2P3 et Universit\'e Paris-Saclay, France};
    \iinstitute{1210798}{State Key Laboratory of Nuclear Physics and Technology, Peking University, China};
    \iinstitute{903302}{Department of Physics and Astronomy, University of California, Irvine, United States};
    \iinstitute{904505}{Department of Physics, Lafayette College, United States};
    \iinstitute{902881}{{INFN Sezione di Genova, Italy}};
    \iinstitute{902729}{Department of Physics, University of Chicago, United States};
    \iinstitute{903094}{Department of Physics, Oklahoma State University, United States};
    \iinstitute{907907}{{Instituto de F{\'i}sica Corpuscular, CSIC-Universitat de Val{\'e}ncia, Spain}};
    \iinstitute{912328}{INFN-TIFPA Trento Institute of Fundamental Physics and Applications, Italy};
    \iinstitute{902990}{Department of Physics, University of Maryland, United States};
    \iinstitute{902858}{{Humboldt-Universit\"at zu Berlin, Institut f\"ur Physik, Germany}};
    \iinstitute{903174}{STFC, United Kingdom};
    \iinstitute{903083}{Department of Physics and Astronomy, Northwestern University, United States};
    \iinstitute{903123}{Institute of High-Energy Physics, China};
    \iinstitute{903156}{Physics \& Astronomy Department, Rice University, United States};
    \iinstitute{903734}{Department of Physics, University of Warwick, United Kingdom};
    \iinstitute{905268}{INFN Sezione di Ferrara, Italy};
    \iinstitute{911708}{The University of Huddersfield, United Kingdom};
    \iinstitute{902882}{Istituto Nazionale di Fisica Nucleare, Italy};
    \iinstitute{903196}{{Department of Physics and Astronomy, University of Sheffield, United Kingdom}};
    \iinstitute{902749}{Columbia University, United States};
    \iinstitute{926589}{CNRS/IN2P3, France};
    \iinstitute{902879}{INFN Sezione di Catania, Italy};
    \iinstitute{912490}{IRFU, CEA, UniversitÃ© Paris-Saclay, France};
    \iinstitute{908012}{University of Minnesota, United States};
    \iinstitute{902660}{{Department of Physics, Universit{\`a} degli Studi di Bari, Italy}};
    \iinstitute{902807}{INFN, Laboratori Nazionali di Frascati, Italy};
    \iinstitute{903307}{University of California, Santa Barbara, United States};
    \iinstitute{903370}{University of Zurich, Switzerland};
    \iinstitute{903304}{University of California-Riverside, United States};
    \iinstitute{1087875}{Universit\`e Paris Saclay, CNRS, CEA, Institut de Physique Th\`eorique, France};
    \iinstitute{907142}{{Laboratori Acceleratori e SuperconduttivitÃ  Applicata (LASA), INFN, Italy}};
    \iinstitute{1743848}{IP2I, Universit\'e Lyon 1, CNRS/IN2P3, France};
    \iinstitute{903085}{University of Notre Dame, United States};
    \iinstitute{903305}{University of California San Diego,  La Jolla, United States};
    \iinstitute{902692}{Brown University, United States};
    \iinstitute{1276460}{Helmholtz-Zentrum Dresden-Rossendorf, Germany};
    \iinstitute{902825}{Chalmers University of Technology, Sweden};
    \iinstitute{903075}{Radboud University and Nikhef, The Netherlands};
    \iinstitute{902948}{{University of Lancaster, Department of Physics, United Kingdom}};
    \iinstitute{1508424}{Laboratoire Charles Coulomb, CNRS and University of Montpellier, France};
    \iinstitute{903889}{Kavli Institute for Theoretical Physics, University of California, United States};
    \iinstitute{903237}{SUNY at Stony Brook, United States};
    \iinstitute{902999}{School of Physics, University of Melbourne, Australia};
    \iinstitute{903299}{UC Berkeley, United States};
    \iinstitute{902893}{Iowa State University, United States};
    \iinstitute{902865}{Illinois Institute of Technology, United States};
    \iinstitute{902912}{Department of Physics and Astronomy, University of Kansas, United States};
    \iinstitute{903940}{{College of Engineering, Design and Physical Sciences, Brunel University, United Kingdom}};
    \iinstitute{903142}{Purdue University, United States};
    \iinstitute{907933}{Theoretische Natuurkunde and IIHE/ELEM, Vrije Universiteit Brussel, Belgium};
    \iinstitute{903009}{{Dipartimento di Fisica Aldo Pontremoli, Universit\'a degli Studi di Milano, Italy}};
    \iinstitute{902804}{Physics Department, University of Florida, United States};
    \iinstitute{905190}{Universitat de Barcelona, Spain};
    \iinstitute{902711}{California Institute of Technology, United States};
    \iinstitute{16750}{Princeton University, United States};
    \iinstitute{902878}{INFN Sezione di Bologna, Italy};
    \iinstitute{902748}{{Department of Physics, University of Colorado, United States}};
    \iinstitute{903326}{University of Vienna, Faculty of Physics, Austria};
    \iinstitute{904961}{JLab, United States};
    \iinstitute{902813}{D\'epartment de Physique Th\'eorique, Universit\'e de Gen\`eve, Switzerland};
    \iinstitute{1189711}{Lawrence Berkeley National Laboratory, United States};
    \iinstitute{903324}{Institut f\"ur Hochenergiephysik der \"Osterreichischen Akademie der Wissenschaften, Austria};
    \iinstitute{903006}{MIchigan State University, United States};
    \iinstitute{903895}{CAS Key Laboratory of Theoretical Physics, Insitute of Theoretical Physics, Chinese Academy of Sciences, P.R.China};
    \iinstitute{902823}{School of Physics and Astronomy, University of Glasgow, United Kingdom};
    \iinstitute{1273509}{University of Oklahoma, United States};
    \iinstitute{912511}{Maryland Center for Fundamental Physics, University of Maryland, United States};
    \iinstitute{902883}{Universit{\'a} di Napoli ``Federico II" and INFN Napoli, Italy};
    \iinstitute{904214}{Physics, SUPA, United Kingdom};
    \iinstitute{902916}{{High Energy Accelerator Research Organization KEK, Japan}};
    \iinstitute{903268}{Tohoku University, Japan};
    \iinstitute{1111512}{Institut f{\"u}r Allgemeine Elektrotechnik, Universit{\"a}t Rostock, Germany};
    \iinstitute{903092}{Ohio State University, United States};
    \iinstitute{907960}{Dipartimento di Fisica, Universit\`a Milano Bicocca, Italy};
    \iinstitute{903119}{LPNHE, Sorbonne Universit\'e, France};
    \iinstitute{905405}{Paul Scherrer Institute, Switzerland};
    \iinstitute{903170}{{Royal Holloway University of London, Department of Physics, United Kingdom}};
    \iinstitute{911212}{Institut de Ciencies del Cosmos (ICC), Spain};
    \iinstitute{904536}{Atominstitut, Technische Universit\"at Wien, Austria}
}

%% file: Section_Intro/section.tex
\section{Introduction}

Muons can be accelerated in rings up to very high energies, without fundamental limitation from synchrotron radiation. The recently formed International Muon Collider Collaboration (IMCC)~\cite{IMCC} targets the design of muon colliders with a center of mass energy $E_{\rm{cm}}$ of 10~TeV or slightly more (10+~TeV), which seem feasible with technologies that can be made available in the near future. The highest $E_{\rm{cm}}$ muon colliders can reach, possibly subject to more radical advances in accelerator technologies, is not yet known and will be assessed.

The physics potential of 10+~TeV muon colliders has been investigated quite extensively over the past two years~\cite{Delahaye:2019omf,Han:2020uid,Costantini:2020stv,Buttazzo:2020uzc,AlAli:2021let,Aime:2022flm,Eichten:2013ckl,Chakrabarty:2014pja,Greco:2016izi,Buttazzo:2018qqp,DiLuzio:2018jwd,Ruhdorfer:2019utl,Capdevilla:2020qel,Han:2020pif,Han:2020uak,Bandyopadhyay:2020otm,Chiesa:2020awd,Dermisek:2020cod,Buttazzo:2020ibd,Yin:2020afe,Capdevilla:2021rwo,Huang:2021nkl,Han:2021lnp,Liu:2021jyc,Chen:2021rnl,Han:2021udl,Capdevilla:2021fmj,Huang:2021biu,Han:2021kes,Dermisek:2021mhi,Bottaro:2021srh,Asadi:2021gah,Bottaro:2021snn,Franceschini:2021aqd,Collamati:2021rbr,Sen:2021fha,Bandyopadhyay:2021pld,Dermisek:2021ajd,Qian:2021ihf,Chiesa:2021qpr,Liu:2021akf,Mastrapasqua:2021uvs,Zaza:2021nnq,Ruiz:2021tdt,Casarsa:2021rud,Capdevilla:2021kcf,deBlas:2022aow,Spor:2022mxl,Bao:2022onq,Forslund:2022xjq,Chen:2022msz,Homiller:123}.
While much is still to be done, the emerging picture~\cite{Aime:2022flm,Delahaye:2019omf,AlAli:2021let,Costantini:2020stv,Buttazzo:2020uzc} is that a 10+~TeV muon collider combines the advantages of proton and of $e^+e^-$ colliders, thanks to the large energy available for direct exploration and to the perspectives for precise measurements within the Standard Model (SM) and beyond. Furthermore the simultaneous availability of energy and precision offers unique opportunities for new physics discovery and characterization. All this at a single collider and on a feasible timescale. The extraordinary physics potential of a 10+~TeV muon collider unquestionably poses the urgency of investing in a complete design study~\cite{Aime:2022flm}.

On the other hand, strategic considerations suggest that a first stage of the muon collider with lower $E_{\rm{cm}}$ could facilitate and accelerate the development of the project. It is worth emphasizing in this context that muon colliders can be built in stages, in spite of being circular colliders. Indeed, the muon production and cooling complex can be used at all energies, and the muon acceleration proceeds through a sequence of rings, which can be reused at higher energy. The final collider ring of the lower energy collider can not be reused for the higher energy stage, but this might have a minor impact on the total cost. The advantage of a low $E_{\rm{cm}}$ first stage mainly stems from the significant reduction of the initial investment. This could give easier and faster access to the necessary financial resources. Furthermore the reduced energy target allows, if needed, to make compromises on technologies that might not yet be fully developed, avoiding potential delays.

When discussing the staging options it should be taken into account that lepton collisions at around 250~GeV can be more easily obtained with circular or linear $e^+e^-$ machines, and with a much higher luminosity than what muon colliders can achieve. So while there is evidently a compelling physics case for a leptonic 250~GeV ``Higgs factory'' at that energy, muon colliders are not the best option. Linear $e^+e^-$ colliders can also reach the TeV scale, up the 3~TeV energy of the last stage of the CLIC project~\cite{CLICdp:2018cto}. The luminosity attainable by a muon collider of 3~TeV is comparable to the one of CLIC. Therefore a muon collider with $E_{\rm{cm}}=3$~TeV, operating at the maximal energy for which an $e^+e^-$ machine has ever been designed, emerges as a natural first stage of the muon collider project.

An alternative for a first muon collider~\cite{Barger:1995hr,Barger:1996jm} is to operate it very close to Higgs pole, $E_{\rm{cm}}\hspace{-2pt}=\hspace{-2pt}m_H\hspace{-2pt}=\hspace{-2pt}125$~GeV, in order to study the lineshape of the Higgs particle. The larger Yukawa coupling of the muon offers in this case a competitive advantage to muon colliders relative to $e^+e^-$ machines at the same energy. However, the Higgs is a rather narrow particle, with a width over mass ratio $\Gamma_H/m_H$ as small as $3\cdot10^{-5}$. The muon beams would thus need a comparably small energy spread $\Delta E/E\hspace{-2pt}=\hspace{-2pt}3\cdot10^{-5}$ for the programme to succeed. Engineering such tiny energy spread might perhaps be possible, however it poses a challenge for the accelerator design that is peculiar of the Higgs pole collider and of no relevance for higher energies, where a much higher permille-level spread is perfectly adequate for physics. 

For this reason, the Higgs pole muon collider is currently not among the targets of the IMCC. Nevertheless in this document we review its physics potential, assuming the feasibility of the small beam energy spread $\Delta E/E\hspace{-2pt}=\hspace{-2pt}3\cdot10^{-5}$. We also assume a relatively large integrated luminosity, to be however collected in a short enough time not to delay the upgrade to higher energy. We also assume that the Beam Induced Backgrounds (BIB) from muon decays can be mitigated, while the BIB impact at the Higgs pole muon collider has never been studied and is expected to be more severe than at 3~TeV.

\newpage

The main goal of the present report is to review the physics potential of a 3~TeV muon collider, with 1~ab$^{-1}$ integrated luminosity if not otherwise specified. Results at 10+TeV energy are also described, occasionally, in order to outline the progression of the physics performances across the stages. The material is collected from different sources, including invited contributions that summarize, adapt and extend recent papers on muon collider physics. Some of these papers were initiated in preparation for this report, and part of the material results from dedicated work and appears here for the first time. 

The physics opportunities of a 3~TeV muon collider overlap in part with those of CLIC, extensively documented in Ref.~\cite{deBlas:2018mhx} and summarized in~\cite{deBlas:2019rxi,EuropeanStrategyforParticlePhysicsPreparatoryGroup:2019qin} in preparation for the 2020 update of the European Strategy for Particle Physics. There are, however, important differences between the two projects that need to be taken into account. 

First, the CLIC stage at 3~TeV is the last of a series of three, which include in particular a stage at 380~GeV that is quite effective for precise measurements of Higgs (and top) properties. The muon collider precision on the determination of the Higgs couplings should thus be reassessed and can not be inferred from CLIC results. 
Second, CLIC targets 5~ab$^{-1}$ luminosity at 3~TeV, while only 1~ab$^{-1}$ is currently foreseen for the 3~TeV muon collider in the baseline design target. This difference is partly compensated by the absence of beamstrahlung at the muon collider, which instead entails a significant reduction of the high-energy luminosity peak at CLIC. However, it can result in a significant degradation of the muon collider performances for those studies that do not rely very strongly on collisions at the highest energy. On the contrary, for studies that do require high energy collisions and that are not strongly sensitive to the integrated luminosity, like direct searches, CLIC sensitivity projections generically apply.

Third, muon colliders pose a novel challenge for detector design, due to the copious BIB from the decay products of the muons in the colliding beams. Since these challenges have never been encountered and addressed before, a design of the muon collider detector and an assessment of its performances is not yet available, unlike for CLIC. Promising preliminary results and directions for further progress, described in Ref.s~\cite{MuonCollider:2022ded,Jindariani:2022gxj}, suggest that reconstruction efficiencies and resolutions comparable to the one of the CLIC detector should be achievable, eventually. Most of the studies we present are based on these assumed performances, encapsulated in the muon collider Delphes card~\cite{deFavereau:2013fsa,delphes_card_mucol}.

Some results are instead obtained with the full simulation of a preliminary muon collider detector, under realistic BIB conditions. In particular, full simulation estimates of Higgs signal-strength measurements are described in Section~\ref{sec:Higgs} and compared with the estimates based on Delphes. Moreover, in Section~\ref{subsec:Stub-tracks} we review a search for disappearing tracks that successfully implements a BIB mitigation strategy for this challenging signal. 

The fourth and most obvious difference with CLIC is that the 3~TeV muon collider collides muons rather than electrons. Engineering muon anti-muon collisions for the first time is in itself a tremendous opportunity in the quest for generic exploration of new physics. Concretely, there are plenty of motivated scenarios where new physics couples more strongly to muons than to electrons. One of them might be waiting for a muon collider to be discovered. 

The current $g$-2 and $B$-physics anomalies offer additional motivations for muon-philic new physics scenarios, that result in several opportunities for the muon collider that are specific of muon collisions, to be reviewed in this document. Obviously, the anomalies could be resolved by new experiments and theoretical calculations in few years, before the muon collider is built. Alternatively, they could be strengthened and become a primary driver of particle physics research. In any case, they provide a concrete motivation to assess the physics potential of a multi-TeV muon colliders, which are found to offer excellent perspectives for progress on the muon anomalies already at 3~TeV, with a very competitive time scale. This further supports the urgency of investing now in a complete muon collider design study.

\newpage

\noindent{}This report illustrates the 3~TeV muon collider physics case under three different perspectives, along the lines described below. A concise summary of our key findings is provided in Section~\ref{sec:KF}.


\subsection*{Higgs and effective field theory}
\hfill {\emph{Editors: J.~de~Blas, P.~Meade and E.~Vryonidou}}

\noindent{}Muon colliders offer several opportunities to perform precise measurements of Standard Model (SM) processes and thus explore new physics Beyond the SM (BSM) indirectly. This is of particular interest for Higgs physics, which is crucial to learn about the microscopic origin of electroweak symmetry breaking. Models where this origin is explained, generically predict deviations from the SM of the Higgs interactions. The HL-LHC will probe Higgs couplings below few percent. A machine capable to improve the sensitivity to the permille or few permille level is needed to make substantial progress on this front.

Permille level Higgs couplings precision is the target of most of the currently proposed $e^+e^-$ Higgs factories, operating typically at hundreds of GeV energies in order to exploit the Higgs–strahlung $e^+e^-\to HZ$ production mechanism. As we will see, such precision or higher can also be obtained via the Higgs measurements at a 10~TeV muon collider. The performances are slightly inferior at the 3~TeV muon collider, but still typically superior to the most optimistic HL-LHC projections. Additionally, muon colliders can directly measure the Higgs trilinear coupling already at the 3~TeV stage, unlike $e^+e^-$ Higgs factories. These measurements are described in Section~\ref{sec:Higgs}, and used to estimate the projected reach in precision for single and triple Higgs interactions. We will also report sensitivity projection for a Higgs-pole muon collider.

Muon colliders also offer many other opportunities than Higgs measurements for indirect new physics exploration. In particular, they enable percent-level measurements of SM electroweak cross-sections at unprecedently high energies. Such high energy measurements give indirect access to new particles at mass $\Lambda\gg E_{\rm{cm}}$, by exploiting the fact that the effect of the these particles scale like $(E_{\rm{cm}}/\Lambda)^2$ relative to the SM. For a 3~TeV muon collider, $\Lambda\sim30$~TeV. For a 10~TeV muon collider, $\Lambda\sim100$~TeV. This is an unprecedented reach for new physics theories endowed with a reasonable flavor structure. Notice in passing that high-energy measurements are also useful to investigate flavor non-universal phenomena, as we will see in Sections~\ref{subsec:2-body-processes},~\ref{gm2},~\ref{ci}~and~\ref{cilfv}.

 In Section~\ref{sec:EFT} of this report we will employ high energy measurements projections, as well as Higgs measurements, for a global assessment of the indirect sensitivity to heavy new physics effects encapsulated in dimension-6 Effective Field Theory (EFT) interaction operators. These results will then be interpreted in terms of the sensitivity to concrete new physics scenarios, among which we include one where a Composite Higgs explains the origin and the scale of electroweak symmetry breaking.


\subsubsection*{Beyond the Standard Model}
\hfill {\emph{Editors: R.~Franceschini, F.~Meloni and S.~Su}}

A muon collider can make substantial progress in the exploration of the most basic BSM questions such as the structure of the Higgs sector and the origin of dark matter, already at the first stage with 3~TeV center of mass energy. 

In Section~\ref{sec:extendedHiggs} we review the muon colliders perspectives to probe extensions of the Higgs sector. In the simplest extension, discussed in Section~\ref{sec:singlet}, only one neutral extra scalar field is added to the SM and all the interactions of the new field proceed through a coupling with the Higgs field. This benchmark scenario exemplifies BSM physics coupled to the SM with ``Higgs portal'' type of interactions, as well as offering a simplified setup for models with a strong first order phase transition, and for the scalar sector of certain Supersymmetric scenarios. We will see that the muon collider can probe the Higgs plus Singlet model by direct production in Vector Boson Fusion (VBF), and also indirectly by single and trilinear Higgs coupling measurements. Extensions of the Higgs sector by a second doublet are considered in Section~\ref{sec:2HDM}. They can be probed at the muon collider by the pair production of the new heavy states, initiated by VBF or by direct muon anti-muon annihilation. A specific two Higgs doublet model, known as Inert Doublet Model, is investigated in Section~\ref{sec:IDM}. The interesting peculiarity of this model is that the lighter neutral BSM Higgs is a possible dark matter candidate, which the 3~TeV muon collider can probe.

Muon colliders can also explore the possibility that dark matter is a weakly interacting particle, with its abundance in the Universe emerging from the most basic and minimal incarnation of the thermal freeze-out mechanism. Among the most studied dark matter candidates there are general admixtures of weak doublets, triplets and uncharged weak singlets. This possibility arises in supersymmetric extensions of the SM. However, a truly minimal model of this class consists of the addition to the SM of a single multiplet of the electroweak group. This results in very sharply defined BSM scenarios, with sharp predictions on the dark matter mass and on its other properties. 

The possibility of probing such minimal dark matter candidates is discussed in Section~\ref{secdm}, by exploiting ``mono-$X$'' analysis that target the detection of SM particles produced in association with the dark matter (see Section~\ref{subsec:monoX}), or by exploiting the effects of the dark matter multiplets on the cross-section of SM processes at high energy. This latter exploration strategy extends its mass reach above the energy threshold for direct production. Finally, in Section~\ref{subsec:Stub-tracks}, we discuss the muon collider sensitivity to the direct production of the charged component of the dark matter multiplet through the detection of disappearing tracks. 

The higgsino (and wino) studies in Section~\ref{subsec:Stub-tracks} provide a first illustration of the muon collider sensitivity to ``unconventional'' manifestations of new physics, such as disappearing tracks and displaced vertices. This rapidly developing domain of investigation delivers at once new BSM targets for the muon collider and requires new ideas and techniques for the mitigation of the BIB effects on these challenging signals. See Section~\ref{secunc} for a discussion.

\subsubsection*{Muon-specific opportunities}
\hfill {\emph{Editors: D.~Buttazzo, R.~Capdevilla and D.~Curtin}}

\noindent{}Lepton flavour universality is not a fundamental property of Nature. It is therefore possible for new physics to exist which couples only or prominently to muons, which we could not discover using only electrons or protons. 
In fact, in many models it is generic for  new physics to couple more strongly to muons than to electrons. Even in the SM lepton flavour universality is violated maximally by the Yukawa interaction with the Higgs field, which is larger for muons than for electrons. New physics associated to the Higgs or to flavour will most likely follow the same pattern, offering a competitive advantage of muon over electron collisions at similar energies. The comparison with proton colliders is less straightforward. By the same consideration one expects larger couplings with second and third-generation quarks, but this has to be folded in with the much lower luminosity for heavier quarks at proton colliders than for muons at a muon collider. The perspectives of muon versus proton colliders are model-dependent and of course very sensitive to the energy of the muon and  proton collider, as we will see in several examples.


Hints of lepton flavour non-universality have been observed in the last decade. Intriguingly enough, most of these hints are observed in processes that involve muons. One is the muon anomalous magnetic moment ($g$-2), that shows an enduring discrepancy with the SM prediction. Another hint comes from $b\to s$ transitions, with several semi-leptonic and leptonic decay rates of $B$ mesons that show a difference between electron and muon final states, thus providing evidence of flavored new interactions that violate lepton flavor universality. 
Realistic models that account for the $B$-physics anomalies agree more with data if new physics couples more strongly to muons than to electrons. One of the reasons for this is because electron interactions have been probed way more extensively than the one of the muons.
Therefore the lepton flavour universality violation in $B$-physics is most likely due to new physics coupled to muons, while the $g$-2 anomaly is certainly due to new physics coupled to muons. Alternatively, the anomalies might not be due to new physics and will be resolved by more precise measurements and theoretical predictions in the next few years.

In Section~\ref{sec:g-2} we assess the sensitivity of a muon collider to probe the new physics that is potentially responsible of the muon $g$-2 anomaly, both from an effective field theory perspective and in the context of specific models. An exhaustive investigation of the possible BSM scenarios allows for the formulation of a no-lose theorem for the muon collider program, in the event that the BSM-nature of the $g$-2 anomaly is confirmed in the coming years.
The first muon collider stage at  3 TeV could probe indirectly several scenarios where new physics interacts mainly with the second generation of fermions. At the same time, all the models with TeV-scale new physics can be probed via direct production, which includes all BSM explanations involving only new singlets generating the new $(g-2)$ contributions.
Further indirect constraints on well-motivated models with heavy new physics come from Higgs physics. The remaining possible new physics interpretations of the muon $g$-2 will be accessible to muon beam dumps experiments, that can efficiently discover light new particles, and to muon colliders of higher energy. A 10~TeV muon collider can fully test new physics in semi-leptonic interactions, and all models that respect minimal flavor violation and do not create a fine tuning problem in the muon mass. Finally, the endgame of this program would a 30~TeV muon collider that can directly probe the dipole operator responsible for the anomalous magnetic moment, closing the window on any possible heavy new physics that might be responsible for the anomaly.

A similar assessment is performed in Section~\ref{sec:Banomaly} for the $B$-physics anomalies. In this case, a muon collider running at an energy of about 7~TeV has the opportunity to provide a complete no-lose theorem, being able to test indirectly the ``nightmare'' scenario where only the four-fermion interactions needed to explain the anomaly are present. This extreme scenario, although not truly motivated from a theoretical perspective, could not be tested by any other collider, including a 100 TeV hadron machine. If some realistic flavor structure is assumed instead, a no-lose theorem along the same lines can be achieved at the 3~TeV muon colliders. This indirect sensitivity is accompanied by a strong direct sensitivity to specific models. 

Finally, in sections \ref{sec:LFV}, \ref{sec:muon-higgs}, and \ref{sec:dark} we will study BSM scenarios that are unrelated with the muon anomalies, namely lepton flavor violation, Higgs physics and extended Higgs sectors, and weakly interacting dark sectors. All these studies focus on scenarios where new physics communicates with the Standard Model through the muon portal, where muon colliders have a clear advantage over any other type of collider.

%% file: Section_Higgs_EFT/section.tex

\newcommand{\cmrule}{\midrule[0.25mm]}
\newcommand{\crowcolor}{\rowcolor[rgb]{0.9,0.9,0.9}}
\newcommand{\ctoprule}{\toprule[0.5mm]}
\newcommand{\cbottomrule}{\bottomrule[0.5mm]}

\newcommand{\crowcolorA}{\rowcolor[rgb]{0.9,0.9,0.9}}

\newcommand{\Dslash}{\cancel{D}~}
\newcommand{\iDslash}{\!~i\!~\cancel{D}\!~}
\renewcommand{\d}{\partial}
\newcommand{\lrD}{~\!\overset{\leftrightarrow}{\hspace{-0.1cm}D}\!}
\newcommand{\lrDa}{~\!\overset{\leftrightarrow}{\hspace{-0.1cm}D}\!^{\!~a}}
\newcommand{\hc}{\mathrm{h.c.}}

\newcommand{\jb}[1]{}
\newcommand{\PM}[1]


\def\thefootnote{\fnsymbol{footnote}}
\setcounter{footnote}{0}


    %
    %

\section{Higgs physics at muon colliders \label{sec:Higgs}}

\begin{figure}[ht]
\centering
\includegraphics[width=0.5\linewidth]{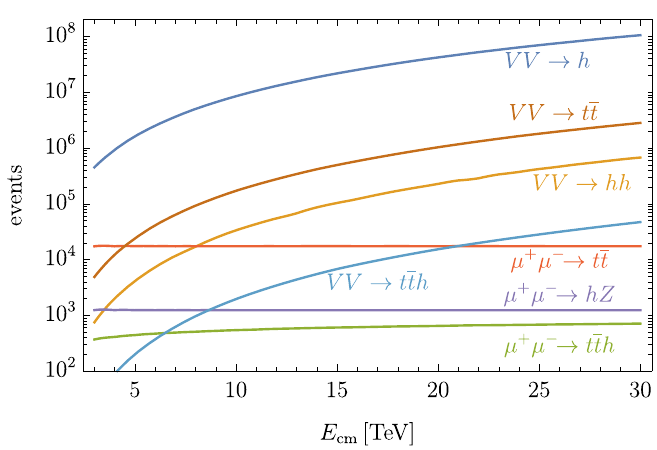}
\caption{\label{fig:xs}
Expected number of events for different processes at a muon collider, as a function of the centre-of-mass energy, for integrated luminosities $L=10~\!{\rm ab}^{-1}(E_{\rm cm} [{\rm TeV}]/10~\!{\rm ~\!TeV})^2$. 
}
\end{figure}

At high-energy muon colliders, as the virtual electroweak gauge boson content of the muon beam becomes relevant, vector boson fusion (VBF) becomes the most important channel for production of SM particles. This is illustrated in Figure~\ref{fig:xs}, which shows how the growth with the energy of VBF Higgs production clearly outmatches the usual higgsstrahlung process dominant at low-energy $e^+ e^-$ Higgs factories.    
An initial estimate for the precision that would be possible for Higgs measurements via W boson fusion (WBF, $\mu^+ \mu^- \to H \overline{\nu} \nu$) and $Z$ boson fusion (ZBF, $\mu^+ \mu^- \to H \mu^+ \mu^-$)  has been recently presented in \cite{AlAli:2021let}. These were obtained including fast detector simulation but they neglect backgrounds, both physics as well as the beam-induced ones. The latter are however suppressed given that the nominal detector only extends to  $|\eta|<2.5$, allowing for a potential tungsten plug to suppress the beam backgrounds within 10 degrees of the beam, as suggested by 1.5 TeV muon collider studies~\cite{Mokhov:2011zzd}.

The original Higgs precision estimates of~\cite{AlAli:2021let} are currently being extended to include the effects of physics backgrounds~\cite{Forslund:2022xjq}, and to characterize the effects of beam in beam backgrounds through the first full simulation studies outlined in~\cite{Jindariani:2022gxj}.  At this point there is only one channel that is available for comparison across full and fast simulation, $H\rightarrow b\bar{b}$, as outlined in section 6 of~\cite{Jindariani:2022gxj}.  One of the main differences between the DELPHES fast simulation and the full simulation results of~\cite{Jindariani:2022gxj} is a reduced precision for jet energy resolution. However, as shown in~\cite{Forslund:2022xjq}, changing the resolution to match the results of full simulation including beam in beam backgrounds does not appreciably alter the precision in the $H\rightarrow b\bar{b}$ channel.  Additionally the results shown in~\cite{Jindariani:2022gxj}, are only the preliminary attempts at showing physics performance.  Therefore the rest of the single Higgs precision numbers used in this section are based on the original DELPHES muon collider detector fast simulation~\cite{deFavereau:2013fsa,delphes_card_mucol}. An additional point to note for Higgs precision at a high energy muon collider, to distinguish between WBF and ZBF, one must be able to tag the forward muons beyond $|\eta|\approx 2.5$.  This capability is included in fast simulation, but further study is needed.  

The projected sensitivities for the main Higgs decays in single H production are estimated at the few percent level at 3 TeV with 1 ab$^{-1}$, whereas at 10 TeV with 10 ab$^{-1}$, sensitivities at the permille level would be possible for the main decay channels ($b\bar{b}$, $WW^*$).
While the 3 TeV numbers could be considered comparable to the HL-LHC, the use of different production mechanisms makes both machines quite complementary, as we will see in the Higgs coupling interpretation presented below.   For example the precision of $y_t$ at a muon collider from $t\bar{t}H$ at 3 and 10 TeV of $35\%$ and $53\%$~\cite{Forslund:2022xjq} is significantly below that of the LHC.  However, there may be additional observables beyond single Higgs precision, such as shown in \cite{AlAli:2021let}, where the process of $W^+W^-\rightarrow t\bar{t}$ was used  to infer an estimated precision on $y_t$ at 10 TeV with 10 ab$^{-1}$, similar to HL-LHC projections. Top physics is particular challenging at a very high energy muon collider and therefore further study is clearly needed.

Relatively high rates are also accesible to high-energy colliders for multi-Higgs processes via VBF production. 
This is particularly the case for $\mu^+ \mu^- \to HH \bar{\nu}\nu$ at 10 TeV 10 ab$^{-1}$, where a total of $3\cdot 10^4$ $HH$ events would be produced. These could
be used to obtain a determination of the triple Higgs coupling $\lambda_3$.
Assuming the uncertainties associated to single Higgs couplings are kept under control by single Higgs processes (see below), 
in Table~\ref{t:H3_sensitivity} we collect the expected precisions for the exclusive determination of the trilinear Higgs coupling, expressed in terms of $\kappa_\lambda\equiv \lambda_3/\lambda_3^{\rm SM}$, obtained using the likelihood from the recent study in  Ref.~\cite{Buttazzo:2020uzc}. This uses the information from the differential distribution in $M_{HH}$ in $\mu^+ \mu^- \to HH \bar{\nu}\nu$. (See also Ref.\cite{Han:2020pif}, which reports similar results.) As can be seen, the trilinear Higgs coupling could be determined at 68\% probability at 3 TeV with a precision of $\sim 15-30\%$ (depending on the luminosity), but still better than the projected error from the HL-LHC of $\sim 50\%$. We also note that the presence of a second minimum in the $\kappa_\lambda$ log-likelihood ``deforms'' the expected 68\% probability intervals with respect to the standard 1$\sigma$ bounds, valid for a Gaussian distribution, which would suggest a more precise result of $\approx 18\%$. The influence of this second minimum could be easily alleviated by an increase in luminosity by roughly a factor of two. This would bring a similar improvement in the bounds, as opposed to the expected $\sqrt{2}$ reduction in the size of the interval, and also single out the solution around $\delta \kappa_\lambda=0$ at 68$\%$ probability, yielding a precision for $\kappa_\lambda$ of 15$\%$. 
On the other hand the higher energy and luminosity of the 10 TeV options would bring a determination at the $\sim 4\%$ level precision, better than CLIC at 3 TeV, and comparable to what would be possible at a 100 TeV hadron collider~\cite{FCC:2018byv}. For comparison, we also report the projected sensitivities at even higher centre-of-mass energies, 14 and 30 TeV, where a one percent level determination could be possible.

\begin{table}[t]
  \centering
    \caption{68$\%$ probability sensitivity to the $H^3$ coupling at a muon collider at different energies. Derived using the likelihood from the study in \cite{Buttazzo:2020uzc}. (We note that the likelihood at 3 TeV is non-Gaussian, with a second minimum for $\delta \kappa_\lambda>0$, so the $68\%$ probability interval is quite different from the 1-$\sigma$ limits computed with reference to the mode of the distribution $\delta \kappa_\lambda^{1\sigma,~3{\rm TeV}}\in [-0.17,0.18]$. See text for details.)
    }
    \label{t:H3_sensitivity}
    {
      \begin{tabular}{c c c c c} 
        \ctoprule
         &
        3 TeV $\mu$-coll. &

        10 TeV $\mu$-coll. &

        14 TeV $\mu$-coll. &

        30 TeV $\mu$-coll. 
        \\
         &
        L$\approx$ 1 ab$^{-1}$ &
        L= 10 ab$^{-1}$ &
        L$\approx$ 20 ab$^{-1}$ &
        L= 90 ab$^{-1}$\\
        &
        \multicolumn{4}{c}{68\% prob. interval }\\
        \cmrule
        $\delta \kappa_\lambda$ &
          [-0.27,0.35]~$\cup$~[0.85,0.94]& 
          
          [-0.035, 0.037] &
          
          [-0.024, 0.025] & 
          
          [-0.011, 0.012] \\ [0.1cm]
          
          &
          $\to$[-0.15,0.16] {\small (2$\times$ L)} & 
          
           & & \\ [0.1cm]
        \cbottomrule
      \end{tabular}
    }
\end{table}

Beyond double Higgs production, a multi-TeV $\mu^+ \mu^-$ collider could use triple-Higgs production to gain sensitivity to the quartic Higgs coupling, $\lambda_4$, as recently explored in Ref.~\cite{Chiesa:2020awd}. The cubic and quartic Higgs interactions are related in the SM and extensions where electroweak symmetry is linearly realized (described at low energies by a SMEFT Lagrangian). If this is not the case, new physics could modify $\lambda_3$ and $\lambda_4$ independently. The quartic coupling is directly tested at leading order via, e.g. $\mu^+ \mu^- \to HHH\bar{\nu}\nu$, which has a cross section of 0.31 (4.18) ab at $\sqrt{s}=3$ (10) TeV~\cite{Chiesa:2020awd}. For realistic luminosities, this makes a 3 TeV option unable to probe the quartic coupling, but this could be tested at 10 TeV to a precision of tens of percent with integrated luminosities of several tens of ab.

Finally, we comment on the possibility of operating at significantly lower energies as a first stage before a high energy muon collider. In particular, one could operate around the Higgs pole $\sqrt{s}=125$ TeV, which also brings the question of what would be the physics benefits of performing $s$-channel Higgs measurements. Indeed, unlike other collider options, a $\sqrt{s}=125$ GeV $\mu^+\mu^-$ collider could perform on-shell Higgs physics directly via $\mu^+ \mu^- \to H$ production which, in particular, brings the opportunity of a direct model-independent measurement of the Higgs width \cite{Han:2012rb} (as opposed to, e.g. $e^+ e^-$ Higgs factories where this could be determined indirectly, by exploiting the measurement of the inclusive $ZH$ cross section in combination with all the other exclusive rates). With a resonant $\mu^+ \mu^- \to H$ cross section of 70 pb, reduced to about 22 pb when taking into account a beam energy spread $R=0.003\%$ together with the effects of initial state radiation~\cite{Greco:2016izi,Franceschini:2021aqd}, a luminosity at the level of several fb$^{-1}$ would yield order $10^{5}$ Higgses, limiting a priori the statistical reach in terms of precision Higgs physics compared to the Higgs factory runs at the different future colliders that have been proposed, where an order of magnitude larger number of Higgs events is expected. The direct measurement of the width at the percent level can partially compensate this loss in terms of pure statistics, though, as it directly normalizes all rates, whereas the normalization at other future $H$ factories comes from a direct measurement of a particular coupling. The expected precision in different channels, together with an optimized study of the determination of the Higgs lineshape from a threshold scan have been recently studied in \cite{deBlas:2022aow}. (See also \cite{Conway:2013lca}.) This includes the main physics backgrounds but ignores the beam-induced ones which, as in the high-energy case, are simply suppressed by a ten degree cut around the beam. 

In what follows we interpret the available projections for single Higgs processes at muon colliders in terms of sensitivity to modifications of the Higgs boson couplings, to illustrate the expected improvements at the different stages, and compared to the knowledge that will be available at the end of the LHC era. 


\subsection{Higgs coupling precision}

To illustrate the potential of the muon collider in measuring the properties of the Higgs boson, we perform here a series of fits to the single Higgs couplings in the so-called $\kappa$ framework~\cite{LHCHiggsCrossSectionWorkingGroup:2012nn,LHCHiggsCrossSectionWorkingGroup:2013rie},\footnote{The fits presented in this section have been performed using the {\tt HEPfit} code~\cite{DeBlas:2019ehy}.} where the cross sections, decomposed as follows
\begin{equation}
( \sigma \cdot \text{BR} ) ( i \to \text{H }\to f ) =  \frac{ \sigma_{i} \cdot \Gamma_{f}}{\Gamma_H},
\label{eq:kappa_zw}
\end{equation}
are parameterized in terms of scaling parameters $\kappa$,
\begin{equation}
( \sigma \cdot \text{BR} ) ( i \to \text{H }\to f ) =  \frac{ \sigma_{i}^{\text{SM}} \kappa_i^2 \cdot \Gamma_{f}^{\text{SM}}\kappa_f^2 }{\Gamma_H^{\text{SM}} \kappa_H^2} 
    = \frac{\kappa_i^2\cdot\kappa_f^2}{\kappa_H^2} \quad ,
\end{equation}
and where we will assume, for the purposes of this section that the Higgs boson decays only into SM final states, i.e. 
$\kappa_{H}^2 \equiv \sum_{j} \kappa_j^2 \Gamma_{j}^\text{SM} /
\Gamma_{H}^\text{SM}\,.
$
Note that the muon collider option operating at 125 GeV offers the possibility of a model-independent measurement of the Higgs width, allowing to close a fit where the Higgs width is a free parameter, $\Gamma_H = \Gamma_{H}^\text{SM} \cdot \kappa_H^2 / (1-\text{BR}_{\rm new})$. A comparison of the results at the different machines releasing the constraint that $\Gamma_H$ contains only SM channels is thus not possible, and here we will restrict our fits to the case $\text{BR}_{\rm new}=0$. As in \cite{deBlas:2019rxi}, we will also assume in the fits that all intrinsic SM theory uncertainties are under control by the time any of these future colliders are built~\cite{Blondel:2019qlh,Freitas:2019bre}.

\begin{table}[t]
  \centering
    \caption{68$\%$ probability sensitivity to modifications on the Higgs coupling from the $\kappa$ fit, assuming no BSM contributions to the Higgs width.  \jb{PRELIMINARY}
    \label{t:Kappa_sensitivity}}
    {\scriptsize
      \begin{tabular}{c|c|c|cc|c} 
        \ctoprule
        &
        HL-LHC &
        HL-LHC &
        HL-LHC &
        HL-LHC &
        HL-LHC  \\
         &
         &
        + 125 GeV $\mu$-coll.  &
        
        + 3 TeV $\mu$-coll. &

        + 10 TeV $\mu$-coll. &
        
        + 10 TeV $\mu$-coll.
        \\
         &
        &
        5 / 20 fb$^{-1}$&
        1 ab$^{-1}$ &
        10 ab$^{-1}$&
         + $e^+e^-$ $H$ fact\\
        Coupling &
        &
        &
        &
        &
        (240/365 GeV)\\
        
        \cmrule
        $\kappa_W~[\%]$ &
          1.7   & 
          
          1.3 / 0.9  & 

          0.4  & 
          
          0.1  & 
          
          0.1    \\ [0.1cm] 
        $\kappa_Z~[\%]$ &
          1.5 & 
          
          1.3 / 1.0 & 

          0.9 & 
          
          0.4 & 
            
          0.1   \\ [0.1cm] 
        $\kappa_g~[\%]$ &
          2.3 & 
          
          1.7 / 1.4 & 

          1.4  & 
          
          0.7  & 
          
          0.6   \\ [0.1cm] 
        $\kappa_\gamma~[\%]$ &
          1.9  & 
          
          1.6 / 1.5 & 

          1.3  & 
          
          0.8  & 
          
          0.8   \\ [0.1cm] 
        $\kappa_{Z\gamma}~[\%]$ &
          10  & 
          
          10 / 10 & 

          9.9  & 
          
          7.2 & 
          
          7.1 \\ [0.1cm] 
        $\kappa_c~[\%]$ &
          -& 
          
          12 / 5.9 & 

          7.4  & 
          
          2.3  & 
          
          1.1  \\ [0.1cm] 

        $\kappa_b~[\%]$ &
          3.6  & 
          
          1.6 / 1.0 & 

          0.9  & 
          
          0.4  & 
          
          0.4  \\ [0.1cm] 
        $\kappa_\mu~[\%]$ &
          4.6  & 
          
          0.6 / 0.3 & 

          4.3  & 
          
          3.4  & 
          
          3.2  \\ [0.1cm] 
        $\kappa_\tau~[\%]$ &
          1.9  & 
          
          1.4 / 1.2 & 

          1.2  & 
          
          0.6  & 
          
          0.4  \\ [0.1cm] 
          \cmrule
        $\kappa_t^\dagger~[\%]$ &
          3.3  & 
          
          3.2 / 3.1 & 

          3.1  & 
          
          3.1  & 
          
          3.1 \\ [0.1cm] 
          \cmrule
          $\Gamma_H^\ddag~[\%]$ &
          5.3  & 
          
          2.7 / 1.7 & 

          1.5  & 
          
          0.5 & 
          
          0.4 \\ [0.1cm]
          \cmrule
          \multicolumn{6}{l}{ {\scriptsize $^\dagger$ No input used for $\mu$ collider.}}\\
          \multicolumn{6}{l}{ {\scriptsize $^\ddag$ Prediction assuming only SM Higgs decay channels. Not a free parameter in the fits.}}\\
        \cbottomrule
      \end{tabular}
    }
\end{table}
\begin{figure}[h!]
\centering
\includegraphics[width=0.875\linewidth]{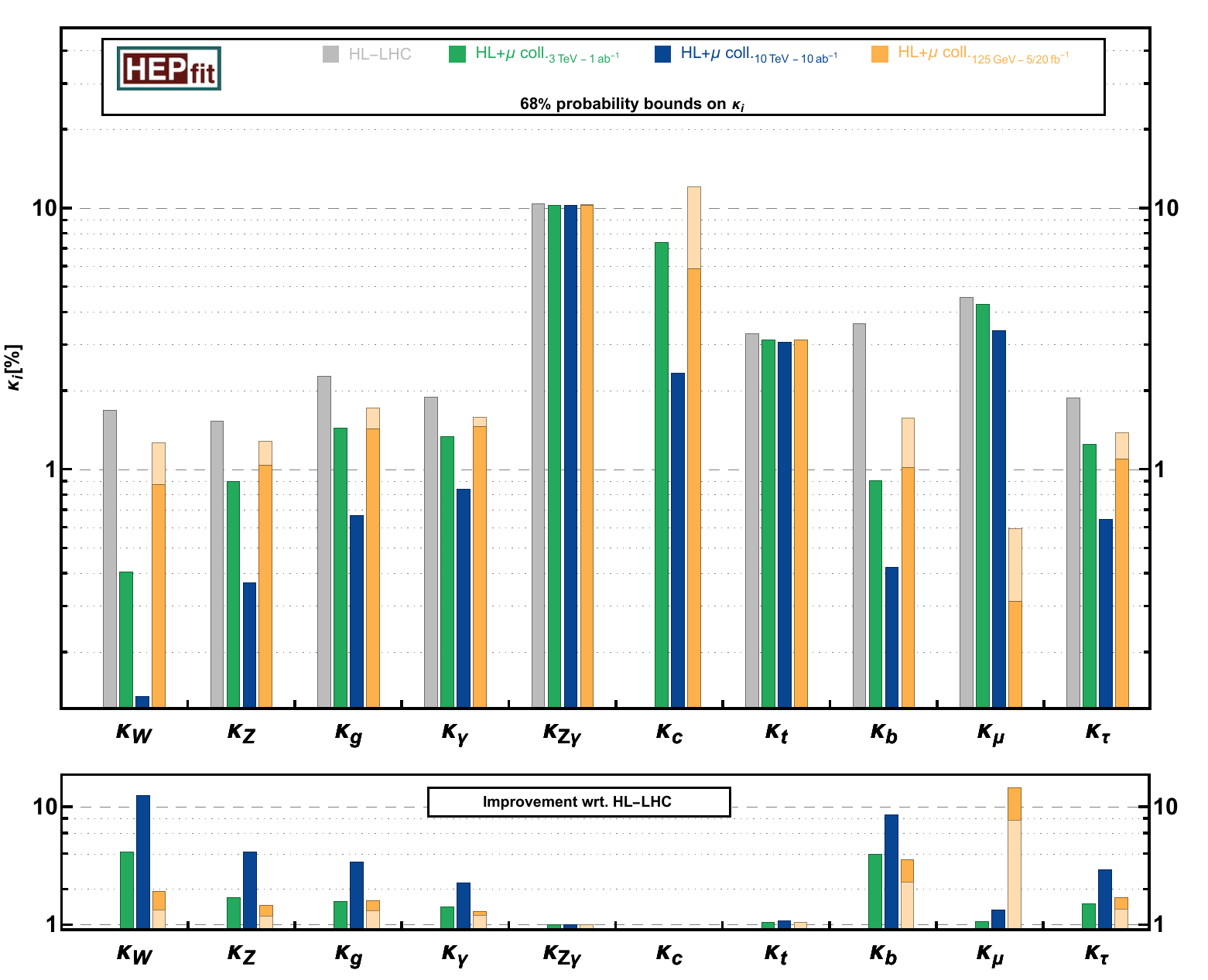}
\caption{\label{fig:kappa}
Sensitivity to modified Higgs couplings in the $\kappa$ framework. We show the marginalized 68\% probability reach for each coupling modifier. For the 125 GeV muon collider, light (dark) shades correspond to a luminosity of 5 (20) fb$^{-1}$.  
}
\end{figure}

The results for the fits at different colliders are presented in Table~\ref{t:Kappa_sensitivity} and illustrated in Figure~\ref{fig:kappa}. We compare with the expected precision at the HL-LHC\footnote{We use the same inputs as in \cite{deBlas:2019rxi}, with the exception of the channels $H \to $invisible. We use the S2 projections for systematic uncertainties, as explained in~\cite{Cepeda:2019klc}.}~\cite{Cepeda:2019klc}, which is also combined with the projections at the different variants of the muon colliders, to show the impact of the muon collider measurements in the knowledge of the different coupling modifiers. In Table~\ref{t:Kappa_sensitivity} we also include the results in combination with the results of a future $e^+e^-$ Higgs factory, using as reference the Higgs precision expected at the FCC-ee~\cite{FCC:2018byv,FCC:2018evy}. 
From the results, it is clear that any incarnation of a muon collider with the considered settings would be able to bring a significant improvement in the knowledge of several of the Higgs couplings with respect to the HL-LHC. This is particularly true for the multi-TeV options for the couplings to vector bosons $Z, W$, where subpercent precision could be achieved, reaching the permille level for the 10 TeV option. Comparatively, the precision of the same couplings for the 125 GeV option is somewhat worse. The main gain from the 125 GeV setup, apart from the measurement of the Higgs width which cannot be truly appreciated in these constrained fits, would be a subpercent determination of the muon Yukawa coupling. For any other coupling, it typically underperforms compared to the 3 TeV results, unless high luminosities are collected. (We show the 125 GeV results for both 5 and 20 fb$^{-1}$, from \cite{deBlas:2022aow}.) 
It is also worth noting the complementarity with future $e^+ e^-$ factories, in particular with the 10 TeV option. Given the different main modes of electroweak production of the Higgs at each facility ($ZH$ at $e^+e^-$ and WBF at a multi-TeV $\mu^+ \mu^-$), each is more sensitive to either $\kappa_Z$ (for $e^+ e^-$) or $\kappa_W$ (for $\mu^+ \mu^-$) and in combination are able to bring both to a precision of one permille (or even below if one assumes custodial symmetry relations).

Finally, and as explained above, we should remind that the estimate for $\kappa_{t}$ presented here might not be representative of the physics potential of the muon colliders for this coupling, and additional study of processes sensitive to the Top Yukawa is needed.


\section{Effective Field Theory interpretations \label{sec:EFT}}

In this section we present a global interpretation of the projections for different types of measurements at a high-energy muon collider in terms of an effective field theory constructed assuming any new degrees of freedom are much heavier than the electroweak scale and that at low energies the particles and symmetries are those of the SM, i.e. the so called SM Effective Field Theory (SMEFT). While a full study in terms of the general SMEFT truncated at the dimension-6 level is not possible with the available set of projections for physics processes at a muon collider, a reasonably global fit can be closed when combining that information with the expected information that will be available by the end of the HL-LHC era, plus making a series of extra assumptions about new physics. 
In particular, following what was done as part of the 2020 European Strategy Group studies \cite{EuropeanStrategyforParticlePhysicsPreparatoryGroup:2019qin,deBlas:2019rxi}, 
we adopt the following dimension-6 EFT Lagrangian~\cite{Giudice:2007fh}:
\begin{equation}
\begin{split}
{\cal L}_{\mathrm{SILH}}=&
\frac{c_\phi }{\Lambda^2} \frac 12 \partial_\mu (\phi^\dagger \phi)\partial^\mu (\phi^\dagger \phi)+ 
\frac{c_T}{\Lambda^2} \frac 12(\phi^\dagger \lrD_\mu \phi)(\phi^\dagger \lrD^\mu \phi) - 
\frac{c_6}{\Lambda^2} \lambda (\phi^\dagger \phi)^3\\
&
+\sum_f\left(\frac{c_{y_f} }{\Lambda^2} y^f_{ij} \phi^\dagger \phi  \bar{\psi}_{Li} \phi \psi_{Rj} + \hc \right)\\
&
+\frac{c_W}{\Lambda^2} \frac{ig}{2}\left(\phi^\dagger \lrDa_\mu \phi\right) D_\nu W^{a~\!\mu\nu} +\frac{c_B}{\Lambda^2} \frac{ig^\prime}{2}\left(\phi^\dagger \lrD_\mu \phi\right) \partial_\nu B^{\mu\nu}\\
&
+\frac{c_{\phi W}}{\Lambda^2} i g D_\mu\phi^\dagger \sigma_a D_\nu \phi W^{a~\!\mu\nu}+\frac{c_{\phi B} }{\Lambda^2} i g^\prime D_\mu\phi^\dagger \sigma_a D_\nu \phi B^{\mu\nu}\\
&+\frac{c_{\gamma} }{\Lambda^2} g^{\prime~\!2}  \phi^\dagger \phi B^{\mu\nu}B_{\mu\nu}+\frac{c_{g}}{\Lambda^2} g^{2}_s  \phi^\dagger \phi G^{A~\!\mu\nu}G^A_{\mu\nu}\\
&-\frac{c_{2W}}{\Lambda^2} \frac{g^2}{2} (D^\mu W_{\mu\nu}^a)(D_\rho W^{a~\!\rho\nu})-\frac{c_{2B}}{\Lambda^2} \frac{g^{\prime~\!2}}{2}(\partial^\mu B_{\mu\nu})(\partial_\rho B^{\rho\nu})\\
&+\frac{c_{3W}}{ \Lambda^2}g^3\varepsilon_{abc}W_{\mu}^{a~\nu}W_\nu^{b~\rho}W_\rho^{c~\mu}. 
\end{split}
\label{eq:LSilh}
\end{equation}
%
While this just contains a subset of the operators of the more general dimension-six SMEFT, the operators in (\ref{eq:LSilh}) are of special relevance for several BSM types of scenarios. For the purpose of this chapter we will focus, in particular,  in the case of the Universal Composite Higgs scenarios and $U(1)$ extensions of the SM. 

In the EFT fits to Eq.~(\ref{eq:LSilh}) we include the following set of experimental inputs and projections:
\begin{itemize}
{\item The complete set of electroweak precision measurements from LEP/SLD~\cite{ALEPH:2005ab}, including the projected measurements of the W mass at the HL-LHC~\cite{Azzi:2019yne}. We also include the aTGC constraints from LEP2.}
{\item The HL-LHC projections for single Higgs signal strengths and double Higgs production from \cite{Cepeda:2019klc}. We assume the S2 scenario for the projected experimental and theory systematics.}
{\item Also from the HL-LHC, the projections from two-to-two fermion processes, expressed in terms of the $W$ and $Y$ oblique parameters, from Ref.\cite{Farina:2016rws}, and the high energy diboson study from~\cite{Franceschini:2017xkh}.}
{\item The expected precision for single-Higgs observables at the 3 and 10 TeV muon colliders from the results of \cite{Forslund:2022xjq}.}
{\item As in the HL-LHC case, we also include the projections from high-energy measurements in two-to-two fermion processes, expressed in terms of $W$ and $Y$ from \cite{Chen:2022msz}, and in diboson processes $\mu^+\mu^-\to ZH, W^+W^-$, $\mu \nu \to WH,WZ$ from Ref.~\cite{Chen:2022msz,Buttazzo:2020uzc}.}
{\item The expected precision for the Higgs self-coupling from the measurement of the di-Higgs invariant mass in $\mu^+ \mu^- \to \bar{\nu}\nu HH$ from Ref.~\cite{Buttazzo:2020uzc}. (See also \cite{Han:2020pif}.)}
\end{itemize}
In all cases we assume the projected experimental measurements to be centered around the SM prediction. 
The assumptions in terms of theory uncertainties follow the same setup as in \cite{deBlas:2019rxi}.

\begin{table}[t]
\caption{\label{tab:eft-silh} 
68\% probability reach on the different Wilson coefficients in the Lagrangian Eq.~(\ref{eq:LSilh}) from the global fit.
In parenthesis we give the corresponding results from a fit assuming only one operator is generated by the UV physics. 
}
\centering
{\footnotesize

\begin{tabular}{c c}

\begin{tabular}{ c | c | c | c | }
\toprule
        &       &\multicolumn{2}{c|}{HL-LHC + $\mu$ collider}  \\
      &HL-LHC   & 3 TeV   & 10 TeV    \\
      &   & (1 ab$^{-1}$)   & (10 ab$^{-1}$)    \\
\midrule
$\frac{c_{\phi}}{\Lambda^2}[\mbox{TeV}^{-2}]$   & $0.52$   & $0.12$   & $0.039$  \\
   & $(0.28)^\dagger$   & $(0.11)$   & $(0.029)$  \\
$\frac{c_{T}}{\Lambda^2}[\mbox{TeV}^{-2}]$   & $0.0056$   & $0.0022$   & $0.0019$  \\
   & $(0.0019)$   & $(0.0019)$   & $(0.0019)$  \\
$\frac{c_{W}}{\Lambda^2}[\mbox{TeV}^{-2}]$   & $0.32$   & $0.0095$   & $0.0011$  \\
   & $(0.021)$   & $(0.0033)$   & $(0.00031)$  \\
$\frac{c_{B}}{\Lambda^2}[\mbox{TeV}^{-2}]$   & $0.33$   & $0.022$   & $0.0022$  \\
   & $(0.026)$   & $(0.0075)$   & $(0.00065)$  \\
$\frac{c_{\phi W}}{\Lambda^2}[\mbox{TeV}^{-2}]$   & $0.31$   & $0.034$   & $0.026$  \\
   & $(0.033)$   & $(0.031)$   & $(0.019)$  \\
$\frac{c_{\phi B}}{\Lambda^2}[\mbox{TeV}^{-2}]$   & $0.32$   & $0.18$   & $0.13$  \\
   & $(0.19)$   & $(0.18)$   & $(0.13)$  \\
$\frac{c_{\gamma}}{\Lambda^2}[\mbox{TeV}^{-2}]$   & $0.0054$   & $0.0047$   & $0.0031$  \\
   & $(0.0041)$   & $(0.0039)$   & $(0.0027)$  \\
$\frac{c_{g}}{\Lambda^2}[\mbox{TeV}^{-2}]$   & $0.0012$   & $0.0011$   & $0.0007$  \\
   & $(0.00052)$   & $(0.00042)$   & $(0.00022)$  \\
\bottomrule
\end{tabular}

&

\begin{tabular}{ c | c | c | c | }
\toprule
        &       &\multicolumn{2}{c|}{HL-LHC + $\mu$ collider}  \\
      &HL-LHC   & 3 TeV   & 10 TeV    \\
      &   & (1 ab$^{-1}$)   & (10 ab$^{-1}$)    \\
\midrule
$\frac{c_{y_{e}}}{\Lambda^2}[\mbox{TeV}^{-2}]$   & $0.25$   & $0.2$   & $0.1$  \\
   & $(0.2)$   & $(0.18)$   & $(0.096)$  \\
$\frac{c_{y_{u}}}{\Lambda^2}[\mbox{TeV}^{-2}]$   & $0.57$   & $0.48$   & $0.3$  \\
   & $(0.24)$   & $(0.19)$   & $(0.089)$  \\
$\frac{c_{y_{d}}}{\Lambda^2}[\mbox{TeV}^{-2}]$   & $0.46$   & $0.15$   & $0.068$  \\
   & $(0.25)$   & $(0.12)$   & $(0.062)$  \\
$\frac{c_{2B}}{\Lambda^2}[\mbox{TeV}^{-2}]$   & $0.087$   & $0.0036$   & $0.00031$  \\
   & $(0.075)$   & $(0.0029)$   & $(0.00026)$  \\
$\frac{c_{2W}}{\Lambda^2}[\mbox{TeV}^{-2}]$   & $0.0087$   & $0.00097$   & $0.000084$  \\
   & $(0.0076)$   & $(0.00078)$   & $(0.00007)$  \\
$\frac{c_{3W}}{\Lambda^2}[\mbox{TeV}^{-2}]$   & $1.7$   & $1.7$   & $1.7$  \\
   & $(1.7)$   & $(1.7)$   & $(1.7)$  \\
$\frac{c_{6}}{\Lambda^2}[\mbox{TeV}^{-2}]$   & $8.4$   & $4.6$   & $0.65$  \\
   & $(7.8)$   & $(4.4)$   & $(0.6)$  \\
\bottomrule 
\multicolumn{4}{c}{}\\[0.35cm]
\end{tabular}
\end{tabular}\\
{ 
\footnotesize 
$^\dagger$ As explained in \cite{deBlas:2019rxi}, due to the treatment of systematics/theory uncertainties in the HL-LHC inputs, this number must be taken with caution, as it would correspond to an effect below the dominant theory uncertainties. A more conservative estimate accounting for 100$\%$ correlated theory errors would give $c_\phi/\Lambda^2\sim 0.42$ TeV$^{-2}$.
}
}
\end{table}

\begin{figure}[ht]
\centering
\includegraphics[width=\linewidth]{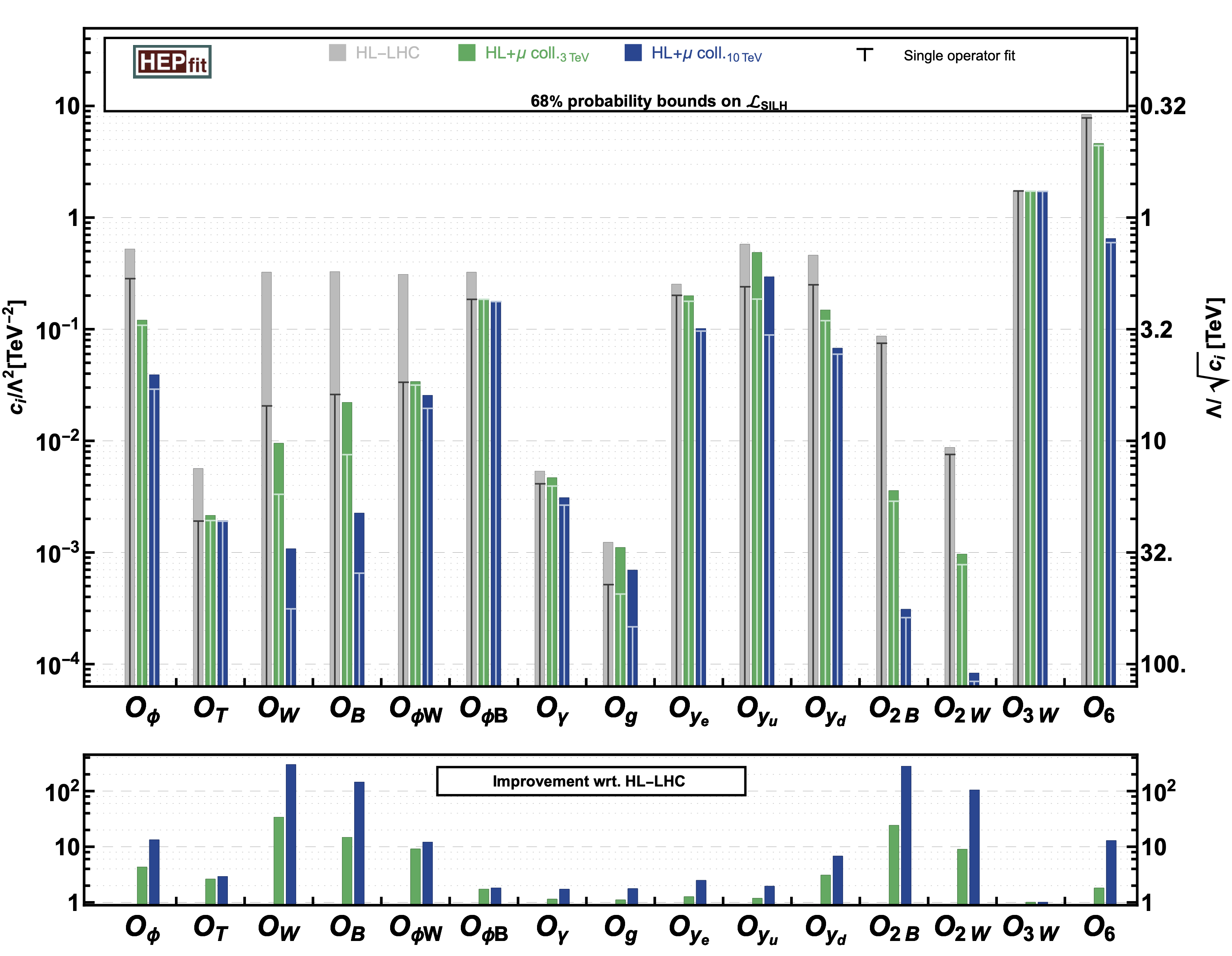}
\caption{\label{fig:SILH}
Global fit to the EFT operators in the Lagrangian (\ref{eq:LSilh}). We show the marginalized 68\% probability reach for each Wilson coefficient $c_i/\Lambda^2$ in Eq.~(\ref{eq:LSilh}) from the global fit (solid bars). The reach of the vertical ``T'' lines indicate the results assuming only the corresponding operator is generated by the new physics.
\jb{UPDATED}
}
\end{figure}

The results of these EFT fits are summarized in Table~\ref{tab:eft-silh} and Figure~\ref{fig:SILH}. We also include in the table and figure the projections obtained from the HL-LHC measurements (also included in the $\mu^+\mu^-$ collider results), to show the improvement in the reach for the different operators, shown explicitly in the lower panel of Figure~\ref{fig:SILH}. This is clear for ${\cal O}_\phi$ due to the increase in precision in the knowledge of the $HZZ$ and $HWW$ interactions and, in particular, for the operators ${\cal O}_{W,B}$ and ${\cal O}_{2B, 2W}$, which induce growing with energy effects in diboson and difermion processes, respectively, and thus benefit from the high energy reach of the 3 and 10 TeV muon colliders. As in the $\kappa$ analysis, we must also note that the improvements in other operators, e.g. ${\cal O}_{y_u}$ which modifies the Top Yukawa coupling, might not represent a fair assessment of the muon collider potential, due to the absence of detailed projections for the processes that would impose the leading constraints on them. For the Top Yukawa this means that not only $ttH$, but, in fact, a combination of different Higgs and Top-quark measurements ~\cite{Franceschini:2021aqd} may be needed to be able to determine the ultimate constraining power on new physics modifying $y_t$. Finally, it should be noted that all projections included here 
correspond to the case where the muon collider beams are unpolarized. The presence of polarization could bring extra information, i.e. allow to test extra directions in the SMEFT parameter space, as it basically doubles the number of observables, e.g. solving flat directions that appear in unpolarized observables due to cancellations (see e.g.~\cite{Barklow:2017suo,DeBlas:2019qco}). In particular, as explained in \cite{Buttazzo:2020uzc}, it would benefit the reach of the ${\cal O}_{W,B}$ operators from the diboson high-energy measurements.

\subsection{Interpretation in terms of BSM benchmark scenarios}

\begin{figure}[t]
\centering
\begin{tabular}{c c}
\includegraphics[width=0.47\linewidth]{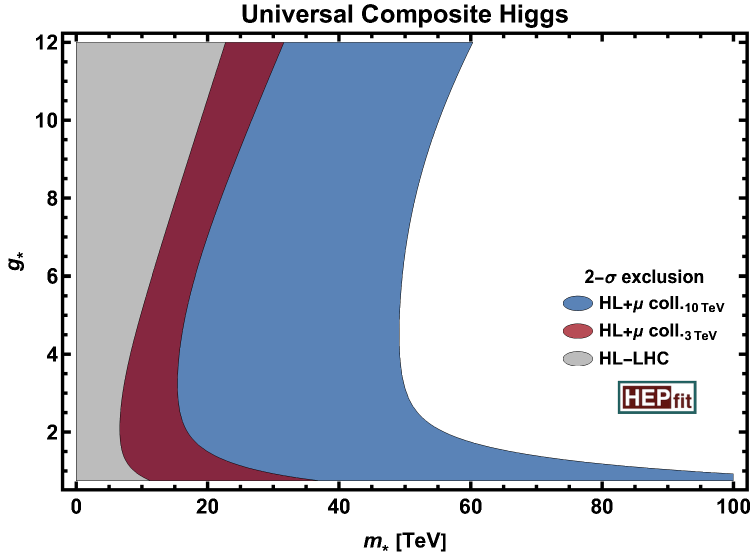}
&
\includegraphics[width=0.47\linewidth]{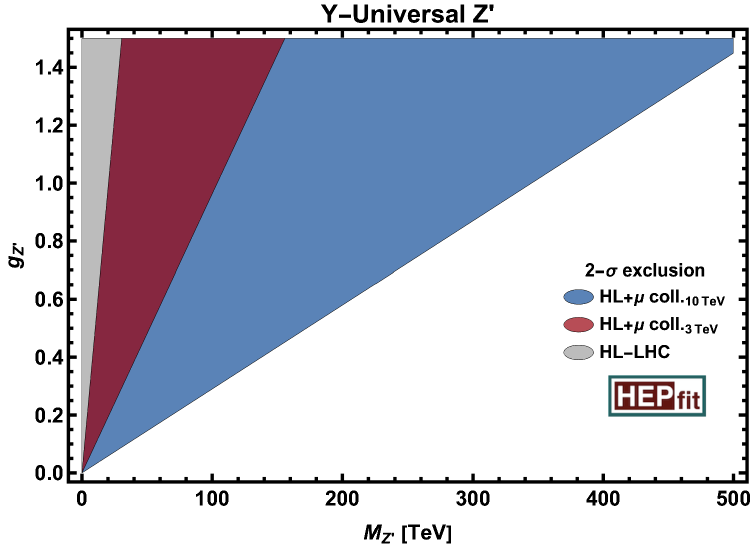}\\
\end{tabular}
\caption{\label{fig:CH_Zp}
(Left) Comparison of the global reach for universal composite Higgs models at the HL-LHC and a high-energy muon collider (combined with the HL-LHC constraints).
The figure compares the 2-$\sigma$ exclusion regions in the $(g_\star, m_\star)$ plane from the fit presented in Figure~\ref{fig:SILH}, using the SILH power-counting described in Eq.~(\ref{eq:SILHpc})
(Right) The same for a BSM extension with a massive replica of the $U(1)_Y$ gauge boson in the $(g_{Z^\prime}, m_{Z^\prime})$ plane from the fit presented in Figure~\ref{fig:SILH}. 
}
\end{figure}

For the case of composite Higgs scenarios we assume the new dynamics is parameterized in terms of a single coupling, $g_\star$, and mass, $m_\star$. As in \cite{deBlas:2019rxi}, we use the following illustrative assumptions for the power counting and contributions of the new physics to the different Wilson coefficients in (\ref{eq:LSilh}):
%
%
%
%
%
%
%
\begin{equation}
\begin{split}
\frac{c_{\phi,6,y_f}}{\Lambda^2}&= \frac{g_\star^2}{ m_\star^2},~~~~~~~~~~~~
\frac{c_{W,B}}{\Lambda^2}= \frac{1}{m_\star^2},
~~~~~~~~~~~
\frac{c_{2W,2B}}{\Lambda^2}= \frac{1}{g_\star^2}\frac{1}{m_\star^2},
\\
\frac{c_{T}}{\Lambda^2}&= \frac{y_t^4}{16\pi^2}\frac{1}{m_\star^2},~~~~
\frac{c_{\gamma,g}}{\Lambda^2}= \frac{y_t^2}{16\pi^2}\frac{1}{m_\star^2},~~~~~~
\frac{c_{\phi W,\phi B}}{\Lambda^2}= \frac{g_\star^2}{16\pi^2}\frac{1}{m_\star^2},~~~~
\frac{c_{3W}}{\Lambda^2}= \frac{1}{16\pi^2}\frac{1}{m_\star^2}.
%
\end{split}
\label{eq:SILHpc}
\end{equation}
and projecting the EFT likelihood onto the $(g_\star,m_\star)$ plane we obtain the exclusion regions in the right panel in Figure~\ref{fig:CH_Zp} for the different muon collider options, combined and in comparison with the HL-LHC reach. We also show the results interpreted in terms of extra vector bosons, using as a representative example the case of a universal $Z^\prime$ coupling to the hypercharge current, also considered in \cite{EuropeanStrategyforParticlePhysicsPreparatoryGroup:2019qin}. In this case the dimension-6 effective Lagrangian only receives tree-level contributions to the operator with coefficient $c_{2B}/\Lambda^2=(g_{Z^\prime}^2/g^{\prime~\!4})/ M_{Z^\prime}^2$. The corresponding indirect constraints in the $(g_{Z^\prime},M_{Z^\prime})$ plane are shown in the left panel of Figure~\ref{fig:CH_Zp}. 

\begin{figure}[h!]
\centering
\begin{tabular}{c c}
\includegraphics[width=0.47\linewidth]{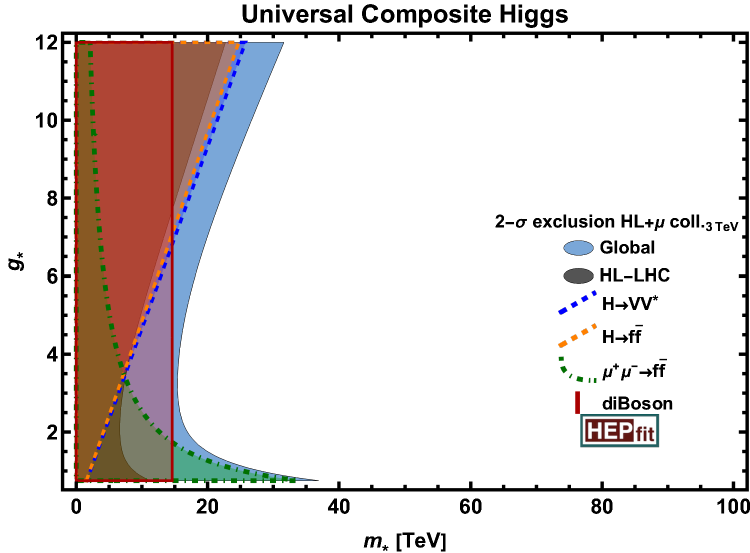}
&
\includegraphics[width=0.47\linewidth]{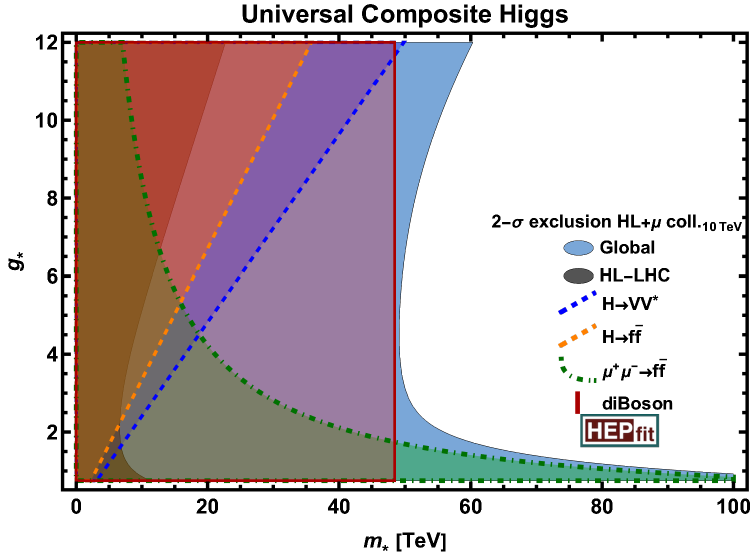}\\
\end{tabular}
\caption{\label{fig:SILH_gstar}
(Left) 2-$\sigma$ exclusion regions in the $(g_\star, m_\star)$ plane from the fit presented in Figure~\ref{fig:SILH}, using the SILH power-counting described in Eq.~(\ref{eq:SILHpc}) and below (solid regions). 
The solid and dashed lines denote the contributions to the constraints from different processes.
The results correspond to the combination of the HL-LHC with the 3 TeV muon collider.
(Right) The same for the 10 TeV muon collider. 
}
\end{figure}

Whereas the bounds on the $Z^\prime$ example considered here are going to be clearly dominated by the high-energy measurements of  $\mu^+ \mu^- \to f \bar{f}$ and the induced constraints on the $Y$ parameter, the situation is more complex for the case of a composite Higgs scenario. The contributions from the different processes in setting the limits are shown separately in Figure~\ref{fig:SILH_gstar}. This highlights the complementarity of the different processes, with the diboson constraints setting the overall mass reach indepedendtly of $g_\star$, extended for low (high) values of $g_\star$ by the difermion (Higgs) bounds.
Going back to Figure~\ref{fig:CH_Zp}, it is clear that, while the 3 TeV option would clearly outperform the HL-LHC, the real leap in terms of indirect sensitivity comes with the 10 TeV option, thanks to the significantly higher energy reach, which boosts the constraining power of difermion and diboson processes on $W,Y$ and $C_{B,W}$, respectively.


%% file: Section_BSM/section.tex
\newcommand\iab{\rm{ab}^{-1}}
\newcommand{\rF}[1]{\textcolor{cyan}{#1}}
\newcommand{\sS}[1]{\textcolor{orange}{#1}}
\newcommand{\fM}[1]{\textcolor{purple}{#1}}
\newcommand*{\pt}{\ensuremath{p_{\text{T}}}\xspace}
\renewcommand*{\pT}{\ensuremath{p_{\text{T}}}\xspace}
\newcommand*{\ninoone}{\ensuremath{\tilde{\chi}^{0}_{1}}\xspace}
\newcommand*{\ninotwo}{\ensuremath{\tilde{\chi}^{0}_{2}}\xspace}
\newcommand*{\chipm}{\ensuremath{\tilde{\chi}^{\pm}}\xspace}
\newcommand*{\chimp}{\ensuremath{\tilde{\chi}^{\mp}}\xspace}
\newcommand*{\SRot}{SR\ensuremath{_{1t}}\xspace}
\newcommand*{\SRtt}{SR\ensuremath{_{2t}}\xspace}
\newcommand*{\SRotp}{SR\ensuremath{^{\gamma}_{1t}}\xspace}
\newcommand*{\SRttp}{SR\ensuremath{^{\gamma}_{2t}}\xspace}
\newcommand*{\GEANT}{\textsc{Geant}\xspace}


\section{Extended Higgs Sectors}
\label{sec:extendedHiggs}

\subsection{SM plus a singlet extension}
\label{sec:singlet}
\begin{figure}[ht]
\begin{centering}
\includegraphics[width=0.67\linewidth]{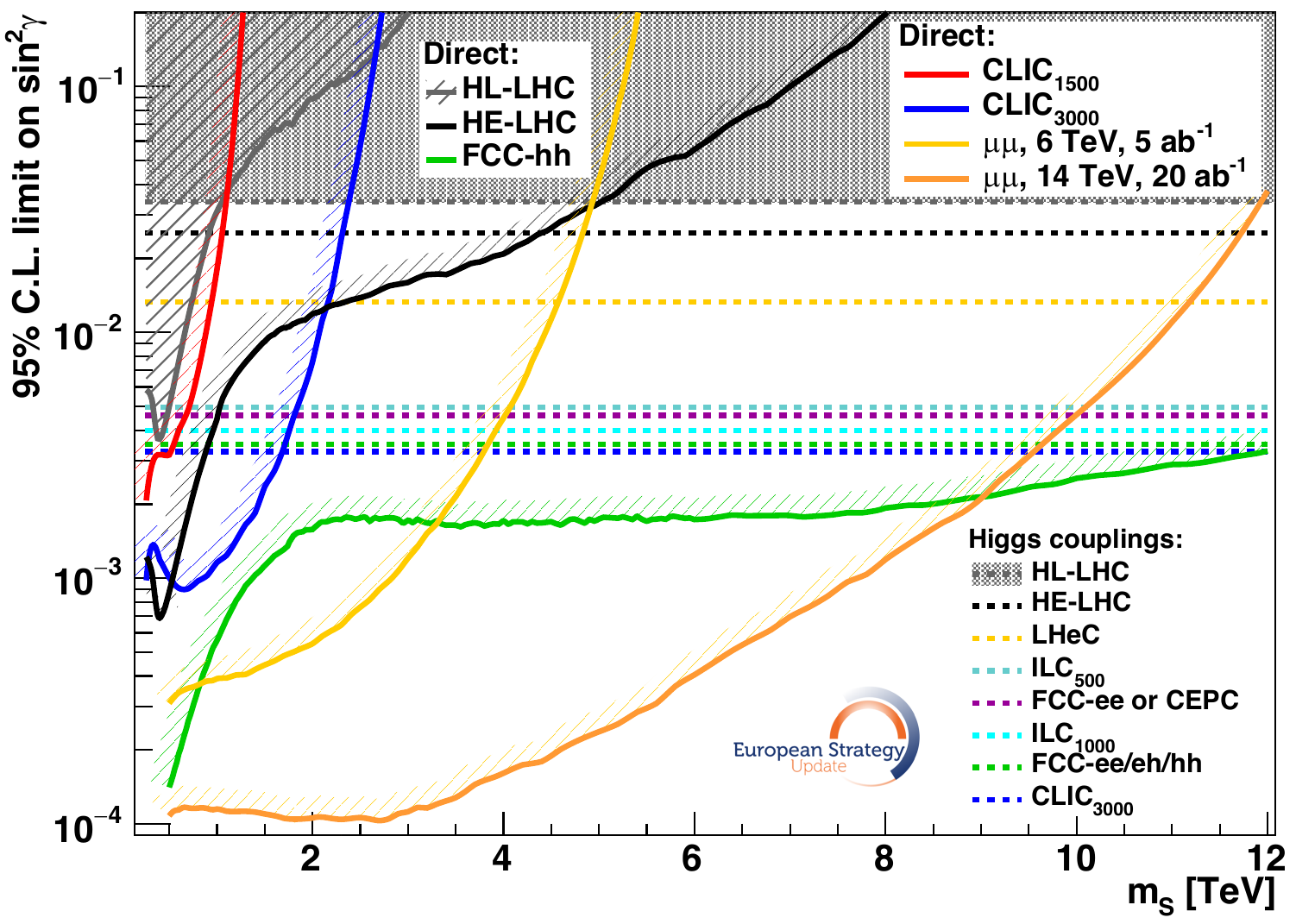}

\caption{
Direct 95~\% 
C.L. reach on heavy singlet mixed with the 
SM Higgs doublet at various muon colliders (adapted from~\cite{Buttazzo:2018qqp}). The direct and indirect reach at other future 
colliders~\cite{EuropeanStrategyforParticlePhysicsPreparatoryGroup:2019qin} is also shown for comparison.\label{fig:Singlet-reach}}
\par\end{centering}
\end{figure}

The simplest extension of the SM Higgs sector is the SM Higgs sector plus an extra real singlet. In the case when the extra singlet mixes with the SM Higgs doublet with mixing parameter $\sin\gamma$, the SM-like Higgs couplings are modified. Through the mixing,  the heavy scalar $S$ can be singly produced and can  decay to a pair of SM gauge bosons or SM-like Higgs bosons.  Considering the Vector Boson fusion production $VV \rightarrow S$, the most sensitive channel at a high energy lepton collider is $S \rightarrow hh \rightarrow 4b$~\cite{Buttazzo:2018qqp}.  {The 95\% C.L. exclusion reach for 
a 3 TeV muon collider with 1 ${\rm ab}^{-1}$ luminosity is shown in Fig.~\ref{fig:Singlet-reach} 
as blue solid curve, which is better than the direct reach of HL-LHC once $\sin^2\gamma<0.1$.} 
Comparing to the sensitivity of indirect measurements of the SM-like Higgs couplings, 
the 3 TeV collider can test new resonances down to mixing angles correlated to a deviation in the Higgs couplings of about 0.1\%.  The sensitivity in $\sin^2\gamma$ is better than that of the Higgs precision measurements at future Higgs factories, which are indicated by the dashed horizontal line in the plot, for $m_S \lesssim  1$ TeV. 
Higher energy muon colliders have better reach in both $\sin^2\gamma$ and $m_S$, surpassing that of Higgs precision measurements for $m_S <  4 (11) $ TeV for 6(14) TeV center-of-mass energy. In the same plot, we also show the direct reach at a 100 TeV hadron collider for comparison: a muon collider of 6 TeV or more has a better reach in the relevant part of parameter space.

\begin{figure}
\centering{}\includegraphics[width=0.47\linewidth]{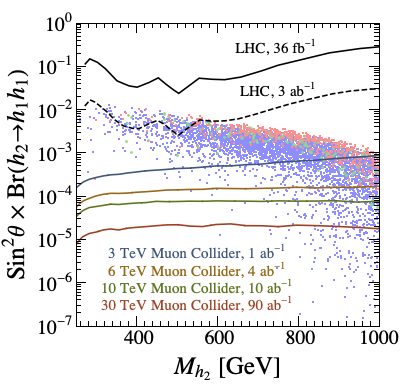}
\includegraphics[width=0.47\linewidth]{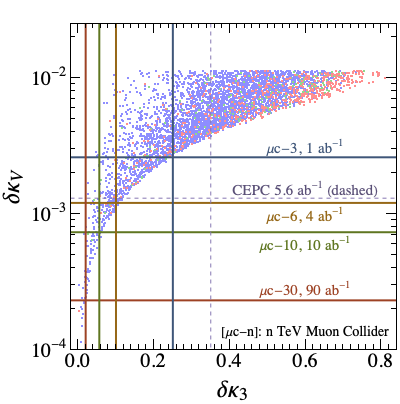}
\caption{\label{fig:EWpt-reach}Direct (left panel) and indirect reach (right panel) on  the SM plus real scalar singlet scenario for muon colliders with various center of mass energy.  Dots indicate points with successful FOEWPT, while red, green and blue dots represent signal-to-noise ratio (SNR) for gravitational eave detection of $[50, +\infty)$, $[10, 50)$ and $[0, 10)$, respectively. Results are taken from \cite{Liu:2021jyc}.
 }\end{figure}

SM plus a real singlet extension can also provide a strong first order electroweak phase transition (FOEWPT), which is essential for the electroweak baryogenesis mechanism to explain the observed cosmological matter-antimatter asymmetry~\cite{Liu:2021jyc, Ruhdorfer:2019utl}.  In the left panel of Fig.~\ref{fig:EWpt-reach}, the colored solid  curves show the muon collider 95\% C.L. exclusion reach for VBF production with di-Higgs decay modes and 4$b$ final states.  A 3 TeV muon collider (1 ${\rm ab}^{-1}$) has a sensitivity more than one order of magnitude better than the HL-LHC (13 TeV, 3 ${\rm ab}^{-1}$).  It also covers most of the points that generate a strong FOEWPT, which are indicated by the dots.  Comparing to the reach of future Gravitational Wave experiment LISA (red and green points), majority of those points falls with the 3 TeV muon collider reach.   Furthermore, the muon colliders also have significant sensitivity to the blue data points which are beyond the reach of the LISA.  Higher energy muon collider can extend the reach further.  The reaches in the SM-like Higgs coupling measurements on $\delta \kappa_3$ and $\delta \kappa_V$ are shown in the right panel of Fig.~\ref{fig:EWpt-reach} for muon collider with various center of mass energy as well as the CEPC option for a Higgs factory, that would be directly sensitive only to $\delta \kappa_V$.  We find that the reach of the 3 TeV muon collider is slightly worse than that of the Higgs factory for $\delta \kappa_V$. However, the reach for muon collider with higher center of mass energy surpasses that of the Higgs factory, plus it can be complemented by the information on the Higgs boson trilinear coupling, offering a better handle to scrutiny any hint of new physics.

\subsection{Two Higgs Doublet Model} 
\label{sec:2HDM}
\begin{figure}[ht]
\centering{}\includegraphics[width=0.45\linewidth]{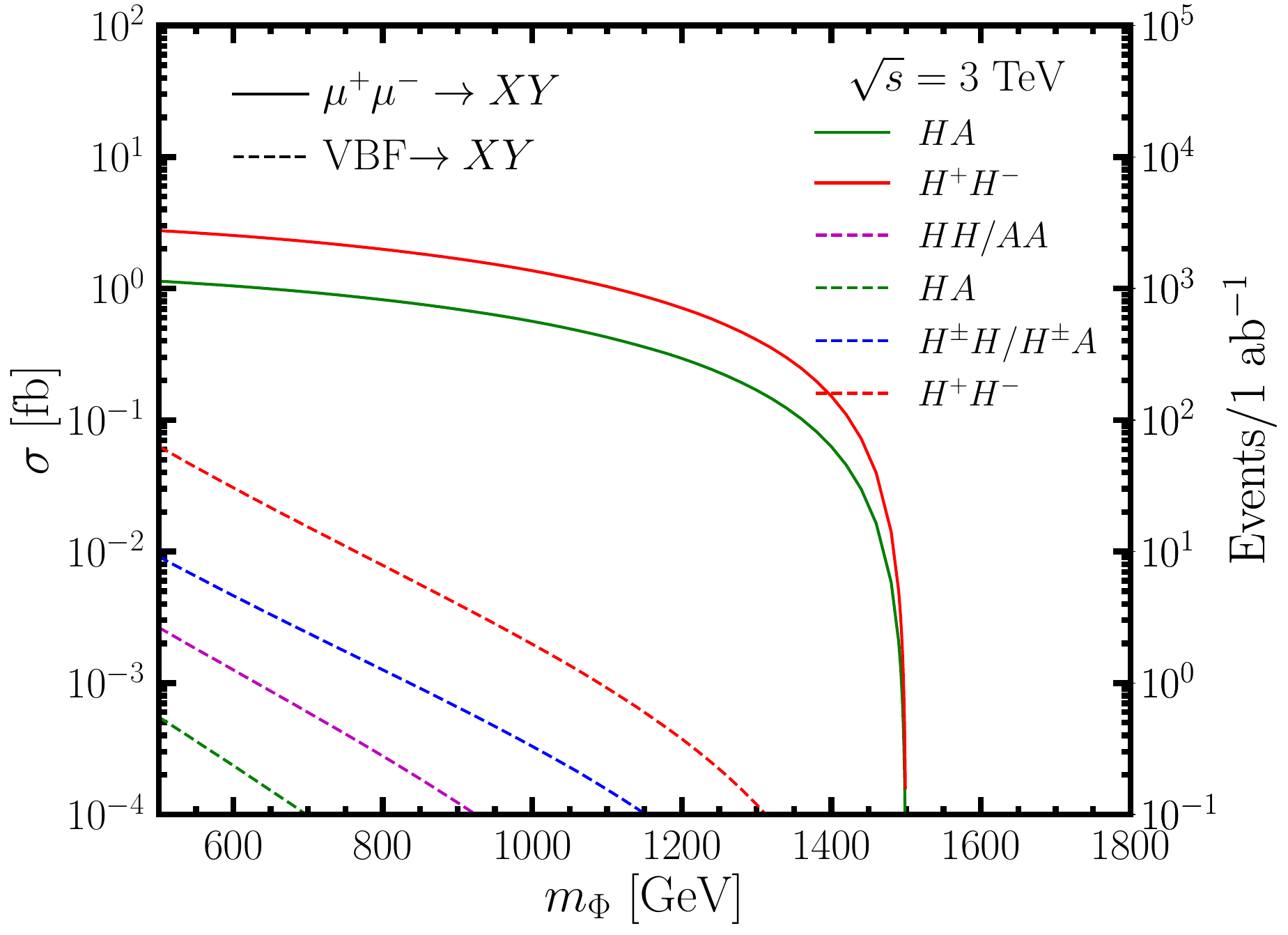}
\includegraphics[width=0.45\linewidth]{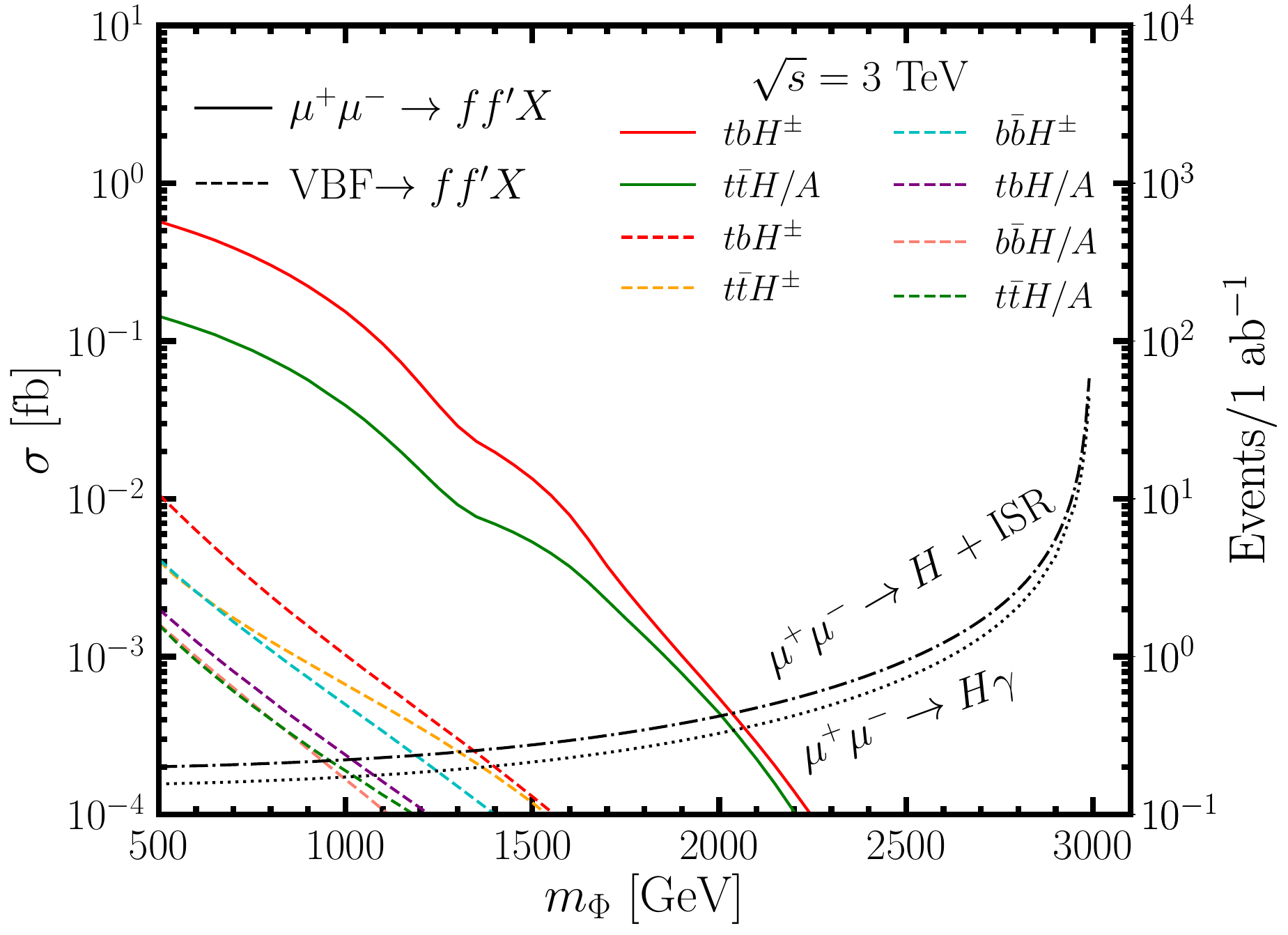}
\caption{\label{fig:Charged-reach}
Cross sections versus the  non-SM Higgs mass for $\sqrt{s}=3$ TeV for pair production (left panel), single production with a pair of fermions and radiative return production (right panel) for $\tan\beta=1$ under the alignment limit of $\cos(\alpha-\beta)=0$.  Plot is produced by authors of Ref.~\cite{Han:2021udl}.}
\end{figure}

In the framework of Two Higgs Doublet Model (2HDM) ~\cite{Branco:2011iw}, the scalar sector consists of 5 physical scalars:  the SM-like Higgs $h$, and the non-SM ones $H,A,H^\pm$ with  $m_h=125$ GeV  after the electroweak symmetry breaking.  The tree-level couplings of Higgs bosons are determined by two parameters: the mixing angle between the neutral CP-even Higgs bosons $\alpha$ and $\tan\beta=v_2/v_1$, with $v_{1,2}$ being the vacuum expectation value for two Higgs doublets.  The un-suppressed gauge couplings of the Higgses with the SM gauge bosons typically involve two non-SM Higgses, for example, $ZHA$ or $W^\pm H^\mp H$.  The Yukawa couplings of the non-SM like Higgses with the SM fermions depends on how the two Higgs  doublets  are  coupled  to  the  leptons  and  quarks, giving rise to four different types of 2HDMs, namely Type-I, Type-II, Type-L and Type-F. 

Once crossing the pair production threshold, the heavy Higgs bosons can be produced in pair via the $\mu^+\mu^-$ annihilation as well as Vector Boson Fusion (VBF): 
\begin{eqnarray}
\mu^+\mu^-\to \gamma^*,Z^*\to H^+ H^-,\quad
\mu^+\mu^- \to Z^* \to HA, && \\
\mu^+\mu^-\to V_1 V_2\ \mu^+(\bar\nu)\mu^-(\nu),\  V_1V_2\to H^+H^-, HA, H^\pm H/H^\pm A, HH/AA, &&
\label{eq:anni}
\end{eqnarray}
The production cross section as a function of the non-SM like Higgs masses for $\sqrt{s}=3$ TeV for various channels are shown in the left panel of Fig.~\ref{fig:Charged-reach} under the alignment limit of $\cos(\alpha-\beta)=0$.  The annihilation processes dominate at  $\sqrt{s}=3$ TeV.   For higher center of mass energies, VBF channels become more and more important~\cite{Han:2021udl}, especially for light scalar masses.   The annihilation process can be separated from the VBF process by comparing the invariant mass distribution of the Higgs pair, which is approximately equal to the collider c.m.~energy $m_{\Phi_1\Phi_2}\approx \sqrt{s}$ for the direct annihilation process, while peaked near the threshold $m_{\Phi_1\Phi_2}\approx m_{\Phi_1}+m_{\Phi_2}$ for the VBF process.  Considering the dominant decay channel of non-SM Higgs into third generation fermions, the SM backgrounds can be sufficiently suppressed.   Reach up to pair production threshold is possible at all $\tan\beta$ region, when all four fermion final states channels are used.  Comparing with HL-LHC reach for Type-II 2HDM, 3 TeV muon collider reach exceeds that of the HL-LHC~\cite{Craig:2016ygr}, except for very small value of $\tan\beta<2$ above the pair production mass threshold.

In the parameter region with enhanced Higgs Yukawa couplings or beyond the Higgs pair production threshold, single production of non-SM Higgs with a pair of fermions could play an important role.  
The production cross section for fermion associated production are shown in the right panel of Fig.~\ref{fig:Charged-reach} for both the annihilation and VBF processes, with $\tan\beta=1$ and $\cos(\alpha-\beta)=0$.  The dominant channel is $tbH^\pm$, followed by $t\bar{t}H/A$.       Note that there are strong $\tan\beta$ dependence on the production cross section, depending on the types of 2HDM~\cite{Han:2021udl}.   

Radiative return $\mu^+\mu^- \to \gamma H$ offers another production channel for the non-SM Higgs, especially in regions with enhanced $H\mu^+\mu^-$ coupling.  The cross section increases as the heavy Higgs mass approaches the collider c.m.~energy, closer to the $s$-channel resonant production. The production cross section is shown as the black curves in the right panel of Fig.~\ref{fig:Charged-reach}. 

In summary, non-SM Higgses can be copiously produced at 3 TeV muon collider.  For pair production,  95\% C.L. exclusion reaches in the Higgs mass up to the production   threshold of $\sqrt{s}/2$ are possible when channels with different final states are combined.  Including single production modes can extend the reach further.  With the combination of both the production mechanisms and decay patterns,  we found that the intermediate and large $\tan\beta$ values offer great discrimination power to separate Type-I and Type-L from Type-II/F.    To further identify either Type-II or Type-F, we need to study the subdominant channels with $\tau$ final states, which could be sizable in the signal rate in Type-II~\cite{Han:2021udl}.

\subsection{ Inert Doublet Model}
 \label{sec:IDM}
 
\begin{figure}
\centering{}\includegraphics[width=0.47\linewidth]{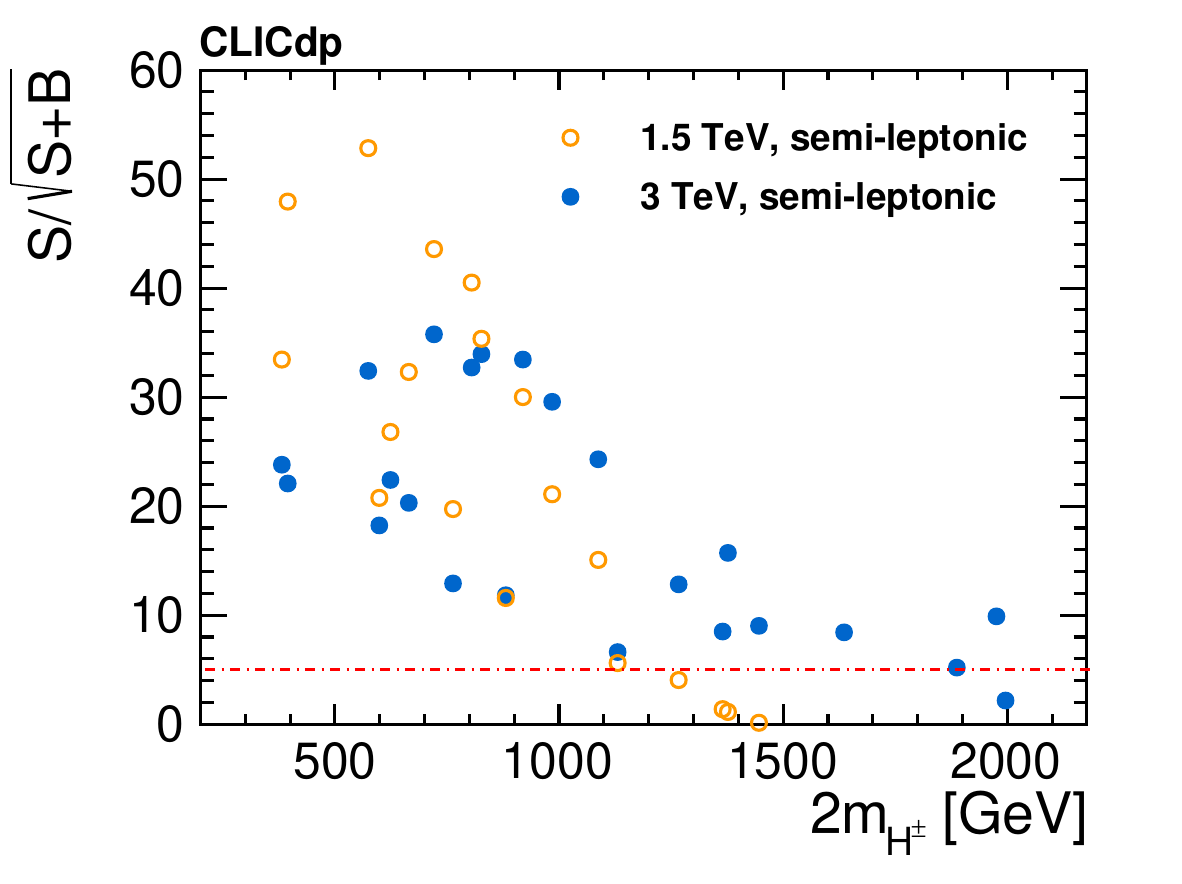}
\includegraphics[width=0.47\linewidth]{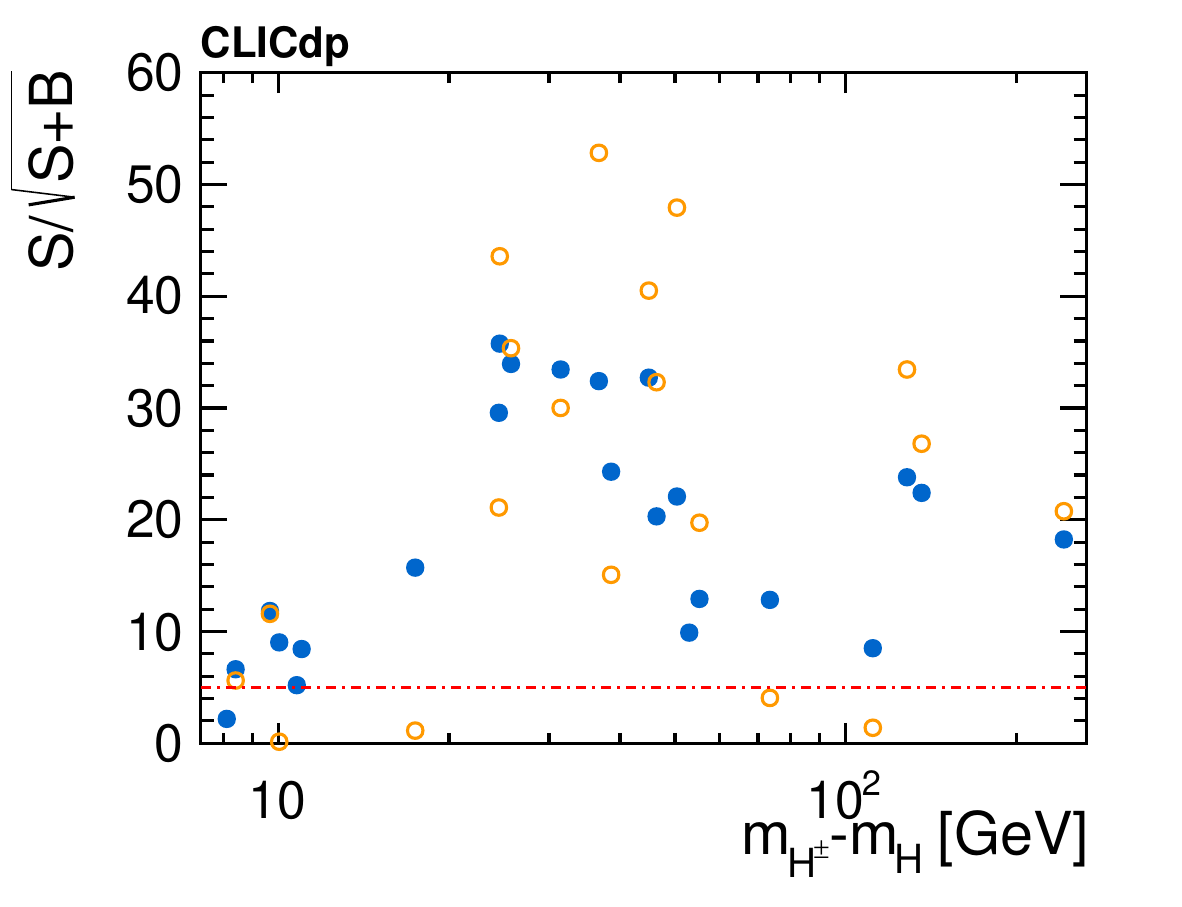}
\caption{\label{fig:iDM-reach}Signal statistical significance for various IDM benchmark points~\cite{Klamka:2022ukx} at high energy lepton collider for charged Higgs pair production and semi-leptonic final states. }
\end{figure}

 Inert Doublet Model (IDM) is an extension of the SM with the second Higgs doublet carries an extra discrete $Z_2$ symmetry and couples to the SM gauge boson only.  The lightest of the extra neutral scalars is a good candidate for a Dark Matter particle.   The production of IDM scalars at lepton colliders is dominated by production of neutral scalar pair $\mu^+\mu^-\rightarrow HA$ or charged scalar pair $\mu^+\mu^-\rightarrow H^+H^-$ via the SM gauge interactions.  The subsequent decay of $A \rightarrow HZ$ and $H^\pm \rightarrow H W^\pm$ leads to $HHZ$ and $HHW^+W^-$ final states, with $H$ being identified as the dark matter particle of missing energy signal.  The leptonic final states have limited reach at high energy lepton colliders.  The discovery reach is only about 500 GeV for scalar mass at 3 TeV collider, given the small leptonic final states branching fractions, and the decreasing of production cross section with the increasing center of mass energy~\cite{Kalinowski:2018kdn,Kalinowski:2020rmb}.   Considering the semi-leptonic final states~\cite{Klamka:2022ukx}, the signal statistical significance for charged Higgs pair production is shown in  Fig.~\ref{fig:iDM-reach} for CLIC 1.5 TeV and 3 TeV with 4~$\iab$ integrated luminosity.   Most of the scenarios considered in the study with $m_H^\pm$ up to about 1 TeV can be be discovery at more than 5 $\sigma$ level for a 3 TeV collider.   The 3 TeV muon collider reach is  similar.

\subsection{MSSM electroweak states }
\label{sec:MSSM}

Electroweak states in supersymmetric models can be pair produced at a muon collider.  The dominant production for Wino-like NLSP with Bino-like LSP are $\mu^+\mu^- \rightarrow \chi^+_1\chi^-_1, \chi^2_0\chi_1^0$, with $\chi^\pm_1$ and $\chi^0_2$ being Wino-like states.  Sensitivity up to pair production mass threshold of $\sqrt{s}/2$  are possible even for $m_{\chi^\pm_1}-m_{\chi^0_1}$ as low as 1 GeV, with no loss in acceptance~\cite{deBlas:2018mhx}.  In comparison, the HL-LHC reach is about 1 TeV  for the Wino NLSP, with Bino-LSP mass up to about 500 GeV~\cite{CidVidal:2018eel}.  

For the case when the higgsino-like states are the NLSP and LSP, the electroweakinos exhibit a compressed spectrum with a production cross section smaller than that of the Wino case.  The high energy lepton collider allow a reach close to the pair production threshold: about 1.3 TeV for CLIC3000 with the mass splitting down to about 0.5 GeV.  The muon collider 3 TeV reach would be similar~\cite{EuropeanStrategyforParticlePhysicsPreparatoryGroup:2019qin}.   In comparison, the HL-LHC reach highly depends on the mass splitting, only about 350 GeV for mass splitting between 1.6 to 50 GeV~\cite{CidVidal:2018eel}.  Searches based on  disappearing charged tracks for pure higgsino states will be covered in Section~\ref{subsec:Stub-tracks}.

The reach for selectron and smuon is about its pair production kinematic threshold of 1.5 TeV for a 3 TeV muon collider.   The reach for stau is slightly worse, given the identification of hadronically decaying $\tau$.  CLIC3000 can reach up to stau of about 1.25 TeV and $\Delta m (\tilde\tau, \chi_1^0)=50$ GeV~\cite{EuropeanStrategyforParticlePhysicsPreparatoryGroup:2019qin}.  The muon collider reach is  similar.


\section{Dark Matter}\label{secdm}
The possibility that Dark Matter is a massive particle charged under electroweak interactions is one of the major themes of research in Dark Matter. Cosmogenic Dark Matter can be observed in ultra-low noise underground detectors into which it is possible to detect directly the DM interaction with the SM matter in the detector. Additionally, DM  can be searched in DM-rich astrophysical environments, where the DM pairs can annihilate and give rise to observables signals in cosmic ray observatories.  These experimental investigation are promising and actively pursued, but suffer few potential roadblocks. Cosmic rays observation can be hampered by   large uncertainties about   astrophysical quantities and astrophysical processes  that can mimic dark matter signals. Furthermore, the unknown  density distribution of the dark matter that undergoes annihilation brings in additional uncertainty. Lab-based direct detection of cosmogenic dark matter has the inherent problem of being a very low momentum transfer process even when Dark Matter is quite heavy, hence background rejection is very challenging. 

\begin{table}
\begin{centering}
\begin{tabular}{c||c|c|c|c}
$n$ & Dirac & Majorana & Complex Scalar & Real Scalar\tabularnewline
\hline 
\hline 
2 & 1.08 & - & 0.58 & -\tabularnewline
\hline 
3 & 2.0 \& 2.4 & 2.86
& 1.6 \& 2.5 & 2.53
\tabularnewline
\hline 
4 & 4.79(9) & - & 4.98(5) & -\tabularnewline
\hline 
5 & 8.8(4) & 13.6(5) & 11.5(7) & 15.4(7)\tabularnewline
\hline 
\end{tabular}
\par\end{centering}
\caption{\label{tab:thermal-masses}Thermal mass, in TeV, for pure SU(2) $n$-plet dark matter WIMP.  Effects of bound states and Sommerfeld enhancement of the annihilation cross-section are included from Ref.~\cite{Bottaro:2021snn,Bottaro:2022}. The neutral component of complex scalars and Dirac fermions can have a tiny electric charge. In some cases it is also possible to assign a non-zero hypercharge consistently with direct searches of dark matter.}
\end{table}

The  possibility to produce dark matter particles in the laboratory and study them with precise particle detectors is a unique capability of particle colliders. The great challenge for particle colliders is to produce these particles with sufficient rate to result in a statistically significant observation. The case of Weakly Interacting Massive Particles (WIMPs) dark matter is particularly useful to gauge the efficacy of particle colliders to test dark matter. In fact WIMPs must feel the weak interactions of the SM, as they use them to be in equilibrium in the early Universe plasma. The WIMP relic abundance is set by the (known) strength of the weak interactions coupling and the (unknown) mass of the WIMP. Therefore, for simple models in which the WIMP is a pure $SU(2)_{W}$ $n$-plet it is possible to sharply predict the mass of the dark matter particle, see Tab.~\ref{tab:thermal-masses} for some examples. As a general rule, the larger the $n$-plet the larger the mass of the WIMP. Smaller masses can be attained for a mixture of an $n$-plet e.g. with a state not charged under $SU(2)_{W}$. Therefore, testing the reach for pure $SU(2)_{W}$ $n$-plet is an excellent benchmark for particle colliders, as it demands to reach the highest mass for a given class of dark matter candidates.

A crucial phenomenological parameter for the detection of WIMP dark matter at colliders is the mass splitting between the neutral component of the dark matter $n$-plet and other electrically charged and neutral components of the multiplet. When this mass splitting is comparable or greater than the detector threshold, typically around 10~GeV, there is a good chance that the production of states furnishing the $n$-plet will give detectable signals, one example is the iDM of Section~\ref{sec:IDM}. 

\subsection{Mono-X \label{subsec:monoX}}

When the mass-splitting between the dark matter particle and the other states of the multiplets is below the detectable threshold, none of the particles in the dark matter multiplet leaves a detectable trace in the detector. This makes the production of dark matter observable only ``by contrast'', e.g. observing a bunch of particles apparently recoiling against nothing. At a muon collider the reaction is
$$
\mu^{+}\mu^{-} \to \chi \chi +X \,,
$$
where $X$ denotes any particle or set of particles allowed by the interactions and $\chi$ is a generic state belonging to the dark matter $n$-plet. 

Searches for general electroweak states have been studied for several types of observables particles $X$ accompanying the production of dark matter. The signal for $X=\gamma,W,Z,\mu^{\pm},\mu\mu$ have been studied in  \cite{Han:2020uak,Bottaro:2021snn}, finding that the a 3 TeV muon collider is in general very sensitive to the production of new electroweak matter.

\begin{figure}
\centering{}\includegraphics[width=0.47\linewidth]{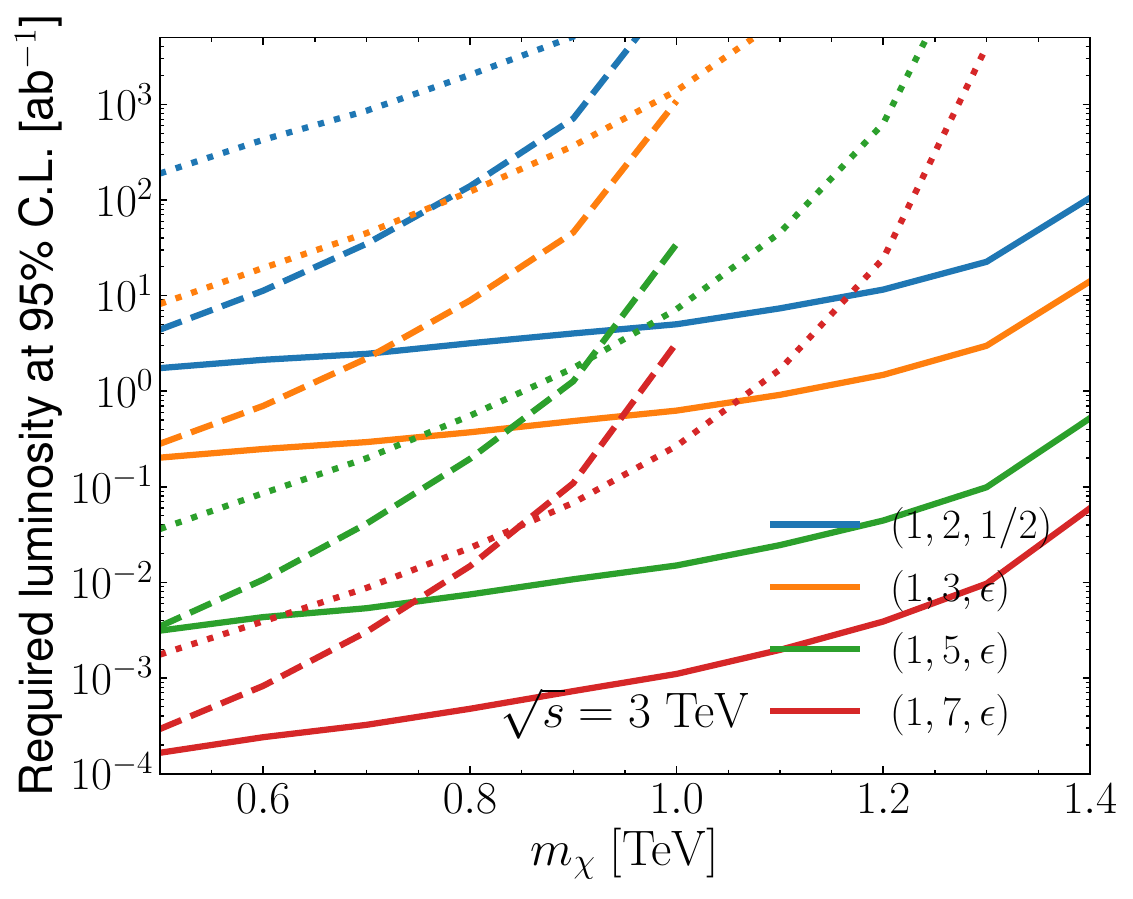}\includegraphics[width=0.47\linewidth]{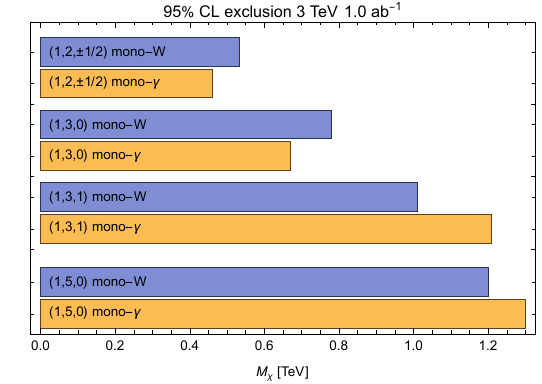}\caption{\label{fig:EWstates}Direct reach on electroweak states in mono-$X$ signals. Left: Luminosity needed to exclude  a Dirac fermion DM candidate for zero systematics~\cite{Han:2020uak} for $X=\gamma$ (solid), $X=\mu$ (dotted), $X=\mu\mu$ (dashed). Right: Mass reach on a fermionic DM candidate (assumed Majorana when $Y=0$, Dirac otherwise) at fixed 1~$\iab$ luminosity for the 3 TeV muon collider for $X=\gamma$ and $X=W$ for 0.1\% systematics~\cite{Bottaro:2021snn,Bottaro:2022}.}
\end{figure}

Figure~\ref{fig:EWstates} summarizes the reach illustrating in the left panel the luminosity needed to reach the 95\% CL exclusion of electroweak matter of a given mass in several production modes $X=\gamma,\mu,\mu\mu$. Among these, the mono-$\gamma$ search is the one placing the best bound for states heavier than about 500~GeV. The right panel shows the mass reach at fixed luminosity 1~$\iab$ and includes the mono-$W$ channel, which is most effective for the same mass range in which mono-$\gamma$ leads the exclusion and in some cases exceeds mono-$\gamma$ results. All in all, the combination of these two channels, especially thanks to different levels of signal-over-background ratio and sources of possible systematics, can provide best mass reach for some DM candidates.
 
\subsection{\label{subsec:2-body-processes}Indirect reach through SM rates}

Pure WIMP DM $n$-plets for $n\geq3$ are too heavy to be directly produced in pairs at the 3 TeV muon collider at their thermal mass, see Tab.~\ref{tab:thermal-masses}. However, these heavy DM candidates can leave observable effects as their off-shell propagation modifies the rate and the distributions of SM processes such as 
\begin{eqnarray}
\mu^{+}\mu^{-} &\to& f \bar{f}\,, \label{proc:ffbar}\\
\mu^{+}\mu^{-} &\to& Z h \,, \label{proc:Zh} \\
\mu^{+}\mu^{-} &\to& W^{+}W^{-}\,, \label{proc:WW}
\end{eqnarray}
and possible higher order processes such as $\mu^{+}\mu^{-}\to WWh$.
Measuring the total rate of eqs.(\ref{proc:ffbar}-\ref{proc:WW}) and using differential information on the angular distribution of the channels in which the charge of the final states $f=e,\mu$  can be tagged reliably, it is possible to put bounds at 95\% CL for the existence of new matter $n$-plets (see Refs.~\cite{DiLuzio:2018jwd,RFXZ} for muon collider specific studies). 

\begin{figure}
\begin{centering}
\includegraphics[width=0.49\linewidth]{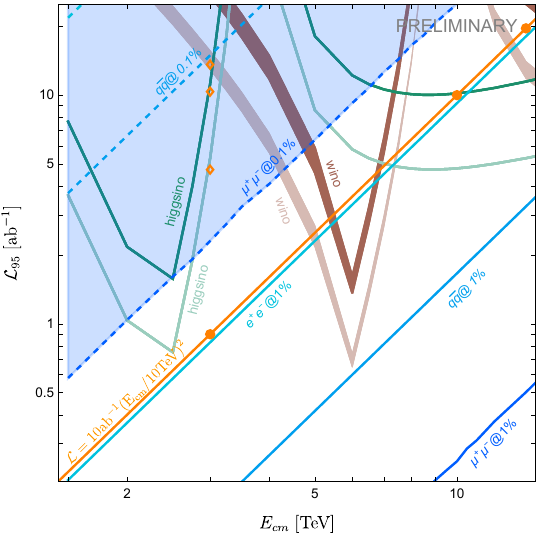}
\includegraphics[width=0.49\linewidth]{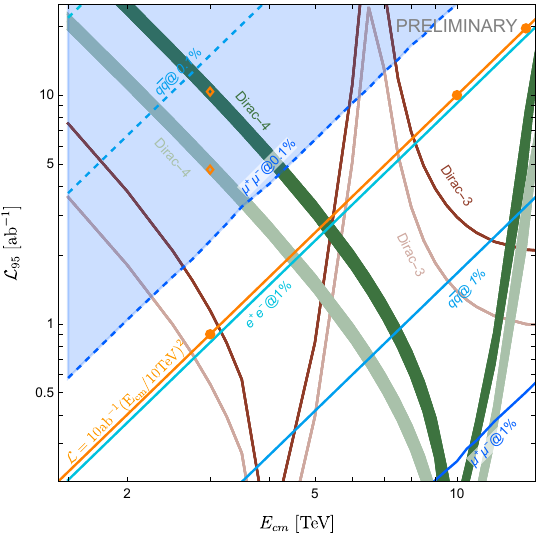}
\par\end{centering}
\caption{\label{fig:Indirect-reach-3plets} Minimal luminosity to exclude a thermal pure higgsino or wino dark matter (left panel) a 2.0 TeV Dirac triplet or 4.8 TeV Dirac 4-plet as function of the collider center of mass energy~\cite{RFXZ} (hyper-charge of the higgsino and Dirac $n$-plets not taken into account). Lighter color lines are for polarized beams. The thickness of the Wino and Dirac 4-plet bands covers the uncertainty on the thermal mass calculations. Diagonal lines mark the precision at which the total rate of the labeled channels are going to be measured. The shaded area indicates that at least one channel is going to be measured with 0.1\% uncertainty and systematic uncertainties need to be evaluated.}
\end{figure}

In Fig.~\ref{fig:Indirect-reach-3plets} we report the minimal luminosity necessary to exclude a thermal pure Wino dark matter (brown bands) as  a function of the collider center of mass energy. These studies are helped by the presence of left-handed fermions initial states, which source larger weak-boson mediated scattering. Therefore it is interesting to study the effect of beam polarization. In the figure the lighter colored lines give the necessary luminosity for an exclusion at a machine capable of 30\% left-handed polarization on the $\mu^{-}$ beam and -30\% for the $\mu^{+}$ beam. Even this modest polarization of the beams can cut significantly the luminosity required for the exclusion.

Figure~\ref{fig:Indirect-reach-3plets} also shows the reach for   a Dirac doublet with zero hyper-charge through the same observables. Neglecting hyper-charge contributions this is the same as the reach for a higgsino.  This reach is complementary to that from direct searches of all sorts, as it does not depend on the higgsino mass splitting and the search final states that it results into. Thus the indirect search can complement the reach discussed in Section~\ref{subsec:Stub-tracks}  from stub-tracks as it  has no dependence on the higgsino lifetime.

The shaded area indicates that the search for new electroweak matter is based on such a luminosity high enough to have statistical uncertainties at the 0.1\% level for some channel. This may require a careful evaluation of possible systematics.  

\FloatBarrier
\ 
\newpage
\section{Unconventional signatures }\label{secunc}

The search for long-lived particles (LLPs) has recently become a priority in the particle physics community~\cite{Curtin:2018mvb, Alimena:2019zri}. LLPs appear in a variety of models and yield a large range of signatures at colliders. Depending on the LLP quantum numbers and lifetime, these can span from LLP decay products appearing in the detector volume, even outside of the beam crossings, to metastable particles with anomalous ionisation disappearing after a short distance.

This wide range of experimental signatures is strongly intertwined with the development of detector technologies and the design of the final detector layout~\cite{Jindariani:2022gxj}. For example, the development of timing-sensitive detectors is crucial both to suppress the abundant beam-induced backgrounds and to detect the presence of heavy, slow-moving, particles that are traveling through the detector. A lively R\&D programme is ongoing to develop the reconstruction algorithms that will profit from these new technologies.

For heavy particles, whose production cross sections are dominated by the annihilation s-channel, there are two main features that make searches for unconventional signatures particularly competitive at a muon collider when compared to other future proposed machines like the FCC-hh. The produced particles tend to be more centrally distributed, impinging on the regions of the detector where reconstruction is comparatively easier, and furthermore tend to have more ``mono-chromatic'' Lorentz boosts which can lead to effectively larger average observed lifetimes for the produced BSM states.

Searches for LLPs that decay within the volume of the tracking detectors (e.g. decay lengths between 1~mm and 500~mm) are particularly interesting as they directly probe the lifetime range motivated by compelling dark matter models. 

\subsection{Search for disappearing tracks}
\label{subsec:Stub-tracks} 

The higgsino is among the most compelling dark matter candidates, with tight connections to the naturalness of the weak scale, which could lead to LLPs being produced in particle collisions.
In scenarios where all other supersymmetric partners are decoupled, the higgsino multiplet consists of an SU(2)-doublet Dirac fermion. Due to loop radiative corrections, the charged state \chipm\ splits from the neutral one \ninoone\ by 344 MeV, giving rise to a mean proper decay length of 6.6 mm for the relic favoured mass of 1.1~TeV~\cite{Hisano:2006nn}. The \chipm\ can then travel a macroscopic distance before decaying into an invisible \ninoone\ and other low-energy Standard Model fermions.

Searches at the LHC are actively targetting this scenario~\cite{ATLAS:2022rme,Aaboud:2017mpt,Aad:2013yna,Sirunyan:2020pjd,CMS:2014gxa}, but are not expected to cover the relic favoured mass~\cite{ATLAS:2018jjf, EuropeanStrategyforParticlePhysicsPreparatoryGroup:2019qin}. A muon collider operating at a multi-TeV centre-of-mass energies could provide a perfect tool to look for these particles.

The production of pairs of electroweakinos at a MuC proceeds mainly via an s-channel photon or off-shell Z-boson, with other processes, such as vector boson fusion, being subdominant. The prospects for such a search were investigated in detail in Ref.~\cite{Capdevilla:2021fmj} exploiting a detector simulation based on \GEANT 4~\cite{Agostinelli:2002hh} for the modelling of the response of the tracking detectors, which are crucial in the estimation of the backgrounds. The simulated events were overlaid with beam-induced background events simulated with the MARS15 software~\cite{Mokhov:2017klc}.

\begin{figure}[h]
\centering
\includegraphics[width=0.8\textwidth]{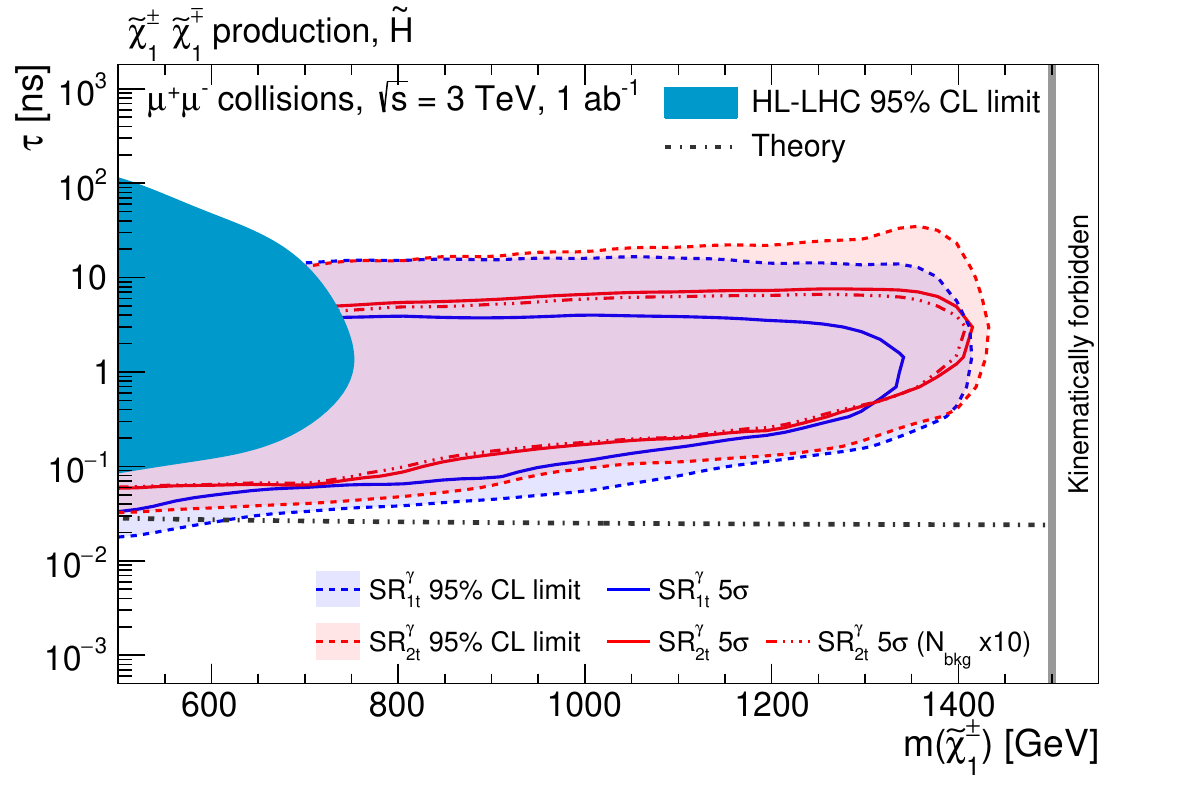}
\caption{Expected sensitivity using 1~ab$^{-1}$ of 3~TeV $\mu^{+}\mu^{-}$ collision data as a function of the \chipm mass and mass difference with the lightest neutral state, assuming a mass-splitting equal to 344~MeV, as per a pure-higgsino scenario~\cite{Capdevilla:2021fmj}.}
\label{fig:HiggsinoStubTracksReachMass}
\end{figure}

The analysis strategy relies on requiring one (\SRotp) or two (\SRttp) disappearing tracks in each event in addition to  a 25~GeV ISR photon. Additional requirements are imposed on the transverse momentum and angular direction of the reconstructed tracklet and on the distance between the two tracklets along the beam axis in the case of events with two candidates. The expected backgrounds are extracted from the full detector simulation and the results are presented assuming a 30\% (100\%) systematic uncertainty on the total background yields for the single (double) tracklet selections.
The corresponding discovery prospects and 95\% CL exclusion reach are shown in Figure~\ref{fig:HiggsinoStubTracksReachMass} for each of the two selection strategies discussed above.

Both event selections are expected to cover a wide range of higgsino masses and lifetimes, well in excess of current and expected collider limits. In the most favourable scenarios, the analysis of 1~ab$^{-1}$ of 3~TeV muon collisions is expected to allow the discovery \chipm masses up to a value close to the kinematic limit of $\sqrt{s}/2$. The interval of lifetimes covered by the experimental search directly depends on the layout of the tracking detector, i.e. the radial position of the tracking layers, and the choices made in the reconstruction and identification of the tracklets, i.e. the minimum number of measured space-points. Considering the current detector design~\cite{CLICdp:2017vju,CLICdp:2018vnx,ILDConceptGroup:2020sfq,ILC:2007vrf}, 1~ab$^{-1}$ of 3~TeV muon collisions would not allow to cover the higgsino thermal target.
An alternative tracking detector design, hard to realise in the presence of the BIB, with tracking layers significantly closer to the beam line would be needed to detect such a signal. Other unconventional signatures, such as soft displaced tracks~\cite{Fukuda:2019kbp} detected in combination with an energetic ISR photon or kinked tracks should be investigated and have the potential to recover sensitivity in this well-motivated scenario.  

%% file: Section_Muon_Specific/section.tex
\newcommand{\kp}[1]{\textcolor{niceblue}{\bf[(Keping) #1]}}
\renewcommand{\todo}[1]{\textcolor{nicered}{\bf[To do: #1]}}
\newcommand{\TeV}{\textrm{TeV}}

\newcommand{\DB}[1]{\textcolor{red}{\bf [#1]}}
\newcommand{\rodolfo}[1]{\textcolor{brown}{#1}}


\def\DC#1{{\bf  \textcolor{magenta}{[{DC: #1}]}}}

\section{The muon anomalous magnetic moment \label{sec:g-2}}

The anomalous magnetic moment of the muon has provided, over the last ten years, an enduring hint for new physics (NP). 
The experimental value of $a_\mu \!=\! (g_\mu \!-\! 2)/2$ from the E821 experiment at the Brookhaven National Lab~\cite{Muong-2:2006rrc} was recently confirmed by the E989 
experiment at Fermilab~\cite{Muong-2:2021vma,Muong-2:2021ojo}, yielding the experimental average $a_\mu^{\scriptscriptstyle \rm EXP} \!=\! 116592061(41) \!\times\! 10^{-11}$. 
The comparison of this value with the Standard Model (SM) prediction $a_\mu^{\scriptscriptstyle \rm SM} \!=\! 116591810(43) \times 10^{-11}$~\cite{Aoyama:2012wk,Gnendiger:2013pva,Keshavarzi:2018mgv,Davier:2019can,Kurz:2014wya,Melnikov:2003xd,Bijnens:2019ghy,Pauk:2014rta,Danilkin:2016hnh,Roig:2019reh,Colangelo:2014qya} 
shows an interesting $4.2\,\sigma$ discrepancy
\begin{equation}
\Delta a_\mu = a_\mu^{\scriptscriptstyle \rm EXP}-a_\mu^{\scriptscriptstyle \rm SM} = 251 \, (59) \times 10^{-11}\,.
\label{eq:gmu}
\end{equation}
In the following, we refer to this as the $g$-2 anomaly. Current and forthcoming plans to confirm the BSM origin of this anomaly include reducing the experimental uncertainty by a factor of four at E989, comparisons between phenomenological and Lattice determinations of the Hadronic Vacuum Polarization contribution to $g$-2 \cite{Gerardin:2019vio,Chao:2021tvp,FermilabLattice:2017wgj,Budapest-Marseille-Wuppertal:2017okr,RBC:2018dos,Giusti:2019xct,Shintani:2019wai,FermilabLattice:2019ugu,Gerardin:2019rua,Giusti:2019hkz,Borsanyi:2020mff}, and new experiments aiming to probe the same physics \cite{Sato:2017sdn,Abbiendi:2016xup}. 
If all of these efforts confirm the presence of NP, then the most urgent task at hand will be to probe this anomaly at higher energies, ultimately in order to discover and study the new BSM particles which give rise to the additional $\Delta a_\mu$ contributions. 
The $\mc$ is a uniquely well-suited machine for this endeavour, not least since it collides the actual particles displaying the anomaly, and hence the only particles guaranteed to couple to the new physics.

There are several ways in which a $\mc$ can provide a powerful high-energy test of the muon $g$-2 and discover the new physics responsible for the anomaly:
\begin{itemize}
\item If the physics responsible for $\Delta a_\mu$ is heavy enough, an Effective Field Theory (EFT) description holds up to the high energies of a $\mc$. This was studied in~\cite{Buttazzo:2020ibd}. In this case, scattering cross-sections induced by the NP effective operators grow at high energies (analogously to the case of weak-interaction cross-sections below the $W$ boson mass), so that a measurement with $\mathcal{O}(1)$ precision at a sufficiently high energy will be sufficient to disentangle NP effects from the SM background. These considerations are completely independent from the specific underlying model.
\item In most motivated models of NP, new particles responsible for $\Delta a_\mu$ are light enough to be directly produced in $\mu^+\mu^-$ collisions at attainable $\mc$ energies. Understanding such opportunities for direct production and discovery at a $\mc$ was the subject of studies~\cite{Capdevilla:2020qel,Capdevilla:2021rwo}. 
It was found that a complete classification of all perturbative BSM models that can give rise to the observed value of $\Delta a_\mu$, and of their experimental signatures, is possible. This motivates muon colliders at the multi-TeV scale.
\item Additional effects in muon couplings to SM gauge and Higgs bosons, correlated with the muon $g$-2, can also be present at a level that can be probed by precision measurements at a $\mc$. Some of these effects can be predicted in a model-independent way, others arise in specific, motivated models.
\end{itemize}

These three strategies together made it possible to formulate a {\it no-lose theorem} for a high-energy $\mc$~\cite{Capdevilla:2020qel,Capdevilla:2021rwo}, assuming that the experimental anomaly in the muon $g$-2 is really due to NP.
The physics case of a high-energy determination of $\Delta a_\mu$, which is unique of a $\mc$, thus represents a striking example of the complementarity and interplay of the high-energy 
and high-intensity frontiers of particle physics, and it highlights the far reaching potential of a $\mc$ to probe NP.

\subsection{High-energy probes of the operators generating the $g$-2}\label{gm2}

In this section, we review the analysis of~\cite{Buttazzo:2020ibd}, which determined that precision measurements at high-energy $\mc$'s can detect deviations in scattering rates that are generated by the same effective operators that give rise to the $g$-2 anomaly, thereby providing a powerful independent verification and detailed examination of the anomaly even if the responsible BSM degrees of freedom are too heavy to be generated on-shell at the collider. 

Heavy NP contributions to $g$-2 arise from the dimension-6 dipole operator 
$\left(\bar\mu_L \sigma_{\mu\nu} \mu_R\right) H F^{\mu\nu},$~\cite{Buchmuller:1985jz} 
where $H$ is the neutral component of the Higgs field
and $F^{\mu\nu}$ is the electromagnetic field strength tensor.
After electroweak (EW) symmetry breaking $H$ is replaced by its vacuum expectation value $v \!=\! 174~$GeV, and one obtains the prediction 
$\Delta a^{\rm\scriptscriptstyle NP}_\mu \sim (g^2_{\scriptscriptstyle\rm NP}/ 16\pi^2) \times (m_\mu v/\Lambda^2)$, 
where $g_{\rm\scriptscriptstyle NP}$ is the typical coupling of the NP sector.
Therefore, the NP chiral enhancement $v/m_\mu \sim 10^3$ with respect to the SM weak contribution, together with the assumption 
of a new strong dynamics with $g_{\rm\scriptscriptstyle NP} \sim 4 \pi$, bring the sensitivity of the muon $g$-2 to NP scales of order $\Lambda \sim 100\,$TeV. Directly detecting new particles at such high scales is far beyond the capabilities of any foreseen collider.
%
Nevertheless, a $\mc$ would still enable to probe NP in the muon $g$-2 in a completely model-independent way.
Indeed, the very same dipole operator that generates $\Delta a_\mu$ unavoidably induces also a NP 
contribution to the scattering process $\mu^+\mu^- \to h \gamma$ \cite{Buttazzo:2020ibd,Yin:2020afe}. Measuring the cross-section for this process would thus be equivalent to measuring $\Delta a_\mu$. 
This would however be a direct determination of the NP contribution, not affected by the hadronic uncertainties that enter the SM prediction of $a_\mu$.

\subsubsection*{Effective interactions}
\begin{figure}
\centering%
\includegraphics[width=0.55\textwidth]{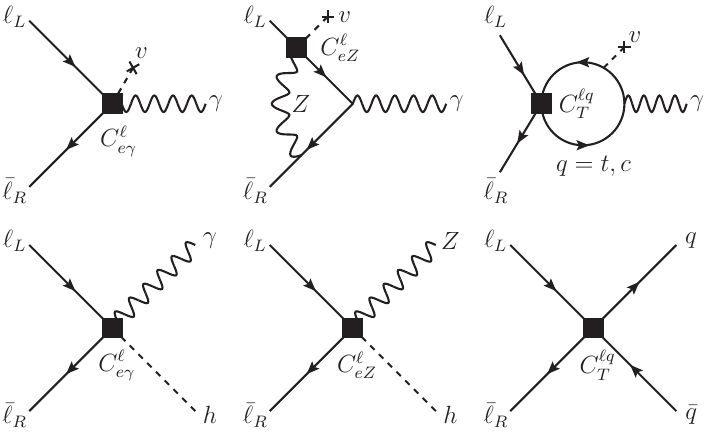}
\caption{{\it Upper row:} Feynman diagrams contributing to the leptonic $g$-2 up to one-loop order in the Standard Model EFT.
{\it Lower row:} Feynman diagrams of the corresponding high-energy scattering processes.
Dimension-6 effective interaction vertices are denoted by a square.}
\label{fig:feyn}
\end{figure}

New interactions emerging at a scale $\Lambda$ larger than the EW scale can be described at energies $E \ll \Lambda$ 
by an effective Lagrangian containing non-renormalizable $SU(3)_c \otimes SU(2)_L \otimes U(1)_Y$ invariant operators.
The relevant effective Lagrangian contributing to $g$-2,
reads~\cite{Buchmuller:1985jz}
\begin{align}
\mathcal{L} = 
\frac{C^\ell_{eB}}{\Lambda^2}
\left( \bar\ell_L \sigma^{\mu\nu}e_{R}\right) \! H B_{\mu\nu} + 
\frac{C^\ell_{eW}}{\Lambda^2}
\left( \bar\ell_L \sigma^{\mu\nu} e_{R} \right) \! \tau^I \! H W_{\mu\nu}^I  + \frac{C^\ell_{T}}{\Lambda^2}( \overline{\ell}_L\sigma_{\mu\nu}e_{R}) (\overline{Q}_L\sigma^{\mu\nu} u_{R}) 
+ h.c.,
\label{eq:L_SMEFT}
\end{align} 
that includes not only the interactions that generate the dipole operator at tree level, but also four-fermion operators that generate the dipole at one loop. The Feynman diagrams relevant for $g$-2 are displayed in figure~\ref{fig:feyn}, top row.
They lead to the following result
\begin{align}
\Delta a_\ell  \simeq \frac{4m_\ell v}{e\Lambda^2} \, 
\bigg(
C^\ell_{e\gamma} - \frac{3\alpha}{2\pi} \frac{c^2_W \!-\! s^2_W}{s_W c_W} \,C^\ell_{eZ} \log\frac{\Lambda}{m_Z}
\bigg) - \sum_{q=c,t} \frac{4m_\ell m_q}{\pi^2} \frac{C_T^{\ell q}}{\Lambda^2}\,
\log\frac{\Lambda}{m_q},
\label{eq:Delta_a_ell}
\end{align}
where $s_W$, $c_W$ are the sine and cosine of the weak mixing angle, $C_{e\gamma}=c_W C_{eB} - s_W C_{eW}$ and
$C_{eZ} = -s_W C_{eB} - c_W C_{eW}$.
Additional loop contributions from the operators $H^\dag H W_{\mu\nu}^IW^{I\mu\nu}$, $H^\dag H B_{\mu\nu}B^{\mu\nu}$, 
and $H^\dag \tau^I H W_{\mu\nu}^I B^{\mu\nu}$ can be neglected because they are suppressed by the small lepton Yukawa couplings.
For simplicity, $C_{eB}$, $C_{eW}$ and $C_{T}$ are assumed to be real. 
The one-loop renormalization effects to $C^\ell_{e\gamma}$ are included
\begin{align}
\!\!\! C^\ell_{e\gamma}(m_\ell) \simeq C^\ell_{e\gamma}\!(\Lambda)\left(1 \!-\! \frac{3y^2_t}{16\pi^2} \log\frac{\Lambda}{m_t}
\!-\! \frac{4 \alpha}{\pi} \log\frac{m_t}{m_\ell}\right).
\label{eq:running_Cegamma}
\end{align}
Numerically, one finds that~\cite{Buttazzo:2020ibd}
\begin{align}
\!\!
\frac{\Delta a_\mu}{3 \!\times\! 10^{-9}} \!\approx \! \left( \frac{250 \, {\rm TeV}}{\Lambda} \right)^{\!\!2} \!\!\!
\left(C^\mu_{e\gamma} \!-\! 0.2 C^{\mu t}_T \!\!-\! 0.001 C^{\mu c}_T \!\!-\! 0.05 C^{\mu}_{eZ}\right). 
\nonumber
\end{align}
%

\noindent A few comments are in order: 
\begin{itemize}
\item The $\Delta a_\mu$ discrepancy can be solved for a NP scale up to $\Lambda\approx 250~$TeV.
This requires a strongly coupled NP sector where $C^\mu_{e\gamma}$ and/or $C^{\mu t}_T \sim g^2_{\rm\scriptscriptstyle NP}/16\pi^2 \sim 1$ 
and a chiral enhancement $v/m_\mu$ compared with the weak SM contribution.
This NP can be tested through high-energy processes such as $\mu^+\!\mu^- \!\to h\gamma$ or 
$\mu^+\!\mu^- \!\to q\bar{q}$ (with $q=c,t$) at a $\mc$.
\item If the underlying NP sector is weakly coupled, $g_{\rm\scriptscriptstyle NP}\lesssim 1$, then
$C^\mu_{e\gamma}$ and $C^{\mu t}_T \lesssim 1/16\pi^2$, implying $\Lambda\lesssim 20~$TeV to solve the $g$-2 anomaly.
In this case, a $\mc$ could still be able to directly produce NP particles~\cite{Capdevilla:2020qel}. 
Even so, the study of the processes $\mu^+\mu^-\to h\gamma$
and $\mu^+\mu^-\to q\bar{q}$ could be crucial to reconstruct the effective dipole vertex $\mu^+\mu^-\gamma$.
\item If the NP sector is weakly coupled, and further $\Delta a_\mu$ scales with lepton masses as the SM weak contribution,
then $\Delta a_\mu \sim m^2_\mu/16\pi^2\Lambda^2$.
Here, the experimental value of $\Delta a_\mu$ can be accommodated only provided that $\Lambda \lesssim 1~$TeV.
For such a low NP scale the EFT description breaks down at the typical multi-TeV $\mc$ energies, and new resonances cannot escape from direct production.
\end{itemize}

\subsubsection*{Dipole operator}
The main contribution to $\Delta a_\mu$ comes from the dipole operator $O_{e\gamma}=\left(\bar\ell_L \sigma_{\mu\nu} e_R\right) H F^{\mu\nu}$. The same operator also induces a contribution to the process $\mu^+\mu^- \to h \gamma$ 
that grows with energy, and thus can become dominant over the SM cross-section at a very high-energy collider. 
Neglecting all masses, the
total $\mu^+\mu^- \!\to  h\gamma$ cross-section is
\begin{align}
\sigma_{h\gamma} \!=\! \frac{s}{48\pi}\frac{|C^{\mu}_{e\gamma}|^2}{\Lambda^4}\!
\approx  0.7\, {\rm ab} \left(\frac{\sqrt{s}}{30\, {\rm TeV}}\right)^{\!2} \!\! \left(\frac{\Delta a_\mu}{3 \times 10^{-9}}, \right)^{\!2}
\label{xsec_HA}
\end{align}
where in the last equation no contribution to $\Delta a_\mu$ other than the one from $C^{\mu}_{e\gamma}$ was assumed, and running effects for $C^{\mu}_{e\gamma}$, see eq.~(\ref{eq:running_Cegamma}), from a scale $\Lambda \approx 100$~TeV have been included.
Notice that there is an identical contribution also to the process $\mu^+\mu^- \!\to\! Z\gamma$ since $H$ contains the longitudinal polarizations of the $Z$.
Given the scaling with energy of the reference integrated luminosity for a $\mc$ \cite{Schulte:2021eyr} one gets about 60 total $h\gamma$ events at $\sqrt{s}=30~$TeV. As it is discussed below, this constitutes a signal that the $\mc$ is sensitive to.

\begin{figure}
\centering%
\includegraphics[width=0.6\textwidth]{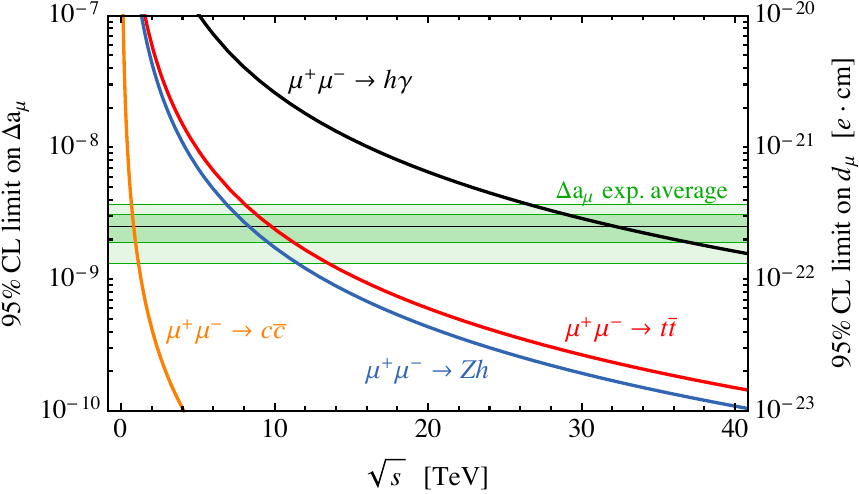}
\caption{Reach on the muon anomalous magnetic moment $\Delta a_\mu$ and muon EDM $d_\mu$, 
as a function of the $\mc$ collider center-of-mass energy $\sqrt{s}$, from the labeled processes. Figure taken from~\cite{Buttazzo:2020ibd}.}
\label{fig:hgamma}
\end{figure}

The SM irreducible $\mu^+\mu^- \to h\gamma$ background is small, $\sigma_{h\gamma}^{\rm SM} \approx 2 \times 10^{-2}\,{\rm ab}\,\big(\frac{30\,{\rm TeV}}{\sqrt{s}}\big)^{\!2}$,
with the dominant contribution arising at one-loop~\cite{Abbasabadi:1995rc} due to the muon Yukawa coupling suppression of the tree-level part.
The main source of background comes from $Z\gamma$ events, where the $Z$ boson is incorrectly reconstructed as a Higgs. 
This cross-section is large, due to the contribution from transverse polarizations.
There are two ways to isolate the $h\gamma$ signal from the background: by means of the different angular distributions of the two processes -- the SM $Z\gamma$ peaks in 
the forward region, while the signal is central -- and by accurately distinguishing $h$ and $Z$ bosons from their decay products, e.g. by precisely reconstructing their invariant mass.
To estimate the reach on $\Delta a_\mu$ a cut-and-count experiment was considered in the $b\bar b$ final state, which has the highest signal yield. The significance of the signal is maximized in the central region $|\!\cos\theta| \lesssim 0.6$. At 30 TeV one gets
\begin{align}
\sigma_{h\gamma}^{\rm cut} &\approx 0.53\, {\rm ab} \,\bigg(\frac{\Delta a_\mu}{3 \times 10^{-9}} \bigg)^{\!2}, & ~~\sigma_{Z\gamma}^{\rm cut} &\approx 82\,{\rm ab}\,.
\end{align}
Requiring at least one jet to be tagged as a $b$, and assuming a $b$-tagging efficiency $\epsilon_b = 80\%$, one finds that a value $\Delta a_\mu = 3\!\times\! 10^{-9}$ can be tested at 95\% C.L.\ at a 30 TeV collider if the probability of reconstructing a $Z$ boson as a Higgs is less than 10\%. The resulting number of signal events is $N_S = 22$, and $N_S/N_B = 0.25$.
Figure~\ref{fig:hgamma} shows as a black line the 95\% C.L.\ reach from $\mu^+\mu^-\to h\gamma$ 
on the anomalous magnetic moment as a function of the collider energy. Note that since the number of signal events scales as the fourth power of the center-of-mass energy, only a collider with $\sqrt{s} \gtrsim 30$~TeV will have the sensitivity to test the $g$-2 anomaly in this channel.

\subsubsection*{Semi-leptonic interactions}
If the magnetic moment arises at one loop from one of the other operators in \eqref{eq:Delta_a_ell}, their Wilson coefficients must be larger to reproduce the observed signal, and the NP will be easier to test at a $\mc$.
We now derive the constraints on the semi-leptonic operators. The operator $O_T^{\mu t}$ that enters 
$\Delta a_\mu$ at one loop can be probed by $\mu^+\mu^-\to t\bar t$ (Fig.~\ref{fig:feyn}). Its contribution 
to the cross-section is
\begin{align}
\label{eq:mumu_tt}
\!\!\!\sigma_{t\bar{t}} =\! \frac{s}{6\pi} \frac{|C^{\mu t}_{T}|^2}{\Lambda^4} N_c 
\approx 
58 \, {\rm ab} \left(\frac{\sqrt{s}}{10 \, {\rm TeV}}\right)^{\!2} \!\! \left(\frac{\Delta a_\mu}{3 \times 10^{-9}} \right)^{\!2}
\end{align}
where the last equality assumes $\Lambda \approx 100~$TeV so that 
$|\Delta a_\mu| \approx 3 \times 10^{-9} \left(100 \,{\rm TeV}/\Lambda\right)^2 |C^{\mu t}_T|$.
We estimate the reach on $\Delta a_\mu$ assuming an overall 50\% efficiency for reconstructing the top quarks, 
and requiring a statistically significant deviation from the SM $\mu^+\mu^-\to t\bar{t}$ background, which has a cross-section
$\sigma_{t\bar{t}}^{\rm SM} \approx 1.7 \,{\rm fb}\,\big(\frac{10\,{\rm TeV}}{\sqrt{s}}\big)^2$.

A similar analysis can be performed for semi-leptonic operator involving charm quarks. If the contrbution from the charm loop dominates, we can probe $|\Delta a_\mu| \approx 3 \times 10^{-9}(10\,{\rm TeV}/\Lambda)^2 |C^{\mu c}_T|$ through the process $\mu^+\mu^- \to \bar{c} c$.
In this case, unitarity constraints on the NP coupling $C_T^{\mu c}$ require a much lower NP scale $\Lambda \lesssim 10$~TeV, 
so that our effective theory analysis will only hold for lower center-of-mass energies.
Combining eq.~(\ref{eq:Delta_a_ell}) and (\ref{eq:mumu_tt}), with $c \leftrightarrow t$, we find
\begin{align}
\sigma_{c\bar{c}}
\, \approx \, 100 \,{\rm fb} \left(\frac{\sqrt{s}}{3 \, {\rm TeV}}\right)^{\!2} \!\! \left(\frac{\Delta a_\mu}{3 \times 10^{-9}} \right)^{\!2}.
\label{eq:mumu_cc_numerics}
\end{align}
The SM cross-section for $\mu^+\mu^- \!\to c\bar{c}$ 
at $\sqrt{s}= 3~$TeV is $\sim 19$~fb. In figure~\ref{fig:hgamma} we show the 95\% C.L.\ constraints on the top and charm contributions to $\Delta a_\mu$ as red and orange lines, respectively, as functions of the collider energy. Notice that the charm contribution can be probed already at $\sqrt{s} = 1$~TeV, while the top contribution can be probed at $\sqrt{s} = 10$~TeV. 

\subsubsection*{Electric dipole moments}
So far, CP conservation has been assumed. If however the coefficients $C_{e\gamma}$, $C_{eZ}$ or $C_T$ are complex, 
the muon electric dipole moment (EDM) $d_\mu$ is unavoidably generated.
Since the cross-sections in eq.~\eqref{xsec_HA} and \eqref{eq:mumu_tt} are proportional to the absolute values of the same coefficients, a $\mc$ offers a unique opportunity to test also $d_\mu$. The current experimental limit $d_\mu < 1.9 \times 10^{-19}\,e\,$cm was set by the BNL E821 experiment~\cite{Muong-2:2008ebm} 
and the new E989 experiment at Fermilab aims to decrease this by two orders of magnitude~\cite{Chislett:2016jau}. 
Similar sensitivities could be reached also by the J-PARC $g$-2 experiment~\cite{Gorringe:2015cma}.

From the model-independent relation~\cite{Giudice:2012ms}
\begin{align}
\frac{d_\mu}{\tan\phi_\mu} =  \frac{\Delta a_\mu}{2 m_\mu} \,e \,\simeq\, 3 \times 10^{-22} \left(\frac{\Delta a_\mu}{3\!\times\! 10^{-9}}\right) e\, {\rm cm}\,,
\label{eq:d_mu}
\end{align}
where $\phi_\mu$ is the argument of the dipole amplitude, the bounds on $\Delta a_\mu$ in figure~\ref{fig:hgamma} can be translated into a model-independent constraint on $d_\mu$. It was found that already a 10~TeV $\mc$ can reach a sensitivity comparable to the ones expected at Fermilab~\cite{Chislett:2016jau} and J-PARC~\cite{Gorringe:2015cma}, while at a 30 TeV collider one gets the bound $d_\mu \lesssim 3\times 10^{-22}\, e$~cm.

\subsection{Direct searches for BSM particles generating the $g$-2}

Here we briefly review the model-exhaustive analyses conducted in~\cite{Capdevilla:2020qel,Capdevilla:2021rwo} and \cite{Capdevilla:2021kcf}, examining all possible perturbative BSM solutions to the $g$-2 anomaly to understand the associated direct production signatures of new states at future $\mc$'s. 
We then summarize the no-lose theorem for discovering NP if the $g$-2 anomaly is confirmed and weakly coupled solutions below $\sim1$ GeV are excluded.

This model-exhaustive analysis first finds the highest possible mass scale of new physics subject only to perturbative unitarity, and optionally the requirements of minimum flavour violation and/or naturalness. The results show that a 3 TeV $\mc$ can discover all new physics scenarios in which $\Delta a_\mu$ is generated by SM singlets with masses above $\sim$ GeV (lighter singlets will be discovered by upcoming low-energy experiments). This includes the case when the singlets decay invisibly, a scenario that can be challenging to probe at hadron colliders and low energy leptons colliders. If new states with electroweak quantum numbers contribute to $g$-2, the minimal requirements of perturbative unitarity guarantee new charged states below $(\sim 100$ TeV), but this is strongly disfavoured by stringent constraints on charged lepton flavour violating (LFV) decays. New physics theories that satisfy LFV bounds by obeying Minimal Flavour Violation (MFV) and that avoid generating a hierarchy problem, not only for the Higgs but also for the muon mass, require the existence of at least one new charged state below $\sim 10$ TeV. This strongly motivates the construction of a high-energy $\mc$.

\begin{figure}[t]
\center
\vspace*{-10mm}
\hspace{-0.75cm}
\includegraphics[width=2.9in]{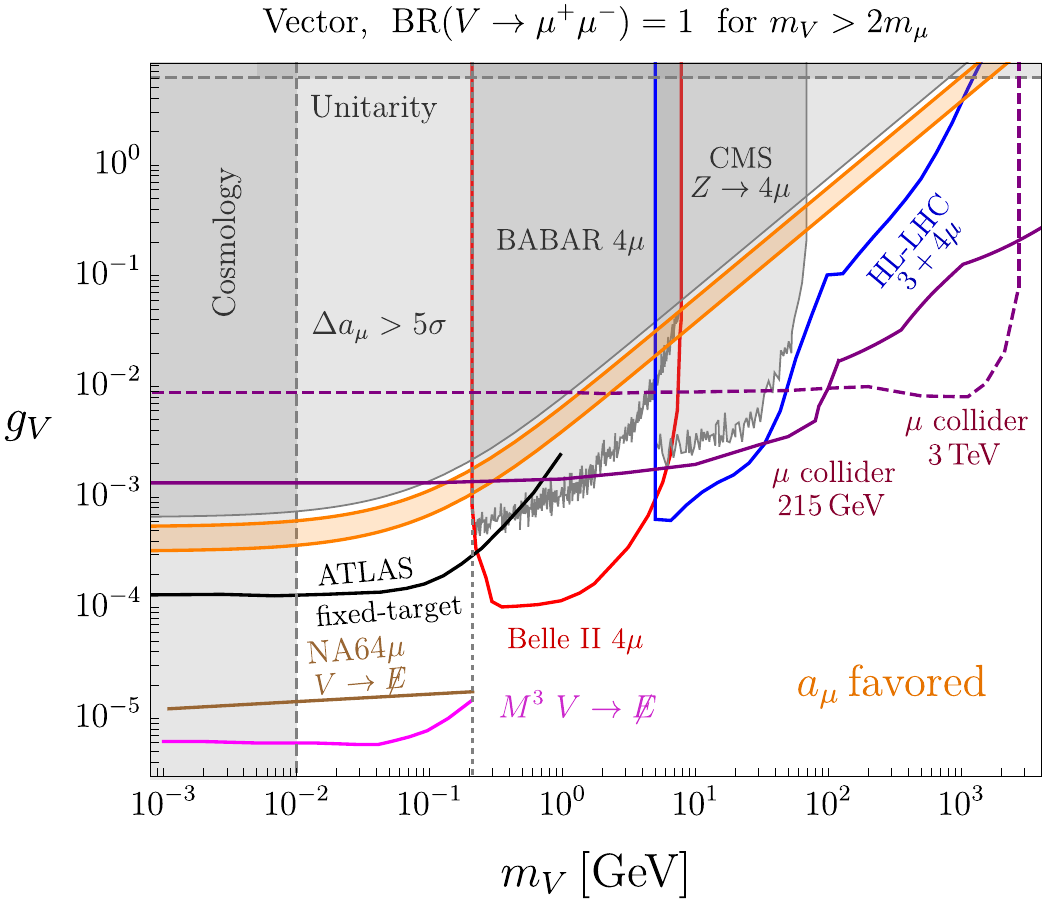} ~
\includegraphics[width=2.9in]{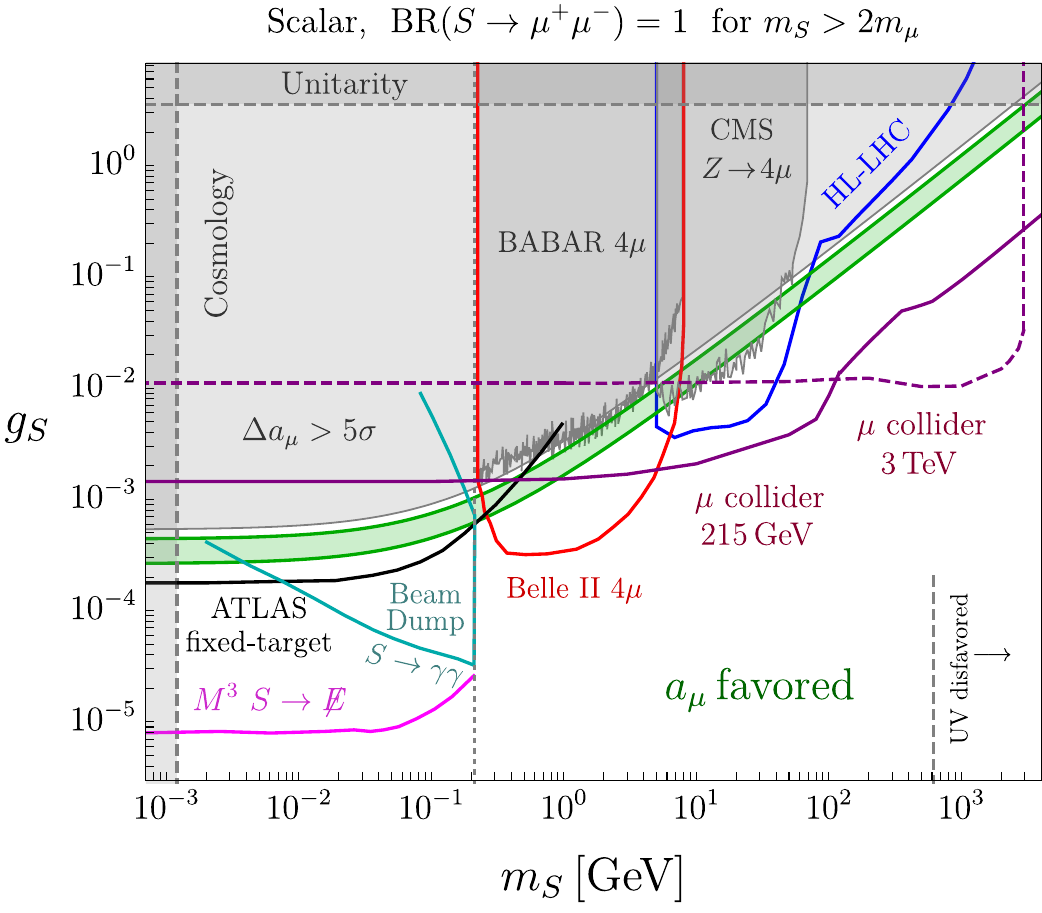}  \\ 
\hspace{-0.75cm}
\includegraphics[width=2.9in]{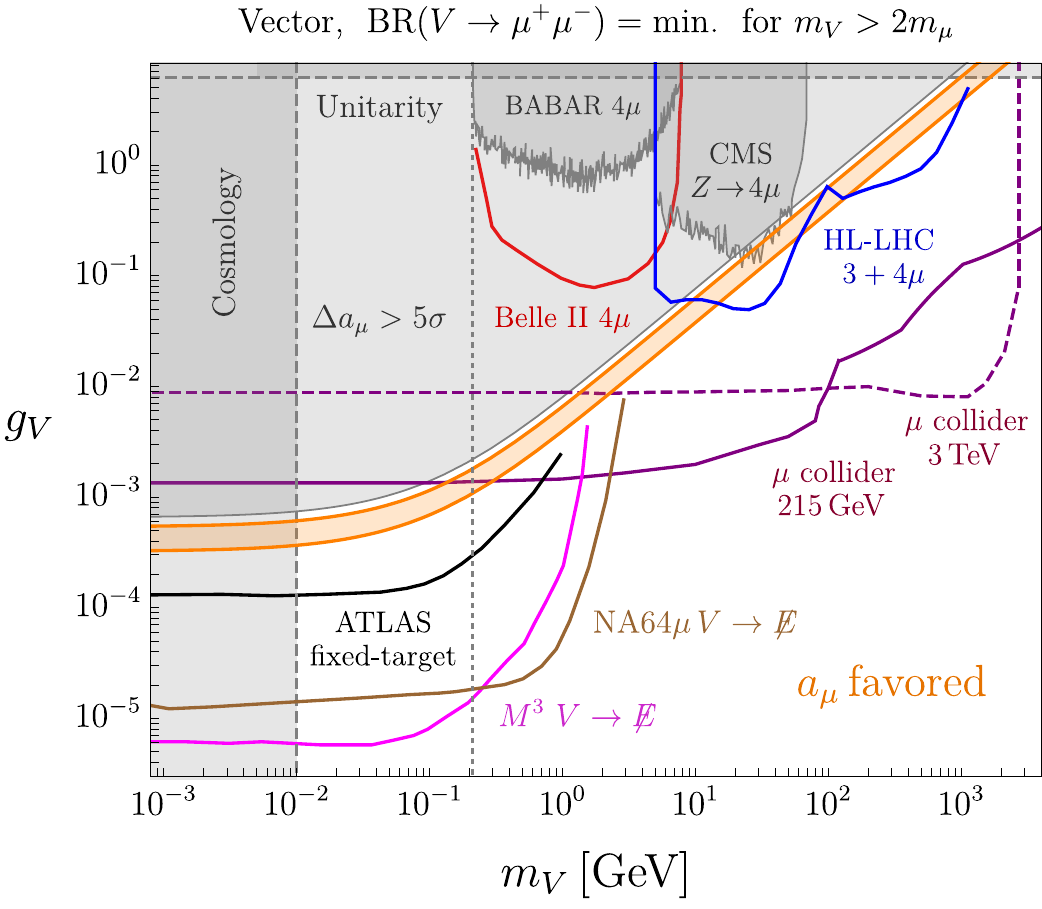} ~
\includegraphics[width=2.9in]{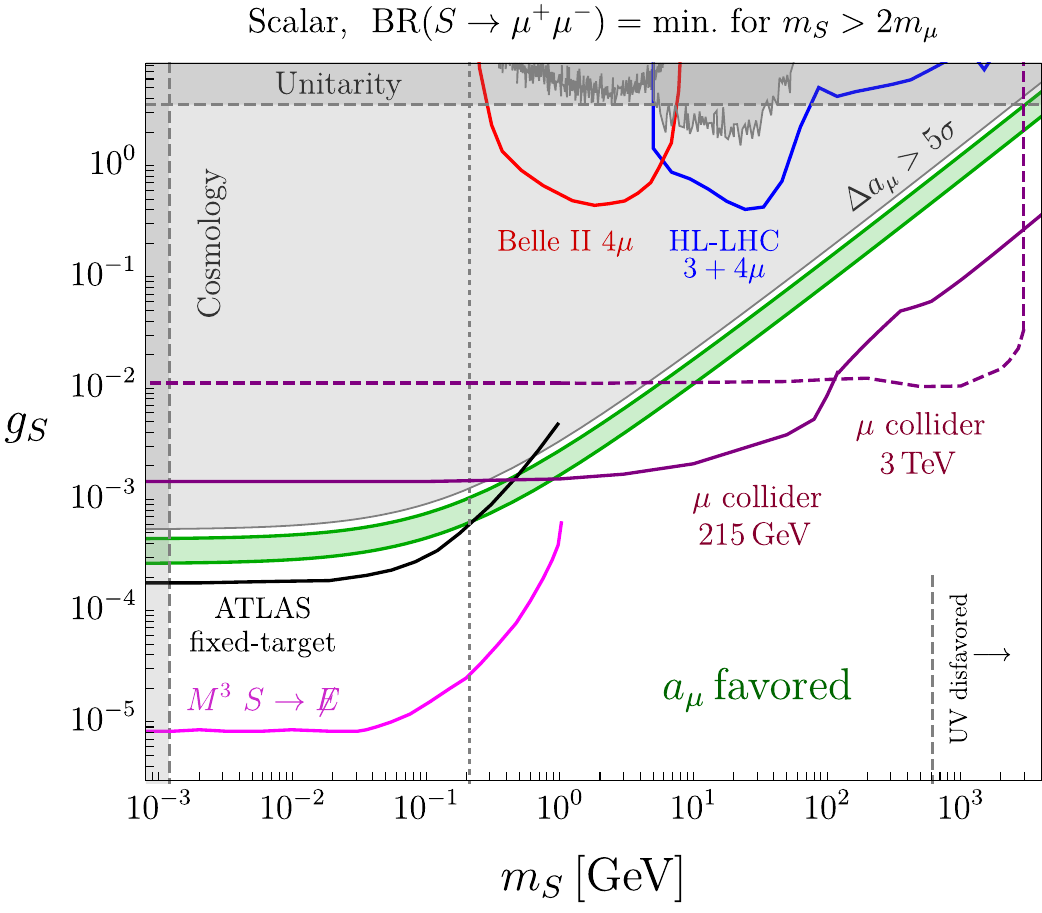} 
\caption{
Singlet models for $g$-2 and their probes at different masses, assuming 100\% branching ratio to di-muons (top) and the minimum branching ratio to di-muon allowed by perturbativity \cite{Capdevilla:2021kcf}.
 \label{babar_singlet}}
\end{figure}

The analysis of~\cite{Capdevilla:2020qel,Capdevilla:2021rwo,Capdevilla:2021kcf} starts with the question {\it What is the highest mass that new particles could have while still generating the measured BSM contribution to $g$-2?} Answering this question is important because knowing the highest mass scale can set the target for the center of mass energy of the $\mc$ needed to detect these new particles. In this analysis, it was assumed that one-loop effects involving BSM states are responsible for the anomaly, since scenarios where new contributions only appear at higher loop order require a lower BSM mass scale to generate the required new contribution.
All possible one-loop BSM contributions to $\Delta a_\mu$ can then be organized into two classes: {\bf Singlet Scenarios:} in which each BSM $g$-2 contribution only involves a muon and a new SM singlet boson that couples to the muon, and {\bf Electroweak (EW) Scenarios:} in which new states with EW quantum numbers contribute to $g$-2. 

\subsubsection*{Singlet mediators}

Throughout this section {\it Singlet Models} refers to the family of models where $\Delta a_\mu$ is generated by a muon-philic singlet, either scalar or vector (where $\mu_L$ and $\mu^c$ are the muon Weyl spinors)
\begin{equation}
 g_S S (\mu_L \mu^c~ + \mu^{c  \dagger} \mu_L^\dagger),~~~~
  g_V  V_\nu (\mu^\dagger_L \bar \sigma^\nu \mu_L + \mu^{c  \dagger}\bar \sigma^\nu \mu^{c})~.
\label{lag-singlets}
\end{equation}

Realizations of these scenarios appear in multiple contexts. For example, vector singlets can be classified either into dark photon or $L_\mu-L_\tau$ like \cite{Greljo:2021npi}. The former are solutions to $g$-2 where couplings between the vector and first generation fermions are generated via loop-induced kinetic mixing. These scenarios are all excluded \cite{Bauer:2018onh,Ballett:2019xoj} or soon to be \cite{Mohlabeng:2019vrz}. The second, $L_\mu-L_\tau$ like scenarios, are vectors that do not couple to first generation fermions. These are highly constrained and a combination of fixed target experiments and muon beam dumps could probe the remaining parameter space \cite{Altmannshofer:2014pba,Krnjaic:2019rsv}. As per singlet scalar UV completinos, one can have models with extra scalars and/or fermions that, after being integrated out, generate the dimension 5 operator $(S/\Lambda)\, H^\dagger L\mu^c$. Once the higgs gets a vev one reproduces the interaction in \ref{lag-singlets}. These models are disfavored for large singlet masses \cite{Capdevilla:2021kcf}.

Figure \ref{babar_singlet} shows the limits and projections on muon-philic vector (left) and scalar (right) singlets,
assuming only di-muon decays where kinematically allowed. The green/orange bands represent the parameter space for which
singlet scalars/vectors resolve $g$-2 within $2\sigma$. Existing experimental limits are shaded in gray, while projections are indicated with colored lines. The  $M^3$ \cite{Kahn:2018cqs}, NA64$\mu$ \cite{Gninenko:2018ter}, and ATLAS fixed-target \cite{Galon:2019owl} experiments 
probe invisibly-decaying singlets; projections here assume a 100\% invisible branching 
fraction. The LHC limits and HL-LHC projections were obtained from $3\mu$/$4\mu$ muon searches.
The purple muon collider projections are obtained from a combination of singlet+photon searches, and from deviations in angular observables of Bhabha scattering~\cite{Capdevilla:2021rwo}.
For scalar singlets whose width is determined entirely by the muon coupling (top right), Fig. \ref{babar_singlet} also shows the projections for a beam
dump search for $S \to \gamma\gamma$ \cite{Chen:2017awl} on the minimal assumption that the scalar-photon 
coupling arises solely from integrating out the muon.
The bottom row shows same as the top row, but assuming that for $m_{S,V}  > 2m_\mu$, the singlets
have the {\it minimum} di-muon branching fraction consistent with unitarity. The curves which
are unaffected by this change of muonic branching fraction correspond to searches that are insensitive to the singlet's decay modes. Projections for $M^3$, NA64$\mu$, and ATLAS fixed-target experiments assume a $\simeq 100\%$ invisible branching fraction for $m_{S/V} > 2m_\mu$, which is model-dependent.


The upshot is that a 3 TeV $\mc$ can directly and indirectly probe the entire space of possible singlet explanations for the $g$-2 anomaly for masses above a few GeV. Muon beamdump experiments, or possibly a lower-energy $\mc$ higgs factory, can probe sub-GeV singlets to provide complete coverage of the possible low-energy solutions to the anomaly.

\subsubsection*{Electroweak mediators}

EW Scenarios can generate the necessary $g$-2 contribution even for NP much above the TeV scale. In particular, the analysis of~\cite{Capdevilla:2020qel,Capdevilla:2021rwo}  carefully studied simplified models featuring new scalars and fermions that yield the \emph{largest possible BSM mass scale} able to account for the anomaly.
By systematically scanning over the entire parameter space of all these models, subject to the constraint that they resolve the $g$-2 anomaly while maintaining perturbative unitarity (as well as other optional constraints), it was possible to derive a model-exhaustive upper bound on the mass of the lightest charged BSM particle that has to exist in order to generate the observed $\Delta a_\mu$. 
The possibility of a high multiplicity of BSM states was also considered by allowing $N_{\rm BSM}$ copies of each BSM model to be present simultaneously. The results show that EW Scenarios must always have at least one new charged state lighter than 

\begin{equation}
\label{e.MchargedmaxXresultpreview}
M^\mathrm{max, X}_\mathrm{BSM, charged} 
\approx
\left( \frac{2.8 \times 10^{-9}}{\Delta a_\mu} \right)^{\frac{1}{2}} \times
\left\{
\begin{array}{rcl}
(100~\TeV)  \ N_{\rm BSM}^{1/2} &   \mathrm{for}  & X = \mbox{(unitarity*)}
\\ \\ 
(20~\TeV)  \ N_{\rm BSM}^{1/2} &   \mathrm{for} & X = \mbox{(unitarity+MFV)} 
\\ \\ 
(20~\TeV)  \ N_{\rm BSM}^{1/6} & \mathrm{for}    & X =  \mbox{(unitarity+naturalness*)}
\\ \\ 
(9~\TeV)  \ N_{\rm BSM}^{1/6} &  \mathrm{for} & X =  \mbox{(unitarity+naturalness+MFV),}
\end{array}
\right.
\end{equation}
where this upper bound is evaluated under four assumptions that the BSM solution to the $g$-2 anomaly must satisfy: perturbative unitarity only; unitarity + MFV; unitarity + naturalness (specifically, avoiding fine-tuning the Higgs and the muon mass); and unitarity + naturalness + MFV. 
The unitarity-only bound represents the very upper limit of what is possible within Quantum Field Theory at the edge of perturbativity, but realizing such high masses requires severe alignment, tuning or another unknown mechanism to avoid stringent constraints from charged lepton flavour-violating (CLFV) decays~\cite{Calibbi:2017uvl,Aubert:2009ag}.
Therefore, every scenario without MFV has been marked with a star (*) above, to indicate additional tuning or unknown flavour mechanisms that have to also be present.

These results, and those from the previous section, have profound implications for the physics motivation of $\mc$. They allow us to formulate a no-lose theorem that can be broken down in chronological progression:

\begin{enumerate}
\item \textbf{Present day:} Confirmation of the $g-$2 anomaly.

\item \textbf{Discover or falsify low-scale Singlet Scenarios $\mathbf{\lesssim}$ GeV:} If Singlet Scenarios with BSM masses below $\sim$ GeV generate the required $\Delta a_\mu$ contribution, multiple fixed-target and $B$-factory experiments are projected to discover new physics in the coming decade.

\item \textbf{Discover or falsify all Singlet Scenarios $\mathbf{\lesssim}$ TeV:} If fixed-target experiments do not discover new BSM singlets that account for $\Delta a_\mu$, a 3 TeV MC with $1~\mathrm{ab}^{-1}$ would be guaranteed to directly discover these singlets if they are heavier than $\sim 10$ GeV. Even a lower-energy machine can be useful: a  215 GeV  muon collider with $0.4~\mathrm{ab}^{-1}$ could directly observe singlets as light as 2 GeV.

\item \textbf{Discover non-pathological Electroweak Scenarios ($\mathbf\lesssim$  10 TeV):} If TeV-scale muon colliders do not discover new physics, the $g$-2 anomaly \emph{must} be generated by EW Scenarios. In that case, all of our results indicate that in most reasonably motivated scenarios, the mass of new charged states cannot be higher than few $\times$ 10 TeV.

\item \textbf{Unitarity Ceiling ($\mathbf\lesssim$  100 TeV):} Even if such a high energy muon collider does not produce new BSM states directly, as we saw in the previous section, a 30 TeV machine would detect deviations in $\mu^+ \mu^- \to h \gamma$, which probes the same effective operator generating $g$-2 at lower energies. This would provide high-energy confirmation of the presence of new physics.
\end{enumerate}

If the $g$-2 anomaly is confirmed, our analysis and the results of the previous section show that finding the origin of this anomaly should be regarded as one of the most important physics motivations for an entire muon collider \emph{program}. Indeed, a  series of colliders with energies from the test-bed-scale $\mathcal{O}$(100 GeV) to the far more ambitious but still imaginable $\mathcal{O}$(10 TeV) scale and beyond has excellent prospects to discover the new particles necessary to explain this mystery.

\subsection{Multi-Higgs Signatures from Vector-like Fermions}
Simple explanations for $g$-2 involve extensions of the SM with new vector-like fermions (VLF) where the corrections to the muon magnetic moment are mediated by the SM Higgs and gauge bosons~\cite{Kannike:2011ng,Dermisek:2013gta}. These models generate effective interactions between the muon and multiple Higgs bosons leading to predictions for di- and tri-Higgs production at a $\mc$ that are directly correlated with the corrections to $\Delta a_{\mu}$. This section reviews the findings of~\cite{Dermisek:2021mhi} on this subject. The authors consider extensions of the SM with VLF doublets, $L_{L,R}$, and singlets $E_{L,R}$ with masses $M_{L,E}$, respectively. It will be assumed that new $L_{L}$ and $E_{R}$ have the same quantum numbers as the SM leptons, but other possibilities will also be commented upon later.

The Yukawa interactions of interest are the following
\begin{flalign}
\mathcal{L}\supset& - y_{\mu}\bar{l}_L\mu_{R}H - \lambda_{E}\bar{l}_{L}E_{R}H  - \lambda_{L}\bar{L}_{L}\mu_{R}H  - \lambda\bar{L}_{L}E_{R}H  - \bar{\lambda}H^{\dagger}\bar{E}_{L}L_{R}  + h.c.,
\label{eq:lagrangian}	
\end{flalign}
where $l_{L}=( \nu_{\mu},  \mu_{L} )^T$,  $ L_{L,R}= ( L_{L,R}^{0}, L_{L,R}^{-})^T$, and $H=(0,\;v + h/\sqrt{2})^{T}$ with $v=174$ GeV.
In the limit $v\ll M_{L,E}$, after integrating out the heavy leptons at tree level, Eq~\ref{eq:lagrangian} becomes
\begin{equation}
\mathcal{L}\supset - y_{\mu}\bar{l}_L\mu_{R}H - \frac{m_{\mu}^{LE}}{v^{3}}\bar{l}_L\mu_{R}H(H^\dagger H) + h.c.,
\label{eq:eff_lagrangian}
\end{equation}
where
\begin{equation}
m_\mu^{LE} \equiv \frac{\lambda_{L} \bar{\lambda} \lambda_{E}}{M_{L}M_{E}} v^3
\label{eq:m^LE}
\end{equation}
is the contribution to the muon mass from mixing with new leptons.
Mixing of the muon with heavy leptons also leads to modifications of the muon couplings to $W$, $Z$, and $h$, and generates new couplings of the muon to new leptons. Assuming that $v\ll M_{L,E}$, the total one-loop correction to $g$-2 induced by these effects is well approximated by~\cite{Kannike:2011ng,Dermisek:2013gta}
\begin{equation}
\Delta a_{\mu}
= - \frac{1}{16\pi^{2}}  \frac{m_\mu m_\mu^{LE}}{v^2}.
\label{eq:dela}
\end{equation}
The explanation of the measured value of $\Delta a_{\mu}$ within $1\sigma$ requires that
\begin{equation}
m_\mu^{LE}/m_\mu = -1.07 \pm 0.25.
\label{eq:mmu_LE_range}
\end{equation}
For couplings of $\mathcal{O}(1)$, Eq (\ref{eq:mmu_LE_range}) can be achieved for new lepton masses even as heavy as 7~TeV while simultaneously satisfying current relevant constraints~\cite{Dermisek:2021ajd}. For couplings close to the limit of perturbativity, $\sqrt{4\pi}$, this range extends to close to 50 TeV. This far exceeds the reach of the LHC and even projected expectations of possible future proton-proton colliders, such as the FCC-hh. However, there are related signals that could be fully probed at, for example, a 3 TeV $\mc$ through the effective interactions generated between the muon and multiple Higgs bosons. These interactions are all generated by Eq.~\ref{eq:eff_lagrangian}~\cite{Dermisek:2021mhi} and they lead to the following predictions
\begin{eqnarray}
\sigma_{\mu^+\mu^- \to hh} &=&  \frac{\left|\lambda^{hh}_{\mu \mu}\right|^2}{64 \pi}   = \frac{9}{64 \pi} \left(\frac{m_\mu^{LE}}{v^2}\right)^2 \label{eq:sigma_hh},\label{eq:EFT_xsections_1}\\
\sigma_{\mu^+\mu^- \to hhh}  &=& \frac{\left|\lambda^{hhh}_{\mu \mu}\right|^2}{6144 \pi^3} s = \frac{3}{4096 \pi^3} \left(\frac{m_\mu^{LE}}{v^3}\right)^2 s \label{eq:sigma_hhh}.
\label{eq:EFT_xsections_2}
\end{eqnarray}
\begin{figure}[t]
\begin{center}
\includegraphics[width=0.46\linewidth]{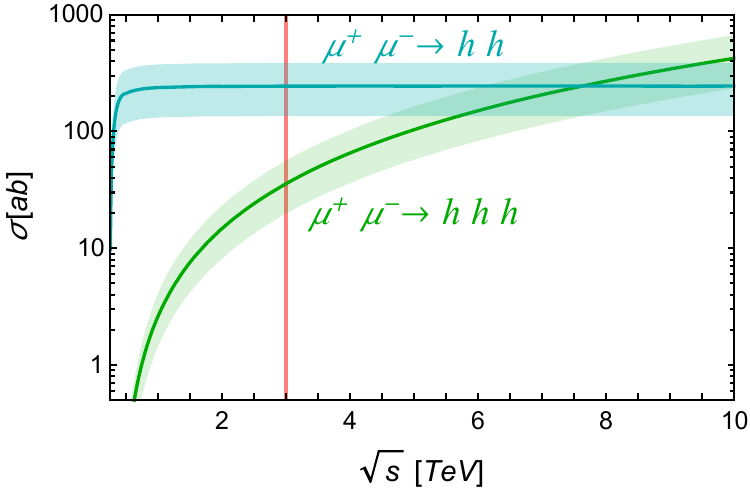}~\includegraphics[width=0.5\linewidth]{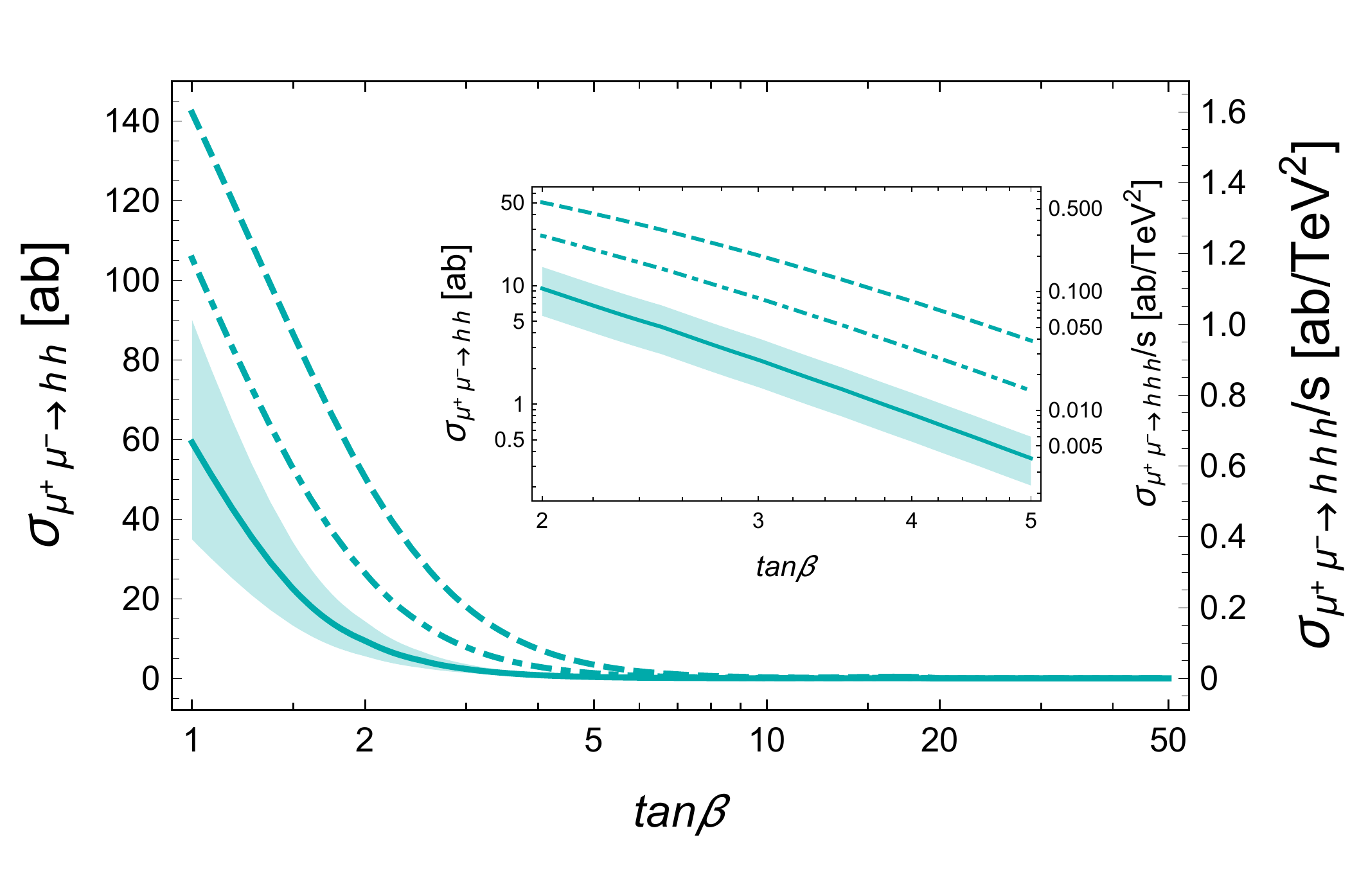}
\end{center}
\vspace{-0.5cm}
\caption{{\it Left:} Cross sections for $hh$ (cyan) and $hhh$ (green) production as a function of $\sqrt{s}$ in models with VLF. {\it Right:} Cross sections for $hh$ (left axis) and $hhh$ (right axis) production as a function of $\tan\beta$ in models with VLF and 2HDM for $M_{L,E}\simeq m_{H,A,H^{\pm}}$. The dot-dashed and dashed lines correspond to the predictions corresponding to the central value of $\Delta a_{\mu}$ and $m_{H,A,H^{\pm}}=3\times M_{L,E}$ and $m_{H,A,H^{\pm}}=5\times M_{L,E}$, respectively. Both panels assume $\Delta a_{\mu}$ is within $1\sigma$ of the measured value (shaded ranges)~\cite{Dermisek:2021mhi}.}
\label{fig:xsection_plot}
\end{figure} 
Thus, considering Eq~\ref{eq:dela}, one can see that the effective interactions of the muon with the Higgs are completely fixed by the muon mass and the predicted value of $\Delta a_{\mu}$.
Fig~\ref{fig:xsection_plot}, shows the total $\mu^{+}\mu^{-}\to hh$ and $\mu^{+}\mu^{-}\to hhh$ cross sections at a $\mc$ as a function of $\sqrt{s}$ calculated from the effective lagrangian and assuming that $\Delta a_{\mu}$ is achieved within $1\sigma$ (shaded ranges). Cross sections for a 3 TeV $\mc$ are highlighted with the red line. One can see that, for example, a $\mc$ running at $\sqrt{s}= 3$ TeV with 1 ab$^{-1}$ of integrated luminosity would see about 240 di-Higgs events and about 35 tri-Higgs events. It should be noted that already at $\sqrt{s}=1$ TeV this is roughly 4 (3) orders of magnitude larger than $\mu^{+}\mu^{-}\to hh$ and $\mu^{+}\mu^{-}\to hhh$ in the SM. Note that di- and tri-Higgs signals produced from vector boson fusion in the SM appear with additional particles in the final state and can be easily vetoed in a dedicated analysis. Similarly, backgrounds involving the $Z$-boson which may be comparable at the level of cross sections, e.g. $\mu^{+}\mu^{-}\to Zh$ or $\mu^{+}\mu^{-}\to ZZ$, can also be easily suppressed via invariant mass cuts on the $Z$-boson masses once the relevant decays are taken into account in a given analysis.

Models with more exotic quantum numbers can also generate a similar correction to $\Delta a_{\mu}$ and, hence, similar predictions for di- and tri-Higgs cross sections. In total there are 5 different combinations of new lepton fields that can lead to mass-enhanced corrections to $\Delta a_{\mu}$ mediated by the SM Higgs. In each case, the correction as given in Eq~\ref{eq:dela} is simply multiplied by a corresponding $c$-factor. The resulting cross sections are then rescaled by a factor of $1/c^{2}$ compared to those in Fig~\ref{fig:xsection_plot}. Table~\ref{table:models} lists the 5 possible models, $c$-factor multiplying Eq~\ref{eq:dela}, and corresponding predictions for di- and tri-Higgs cross sections for a $\mc$ running at $\sqrt{s}=3$ TeV, assuming $\Delta a_{\mu}\pm 1\sigma$. A $\mc$ running even at moderate center-of-mass energies, $\sqrt{s}\sim 1-3$ TeV, can fully probe these scenarios. 

\subsubsection*{Vector-like fermions and Two-Higgs-Doublet models}

It is straightforward to extend the discussion from the previous section to a 2HDM (or any model where the Higgs acts as one component of the sector triggering EWSB)~\cite{Dermisek:2020cod,Dermisek:2021ajd}. For instance, in a type-II 2HDM where charged leptons couple exclusively to one Higgs doublet, $H_{d}$, (which can be achieved by assuming a $Z_{2}$ symmetry) the lagrangian in Eq.~\ref{eq:lagrangian} from the previous section, is simply modified with the replacement $H\to H_{d}$. In this case both Higgs doublets develop a vev $\left\langle H_{d}^{0} \right\rangle=v_{d}$ and $\left\langle H_{u}^{0} \right\rangle=v_{u}$, where $\sqrt{v_{d}^{2}+v_{u}^{2}}=v=174$ GeV and $\tan\beta = v_{u}/v_{d}$. The effective interactions generated by integrating out heavy leptons is then
\begin{equation}
\mathcal{L}\supset y_{\mu}\bar{\mu}_{L}\mu_{R}H_{d} - \frac{m_{\mu}^{LE}}{v_{d}^{3}}\bar{\mu}_{L}\mu_{R}H_{d}(H_{d}^{\dagger} H_{d}).
\label{eq:eff_lagrangian_2HDM}
\end{equation}
Similar modifications to $Z$, $W$, and the SM-like Higgs couplings to the muon are also generated after EWSB.
Including the additional corrections to $\Delta a_{\mu}$ from heavy charged and neutral Higgs bosons leads to~\cite{Dermisek:2020cod,Dermisek:2021ajd}
\begin{equation}
\Delta a_{\mu}=-\frac{1+\tan^{2}\beta}{16\pi^{2}}\frac{m_{\mu}m_{\mu}^{LE}}{v^{2}},\;\;\;\;\;\;\;m_\mu^{LE} \equiv \frac{\lambda_{L} \bar{\lambda} \lambda_{E}}{M_{L}M_{E}} v_{d}^3,
\label{eq:gm2}
\end{equation}
where $M_{L,E}\simeq m_{H,A,H^{\pm}}$ is assumed for simplicity. The first term in Eq~\ref{eq:gm2} results from the same loops as in the SM, i.e. involving the $Z$, $W$, and SM-like Higgs, whereas the second term, enhanced in comparison by $\tan^{2}\beta$, results from the additional contributions from the heavy Higgses. The corresponding requirement to satisfy $\Delta a_{\mu}$ within $1\sigma$ then becomes
\begin{equation}
m_{\mu}^{LE}/m_{\mu}=(-1.07\pm0.25)/(1+\tan^{2}\beta).
\label{eq:gm2_range}
\end{equation}
Just as in the previous section, effective interactions between the muon and multiple Higgs bosons are generated via the single dimension-six operator in Eq~\ref{eq:eff_lagrangian}. Thus, predictions for di- and tri-Higgs cross sections follow in the same way simply by replacing $m_{\mu}^{LE}$ with the corresponding definition in Eq~\ref{eq:gm2}. Considering Eq~\ref{eq:gm2_range}, it follows that $\sigma_{\mu^{+}\mu^{-}\to hh}$ and $\sigma_{\mu^{+}\mu^{-}\to hhh}$ cross sections in a type-II 2HDM decrease as $1/\tan^{4}\beta$. 

\begin{table}[t]
\begin{center}
\begin{tabular}{ |c||c||c|c| } 
\hline
$SU(2)\times U(1)_{Y}$& c & $\sigma_{hh}(3\;\text{TeV})$[ab] & $\sigma_{hhh}(3\;\text{TeV})$[ab] \\
\hline
$\mathbf{2}_{-1/2}\oplus\mathbf{1}_{-1}$ & 1 & $244^{+141}_{-109}$& $35.8^{+20.8}_{-15.9}$ \\
\hline
$\mathbf{2}_{-1/2}\oplus\mathbf{3}_{-1}$  & 5 &$10^{+6}_{-4}$&$1.43^{+0.8}_{-0.6}$\\
\hline
$\mathbf{2}_{-3/2}\oplus\mathbf{1}_{-1}$ & 3 & $27^{+16}_{-12}$ & $4.0^{+2.3}_{-1.8}$\\
\hline
$\mathbf{2}_{-3/2}\oplus\mathbf{3}_{-1}$ & 3 &$27^{+16}_{-12}$ & $4.0^{+2.3}_{-1.8}$\\
\hline
$\mathbf{2}_{-1/2}\oplus\;\mathbf{3}_{0}$ &1 & $244^{+141}_{-109}$& $35.7^{+20.7}_{-15.9}$\\
\hline
\end{tabular}
\caption{Quantum numbers of $L_{L,R}\oplus E_{L,R}$ under $SU(2)\times U(1)_{Y}$, corresponding $c$-factor for $\Delta a_{\mu}$, and predictions for di- and tri-Higgs cross sections running at $\sqrt{s}=3$ TeV, assuming $\Delta a_{\mu}\pm 1\sigma$.}
\label{table:models}
\end{center}
\end{table}

Fig~\ref{fig:xsection_plot} shows the $\tan\beta$ dependence of $\sigma_{\mu^{+}\mu^{-}\to hh}$ and $\sigma_{\mu^{+}\mu^{-}\to hhh}/s$ calculated from the effective lagrangian when $\Delta a_{\mu}$ is achieved within $1\sigma$ (shaded range) and $M_{L,E}\simeq m_{H,A,H^{\pm}}$. The dot-dashed and dashed lines correspond to the predictions corresponding to the central value of $\Delta a_{\mu}$ and $m_{H,A,H^{\pm}}=3\times M_{L,E}$ and $m_{H,A,H^{\pm}}=5\times M_{L,E}$, respectively. Its expected that future measurements of $h\to \mu^{+}\mu^{-}$ will probe $\tan\beta$ up to $\sim 5$ and the inset zooms into this region~\cite{Dermisek:2021mhi}.

For a $\mc$ running at center-of-mass energy of 3 TeV with, for example, 1 ab$^{-1}$ of luminosity 3 di-Higgs events are expected in these scenarios for $\tan\beta \simeq 3$. For tri-Higgs the same sensitivity does not extend much above $\tan\beta \simeq 1$. When $m_{H,A,H^{\pm}}=5\times M_{L,E}$, the corresponding sensitivities to $\tan\beta$ increase to about $\tan\beta\simeq 5$ and 2.5 for di-Higgs and tri-Higgs signals, respectively.

These conclusions also extend to models with additional scalars where the SM Higgs is only one component of the scalar sector responsible for EWSB. Mixing within the Higgs sector (e.g. $\tan\beta$ in a 2HDM) introduces a free parameter to the predictions and correlations between the muon magnetic moment and effective Higgs couplings. Thus, the corresponding predictions for di- and tri-Higgs signals at a $\mc$ are not as sharp in these scenarios as compared to the SM. Though in a 2HDM the observables parametrically interpolate between the SM and models with scalars that do not participate in EWSB.

\section{Lepton Flavour Universality and B physics \label{sec:Banomaly}}

The rich set of observed deviations from SM predictions in rare semileptonic $B$-meson decays, induced by the $b \to s \mu^+ \mu^-$ partonic transition, represents a compelling hint for new physics.
If confirmed by forthcoming experimental investigation, these observations would not only constitute the first evidence for physics beyond the SM, but would also signal a breaking of Lepton Flavor Universality (LFU) beyond the Yukawa interactions.

In particular, LFU ratios are a very clean and robust probe of new physics, due to the highly suppressed hadronic uncertainties that cancel out in the ratios
\begin{equation}
R_X = \frac{{\rm BR}(B \to X \mu^+\mu^-)}{{\rm BR}(B\to X e^+e^-)},
\end{equation}
where $X$ is a final state involving an $s$ quark. All these ratios are predicted to be equal to 1 with very high accuracy in the SM, in the limit where the lepton masses can be neglected.

The LHCb collaboration has measured LFU ratios that deviate from the SM expectation in various channels. The most precisely measured decays are $B^+\to K^+\ell^+\ell^-$ and $B^0 \to K^{*0} \ell^+\ell^-$ in the di-lepton invariant mass region $q^2 \in [1.1,6]\, {\rm GeV}^2$, which yield~\cite{Aaij:2017vbb,Aaij:2021vac}
\begin{equation}
R_K = 0.846^{+0.042+0.013}_{-0.039-0.012},\qquad R_{K^*} = 0.685^{+0.113}_{-0.069}\pm 0.047,
\end{equation}
showing a discrepancy with the SM predictions~\cite{Bobeth:2007dw,Bordone:2016gaq,Capdevila:2017bsm,Alok:2017sui} of 3.1$\sigma$ and 2.4$\sigma$, respectively. Further LFU measurements with the isospin partners $B^0\to K_S\ell^+\ell^-$ and $B^+\to K^{*+}\ell^+\ell^-$, and in the baryonic decay $\Lambda_b \to pK\ell^+\ell^-$, have a lower significance, but show similar deviations. While the individual significance of these measurements is still low, the consistency of the deviations in many different clean channels can be interpreted as a strong hint for new physics.

At the same time, deviations from SM predictions are observed in various $b\to s \mu^+\mu^-$ decays, both in absolute branching ratios and in angular distributions. While more affected by hadronic uncertainties than the LFU ratios, these measurements are clean from an experimental point of view, and strengthen the evidence for new physics coupled to heavy quarks and muons (rather than electrons). Global fits of the various $b\to s\mu\mu$ anomalies have been performed~\cite{Altmannshofer:2021qrr,Aebischer:2019mlg}, reporting strong evidence for BSM interactions.

The effective Lagrangian responsible for semi-leptonic $b \rightarrow s \mu^{+} \mu^{-}$-transitions can be expressed as ($V$ denotes the CKM matrix)
\begin{equation}\label{eq:Leff}
\mathcal{L}_{b \rightarrow s \mu \mu}^{\mathrm{NP}} \supset \frac{4 G_{\rm F}}{\sqrt{2}} V_{t b} V_{t s}^{*}\left(C_{9}^{\mu} O_{9}^{\mu}+C_{10}^{\mu} O_{10}^{\mu}\right)+\mathrm{h.c.} 
\end{equation}
with the relevant operators 
\begin{equation}
\begin{aligned}
O_{9}^{\mu} &=\frac{\alpha}{4 \pi}\left(\bar{s}_{\rm L} \gamma_{\mu} b_{\rm L}\right)\left(\bar{\mu} \gamma^{\mu} \mu\right) ,\\
O_{10}^{\mu} &=\frac{\alpha}{4 \pi}\left(\bar{s}_{\rm L} \gamma_{\mu} b_{\rm L}\right)\left(\bar{\mu} \gamma^{\mu} \gamma_{5} \mu\right) .
\end{aligned}
\end{equation}
Using these operators to explain the anomalies leads to best-fit values of the Wilson-coefficients  $ C_9 = -C_{10} = -0.43$, with the $1\sigma$ range being  $ [-0.50,-0.36]$~\cite{Altmannshofer:2021qrr,Aebischer:2019mlg}. This corresponds to a new physics scale of $\Lambda = 39$\,TeV.  Perturbative unitarity analysis suggests new mass thresholds below $\lesssim 100$\,TeV.

Should these hints for Lepton Flavour Universality be confirmed by upcoming measurements, a major goal of HEP will be to understand the nature of the underlying new physics. Given the high EFT scale required to fit the deviation it is possible, and likely, that such NP is too heavy to be observed at the LHC. A more powerful collider would therefore be needed. In this Section we find the reach of a $\mc$ on the NP responsible for the $B$-anomalies, both from the EFT perspective as well as considering some of the NP scenarios more commonly known in the literature.

\subsection{\emph{Nightmare} scenario: contact interactions}\label{ci}

In this Section we consider the pessimistic scenario where the new physics states responsible for the anomalies are much heavier than the colliders' energy reach for on-shell production even at future colliders.\footnote{The set of such models is not any empty set. To name one explicit example, a scalar leptoquark mediator $S_3$~\cite{Gherardi:2020qhc} with a conserved baryon and a muon number which would explain almost a minimal set of couplings needed to fit the anomaly~\cite{Greljo:2021xmg} can be as heavy as 69\,TeV and still pass all the complementary experimental bounds and perturbative unitarity~\cite{DiLuzio:2019jyq}. This is far beyond the reach for on-shell production at any considered future collider.} Nonetheless, the effect of these new states can be captured by contact interactions that would leave a trace in the high-invariant mass tails at the energy frontier providing a complementary information about the new physics~\cite{Greljo:2017vvb}. For example, measuring such interactions and establishing a correlation with the low-energy observables would exclude light mediators and potentially uncover other properties of new physics. 

\begin{figure}[p]
\centering
\includegraphics[height=8cm]{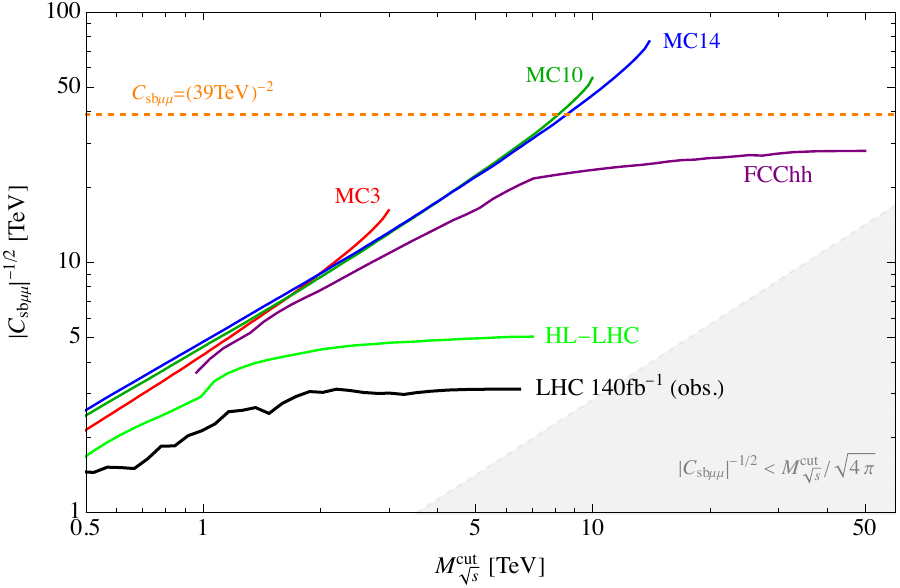} \\ \vspace{0.5cm}
\includegraphics[height=8cm]{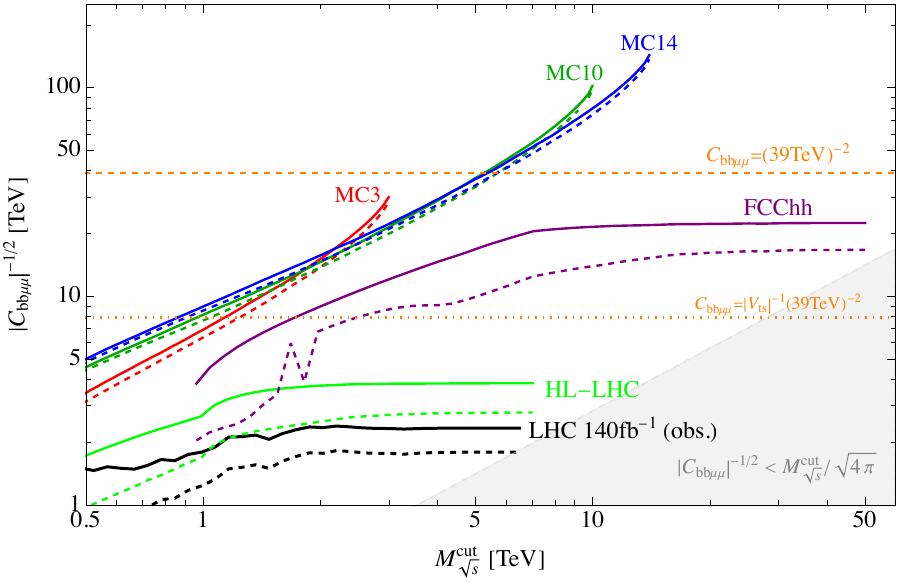}
\caption{\label{fig:EFTlimits_RKMC} Sensitivity reach (at 95\% CL) for the $(\bar{s}_L \gamma_\alpha b_L) (\bar{\mu}_L \gamma^\alpha \mu_L)$ (top) and  $(\bar{b}_L \gamma_\alpha b_L) (\bar{\mu}_L \gamma^\alpha \mu_L)$ (bottom) contact interactions as a function of the upper cut on the final-state invariant mass for various $\mc$, HL-LHC, FCC-hh, and the present LHC bounds. These are compared with values required to fit $b \to s \mu^+ \mu^-$ anomalies without (dashed orange line) or with (dotted orange line) a flavor enhancement of the $bb$ operator compared to the $bs$ one. For the bottom plot solid (dashed) lines represent the limit for positive (negative) values of $C_{bb\mu\mu}$. The gray area represents a region where the EFT bounds are not valid (for a strongly coupled UV completion, for weakly coupled ones the area is larger).}
\end{figure}

The most pessimistic case would be to assume that only the contact interaction behind the anomalies, $(\bar{s}_L \gamma_\alpha b_L)(\bar \mu_L \gamma^\alpha \mu_L)$, is important at high-$p_T$. However, realistic models in general also induce contributions to quark flavor conserving operators. We thus also consider the four-fermion operator $(\bar{b}_L \gamma_\alpha b_L)(\bar \mu_L \gamma^\alpha \mu_L)$. To summarise, the contact interactions we consider are:
\begin{equation}\label{eq:nightmare_EFT}
	\mathcal{L}_{\rm EFT} = C_{bb\mu\mu} \; (\bar{b}_L \gamma_\alpha b_L)(\bar \mu_L \gamma^\alpha \mu_L) + \left[ C_{sb\mu\mu} \; (\bar{s}_L \gamma_\alpha b_L)(\bar \mu_L \gamma^\alpha \mu_L)  + \text{h.c.} \right]
\end{equation}

Here we calculate and compare the reach on these intereactions at the following colliders
\begin{center}
\begin{tabular}{ | c | c | c | c | }
\hline
\textbf{Collider} & \textbf{C.o.m. Energy} & \textbf{Luminosity} & \textbf{Label} \\
\hline
LHC Run-2 \cite{CMS:2021ctt} & 13 TeV & 140 fb$^{-1}$ & LHC \\
HL-LHC & 14 TeV & 6 ab$^{-1}$ & HL-LHC \\
FCC-hh & 100 TeV & 30 ab$^{-1}$ & FCC-hh \\
$\mc$ & 3 TeV & 1 ab$^{-1}$ & MC3 \\
$\mc$ & 10 TeV & 10 ab$^{-1}$ & MC10 \\
$\mc$ & 14 TeV & 20 ab$^{-1}$ & MC14 \\
\hline
\end{tabular}
\end{center} 

For the hadron colliders we study the high-energy di-muon production, $p p \to \mu^+ \mu^-$, while, for the $\mc$ we consider inclusive high-energy di-jet production via $\mu^- \mu^+ \to j j$. For $\mc$ we take into account the full EW PDF of the muon, obtained by numerically solving the DGLAP evolution of the partonic distribution functions inside the muon using QED+QCD interactions below the EW scale and the full unbroken SM interactions above \cite{Han:2020uid,Han:2021kes,muPDF}. We checked that the purely QCD dijet cross section, initiated by quarks and gluons inside the muon, is always completely negligible with respect to the muon-initiated Drell-Yan one. On top of the statistical uncertainty, we include a 2\% systematic uncertainty in each bin. While some improvement in sensitivity is expected by requiring one (or both) jet to be $b$-tagged, the overall picture will not change drastically for both hadron and $\mc$ \cite{Afik:2019htr,Marzocca:2020ueu}, therefore we just consider the inclusive cross section at this point.
For more details we refer to \cite{RKMC}.

Our results are collected in Fig.~\ref{fig:EFTlimits_RKMC}, where we show the expected 95\%CL sensitivity as a function of the upper cut on the invariant mass of the final state for different colliders. The present LHC bounds with 140 fb$^{-1}$ of luminosity are shown in black \cite{CMS:2021ctt}. The dashed orange line is the reference value for $C_{sb\mu\mu}$ required to fit the anomalies, while for $C_{bb\mu\mu}$ we also show as a dotted orange line a reference value where this flavor conserving interaction is enhanced by a factor of $1/|V_{ts}| \approx 25$ with respect to the flavor violating one, as expected in many realistic scenarios~\cite{Buttazzo:2017ixm}.

\subsection{$Z'$ models}

A few explicit mediators can give rise to the effective interactions of Eq.~\ref{eq:nightmare_EFT} (see e.g.~\cite{Buttazzo:2017ixm}). The simplest possibility is perhaps a new vector boson coupled to quarks and muons. This scenario was studied at a muon collider in~\cite{Huang:2021biu}, whose key findings are reviewed here.
The authors consider a $Z^{\prime}$ which couples non-universally to a left-handed lepton current, and to the left-handed flavor-changing $(bs)$ quark current.
The Lagrangian relevant for $b \rightarrow s \mu^{+} \mu^{-}$ transitions is
\begin{equation}
\mathcal{L}_{Z^{\prime}} \supset \left(\lambda_{i j}^{\rm Q} \bar{d}_{\rm L}^{i} \gamma^{\mu} d_{\rm L}^{j}+\lambda_{\alpha \beta}^{\rm L} \bar{\ell}_{\rm L}^{\alpha} \gamma^{\mu} \ell_{\rm L}^{\beta}\right) Z_{\mu}^{\prime} \;,
\end{equation}
where $\ell^{i}$  and $d^{i}$ represent the corresponding generations of charged leptons and down-type quarks.

Integrating out the $Z'$ field yields the following effective Lagrangian:
\begin{equation}
\begin{aligned}
\mathcal{L}_{Z^{\prime}}^{\mathrm{eff}} &=-\frac{1}{2 M_{Z^{\prime}}^{2}}\left(\lambda_{i j}^{\rm Q} \bar{d}_{\rm L}^{i} \gamma_{\mu} d_{\rm L}^{j}+\lambda_{\alpha \beta}^{\rm L} \bar{\ell}_{\rm L}^{\alpha} \gamma_{\mu} \ell_{\rm L}^{\beta}\right)^{2} \\
& \supset-\frac{1}{2 M_{Z^{\prime}}^{2}}\left[\left(\lambda_{23}^{\rm Q}\right)^{2}\left(\bar{s}_{\rm L} \gamma_{\mu} b_{\rm L}\right)^{2} 
+2 \lambda_{23}^{\rm Q} \lambda_{22}^{\rm L}\left(\bar{s}_{\rm L} \gamma_{\mu} b_{\rm L}\right)\left(\bar{\mu}_{\rm L} \gamma^{\mu} \mu_{\rm L}\right)+\mathrm{h.c.}\right].
\end{aligned}
\end{equation}
Now one can obtain the relevant Wilson coefficients at  tree-level by matching onto the effective Lagrangian for the low-energy observables, Eq.~\ref{eq:Leff}, at the scale $\mu=M_{Z^{\prime}}$ as
\begin{equation} \label{eq:wc_zp}
C_{9}^{\mu}=-C_{10}^{\mu}=-\frac{\pi}{\sqrt{2} G_{\rm F} M_{Z^{\prime}}^{2} \alpha}\left(\frac{\lambda_{23}^{\rm Q} \lambda_{22}^{\rm L}}{V_{t b} V_{t s}^{*}}\right).
\end{equation}

At a $\mc$, the vector boson $Z'$ can mediate $bs$ production via $s$-channel. This process is related to the simple $Z'$-mediated process $b \to s \mu^+ \mu^-$ needed for the ${R_{K^{(*)}}}$ anomaly by a crossing symmetry, and it enables a robust direct test to the $Z'$ interpretation~\cite{Huang:2021biu}.

Because of the limited power of flavor reconstruction, the major background of the $bs$ final state comes from the SM dijet signals, namely $\mu^+ \mu^- \to j  j$ with $j$ being $u$, $d$, $s$, $c$ and $b$.
The final sensitivity is subject to the $b$-jet tagging efficiency and the mistag rate.
This study assumed a conservative experimental performance, with the $b$-jet tagging efficiency being $\epsilon^{}_{b} = 70\%$~\cite{Liu:2021jyc} and mistag rates being $\epsilon^{}_{uds} = 1\%$ and $\epsilon^{}_{c} = 10\%$.
While counting the signal events, it is required that one of the jets is successfully tagged as a $b$ jet, while the other is not. 
The $t$-jet should be able to be clearly separated from $b$-jet with proper cuts on the jet structure.

The sensitivity is studied at the parton level for the $\mc$ setup $\sqrt{s} = 3~{\rm TeV}$ and $L=1~{\rm ab}^{-1}$ by counting the event number with respect to the polar angle.
For this purpose, the following chi-square
\begin{eqnarray} \label{eq:}
	\chi^2 = \sum^{}_{i}\frac{(N^{}_{i}-\widetilde{N}^{}_{i})^2}{N^{}_{i} + \epsilon^2 \cdot N^{2}_{i}}
	\;,
\end{eqnarray}
is defined, where $i$ sums over polar angles with a bin size of $\cos{\theta}=0.1$,
$N^{}_{i}$ is the predicted total event number of signal plus SM backgrounds, $\widetilde{N}^{}_{i}$ is SM only event number, and we fix the possible systematic error $\epsilon$ as  $0.1\%$.

\begin{figure}[t!]
\begin{center}
\includegraphics[width=0.51\textwidth]{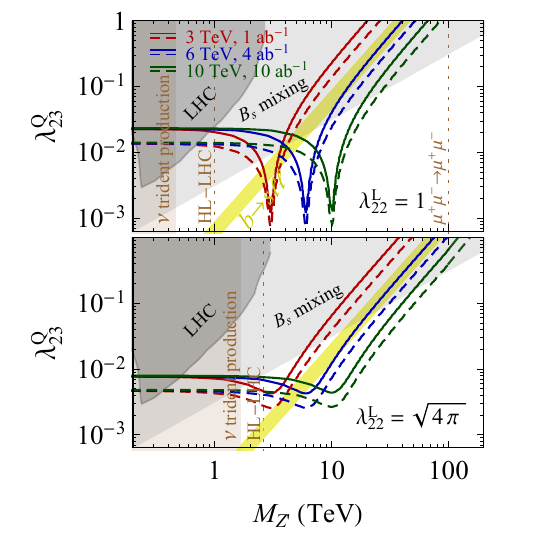}\hspace{-1cm}\includegraphics[width=0.51\textwidth]{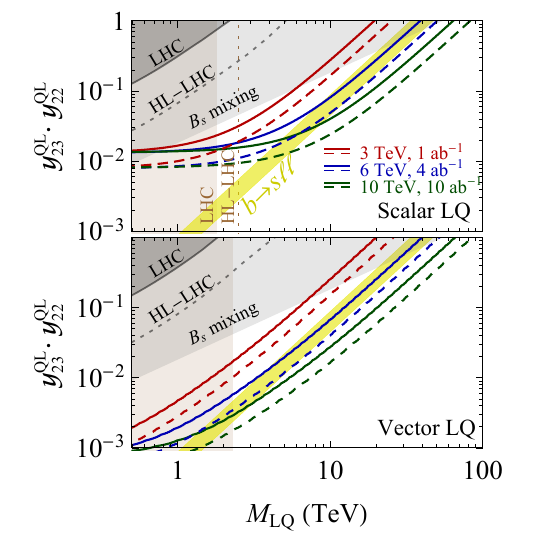} 
\vspace{-0.5cm}
\end{center}
\caption{{\it Left:} Sensitivities to the $Z'$ model with $\lambda^{\rm L}_{22} = 1$ (upper panel) and $\lambda^{\rm L}_{22} = \sqrt{4\pi}$ (lower panel) via  $\mu^{+}\mu^{-} \to b \bar s$ at a $\mc$ with $\sqrt{s} = 3,6,10~{\rm TeV}$ (red, blue, green). Other limits include the neutrino trident production~\cite{Altmannshofer:2014pba}, LHC~\cite{Allanach:2015gkd}, HL-LHC~\cite{delAguila:2014soa}, and $B^{}_{s}$ mixing~\cite{DiLuzio:2019jyq}. {\it Right:} Sensitivities to the LQ model via  $\mu^{+}\mu^{-} \to b \bar s$ at a $\mc$ for scalar (upper) and vector (lower) LQ. Figures from Ref.~\cite{Huang:2021biu}.}
\label{fig:contours_Zp}
\end{figure}

The final sensitivity to $Z'$ connecting the $\mu\mu$ and $bs$ currents is shown in the left panels of Fig.~\ref{fig:contours_Zp}.
The red curves mark the sensitivity of the $\mc$ with $\sqrt{s} = 3~{\rm TeV}$ and $L=1~{\rm ab}^{-1}$ if we take $\lambda^{\rm L}_{22} = 1$ (upper panel) or $\lambda^{\rm L}_{22} = \sqrt{4\pi}$ (lower panel). 
Note that large $\lambda^{\rm L}_{22}$ is needed because 
$\lambda^{\rm Q}_{23}$ is strongly constrained by $B_s-B_{\bar s}$ mixing. 
The solid and dashed curves represent the cases without and with flavor tagging, respectively.
The parameter space of $Z'$ explaining the ${R_{K^{(*)}}}$ anomaly is given as the yellow band, which is actually limited by neutrino trident production and $B^{}_{s}$ mixing.
If $\lambda^{\rm L}_{22} = 1$ is assumed, the $Z'$ parameter space which survives in explaining the ${R_{K^{(*)}}}$ anomaly (yellow bands) can be largely covered. Even though it is not shown here, it is expected that the radiative return process, $\mu^+ \mu^- \to b s \gamma$, will explore the rest of the surviving parameter space.
Moreover, it is clear that a higher energy collider can probe higher $Z'$ masses. This is helpful to probe the ${R_{K^{(*)}}}$ anomaly when a larger $\lambda^{\rm L}_{22}$ is taken. For instance, 
for $\lambda^{\rm L}_{22} = \sqrt{4\pi}$ the $\mc$ with $\sqrt{s} = 6~{\rm TeV}$ will rule out most of the favored parameter space.

\subsection{Scalar Leptoquarks}

In order to address the $R_{K^{(*)}}$ anomaly, there is another popular class of models in which leptoquarks (LQ) are applied.  Here we briefly review the findings of \cite{Huang:2021biu} regarding scalar LQ. There are only four scalar LQ  which can interact with the SM-fermions at renormalizable level.   Interestingly, $S_3 \sim (3,3,-1/3)$  can simultaneously address $R_K$ and $R_{K^*}$ and its constraints are not in conflict with the experimental data \cite{Queiroz:2014pra,Angelescu:2018tyl}.
Similarly, the vector LQ $U_{1} \sim (3,1,2/3)$ can also provide a good fit for the $R_{K^{(*)}}$-anomaly. Note that it requires a proper UV-completion for theoretical consistency. 

The relevant Lagrangian for $S_3$ can be written as:
\begin{equation}
\mathcal{L}_{S_{3}}=-M_{S_{3}}^{2}\left|S_{3}^{a}\right|^{2}+y_{i \alpha}^{\rm LQ} \overline{Q^{\rm c}}^{i}\left(\epsilon \sigma^{a}\right) L^{\alpha} S_{3}^{a}+\mathrm{h.c.},
\end{equation}
with lepton and quark-doublets $L^{\alpha}=\left(\nu_{\rm L}^{\alpha},~ \ell_{\rm L}^{\alpha}\right)^{\rm T}$ and $Q^{i}=\left(V_{j i}^{*} u_{\rm L}^{j}, ~ d_{\rm L}^{i}\right)^{\rm T}$, and Pauli-matrices   $\sigma^{a}$ ($a=1,2,3$; $\epsilon=i \sigma^{2}$). 
The LQ contributes to the Wilson-coefficients at  tree-level [cf.\ Fig.~\ref{fig:feyn}] and one can identify: 
\begin{equation} \label{eq:wc_lq}
C_{9}^{\mu}=-C_{10}^{\mu}=\frac{\pi}{\sqrt{2} G_{\rm F} M_{S_{3}}^{2} \alpha}\left(\frac{y_{32}^{\rm LQ} y_{22}^{\rm LQ *}}{V_{t b} V_{t s}^{*}}\right). 
\end{equation}

In contrast to $Z'$ scenario, the process mediated by LQ is $t$-channel~\cite{Huang:2021biu}. Hence, a different event distribution is expected if the mediator mass is reachable at the colliding energy.
If the mediator mass is large, one can still test the ${R_{K^{(*)}}}$ anomaly at the $\mc$ but can no longer differentiate various models. 
In this regard, it is convenient to describe with an effective theory in terms of the Wilson coefficients $C^{\mu}_9$ and $C^{\mu}_{10}$.
It is easy to find the cross section of $\mu^+ \mu^- \to b  \bar s$  to be
\begin{eqnarray}
	\sigma(s) =  \frac{G^{2}_{\rm F} \alpha^2 |V^{}_{tb} V^{*}_{ts}|^2s}{8 \pi^3}\left( |C^{\mu }_{9}|^2 + |C^{\mu }_{10}|^2\right).
\end{eqnarray}
When the mediator mass is very large, the signal event number is fixed by the Wilson coefficients, regardless of the details of the UV completion. 
If one takes the best-fit scenario of $B$ anomaly fit, i.e., $ C^{\mu}_{9} = - C^{\mu}_{10} = -0.43$, one obtains the event number of $bs$ as
\begin{eqnarray}
	\#{\rm signal} \simeq 10^{3} \left(\frac{\sqrt{s}}{6~{\rm TeV}}\right)^2 \left(\frac{L}{4~{\rm ab^{-1}}} \right) .
\end{eqnarray}
The SM background of quark dijets without flavor tagging reads
$1.2\times 10^{5} \cdot (6~{\rm TeV}/ \sqrt{s})^2 \cdot (L/4~{\rm ab^{-1}})$. The signal is found to exceed the SM background uncertainty at around $3\sigma$ confidence level.

The sensitivity to the $S^{}_{3}$ LQ model is shown in the upper-right panel of Fig.~\ref{fig:contours_Zp}.
The $\mc$ with $\sqrt{s} = 3~{\rm TeV}$ and $L=1~{\rm ab}^{-1}$ will reach the red curves. 
The solid and dashed curves stand for the cases without and with the flavor tagging procedure.
For $\sqrt{s} = 3~{\rm TeV}$, an upgrade of the luminosity $L=1~{\rm ab}^{-1}$ by a factor of 4 to 8 or a better tagging efficiency is required to cover the LQ parameter space indicated by the ${R_{K^{(*)}}}$ anomaly.

Nevertheless, it is interesting to discuss the potential of $\mc$ with other options. For the setup $\sqrt{s} = 6~{\rm TeV}$ and $L = 4~{\rm ab^{-1}}$, we find most of the parameter space suggested by the ${R_{K^{(*)}}}$ anomaly will be probed.
For demonstration, we also show the case of $U^{}_{1}$ vector LQ in the lower-right panel of Fig.~\ref{fig:contours_Zp}, for which the setup $\sqrt{s} = 6~{\rm TeV}$ and $L = 4~{\rm ab^{-1}}$ can fully cover the indicated parameter space.

\subsection{Vector Leptoquarks}

We now focus on the phenomenology of the vector LQ known in the literature as $U_1^{\mu}$, at a $\mc$. As a proof-of-principle, the authors in~\cite{Asadi:2021gah} explore the reach of two benchmark $\mc$ facilities (1 ab$^{-1}$ at 3 TeV and 20 ab$^{-1}$ at 14 TeV) for $U_1^{\mu}$ production and contribution to Lepton Flavor Universality Violation. The Lagrangian of this model includes 
\begin{equation}
\mathcal{L}_\text{LQ} \supset \frac{g_U}{\sqrt{2}}
U_1^{\mu}
\beta^{ij}_L \bar{Q}_L^i \gamma_{\mu} L_L^j 
 + \textrm{h.c.},
\label{eq:U1_lag}
\end{equation}
where $g_U \beta_L^{ij}$ parametrizes the coupling of the vector LQ $U_1$ to a left-handed $i$-generation quark and $j$-generation massive lepton. This model can explain the observed anomaly if
\begin{equation}
     \frac{\beta_L^{22}\beta_L^{32}}{m_\text{LQ}^2} \approx 1.98 \times 10^{-3} \text{ TeV}^{-2}.
     \label{eq:RKline}
\end{equation} 

Note that each $\beta_L^{ij}$ is a parameter of the theory. For concreteness, a multitude of coupling scenarios are considered, such as
\begin{equation}
\label{eq:betas}
\beta_L^{ij} = \begin{pmatrix}
0 & 0 & 0 \\
0 & \beta_L^{22}=\beta_L^{32} & 0\\
0 & \beta_L^{32} & 0
\end{pmatrix},
\end{equation}
i.e. and equal coupling of $U_1$ to $\mu s$ and $\mu b$, and zero coupling to other flavors of quarks and leptons. Other coupling schemes are considered to explore the phenomenological consequences, but the choice given in Eq.~\ref{eq:betas} provides the minimal structure to address the flavor anomalies. 

To generate the events, three production mechanisms were considered: pair production, single production, and Drell-Yan. 

\textbf{Pair Production.}
This channel is dominated by producing two on-shell $U_1$. These processes are initiated either by direct muon collisions or initial state vector boson fusion. A cut on the invariant mass of the bottom quark pair in the final state, $m_{bb}$, can significantly reduce the background. Note that pair production of $U_1$ from initial vector bosons is determined by its gauge interactions and it is independent of the $\beta_L$ couplings to SM fermions.

\textbf{Single Production.}
This channel has distinct phenomenology from the pair production one. While pair production falls off steeply once the two $U_1$s are not produced on-shell, the single production channel doesn't fall off until the mass threshold ($m_\text{LQ} = \sqrt{s}$). Additionally, the single-production diagrams all depend on $\beta_L$ and lose sensitivity in the weak coupling region of parameter space. The background diagrams of this channel are similar to that from the pair-production channel, with one of the final state particles missing. Again one can leverage the different topology of the background and signal diagrams to impose appropriate cuts. For example, a cut on the angular distance between the two final $b$ quarks and on the pseudorapidity of the final $\mu$ can significantly improve the signal-to-background ratio for this channel ~\cite{Asadi:2021gah}.

\textbf{Drell-Yan.}
Finally, a $t$-channel exchange of the LQ can give rise to a final state with $b$-quark jets. This interferes with the $s$-channel SM signal. Depending on the mass of the $U_1$, the distribution of events in kinetic variable (e.g. $\eta$ or $p_T$) can be very different. By binning the events in different $\eta$ bins and fitting the distribution, the background and signal events can be more easily separated. 

\begin{figure}
\resizebox{\columnwidth}{!}{
\includegraphics[scale=0.6]{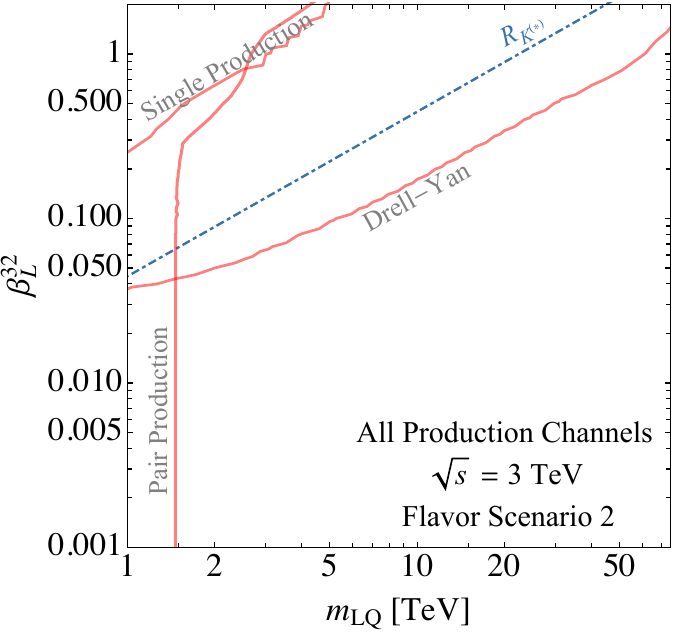} 
\includegraphics[scale=0.6]{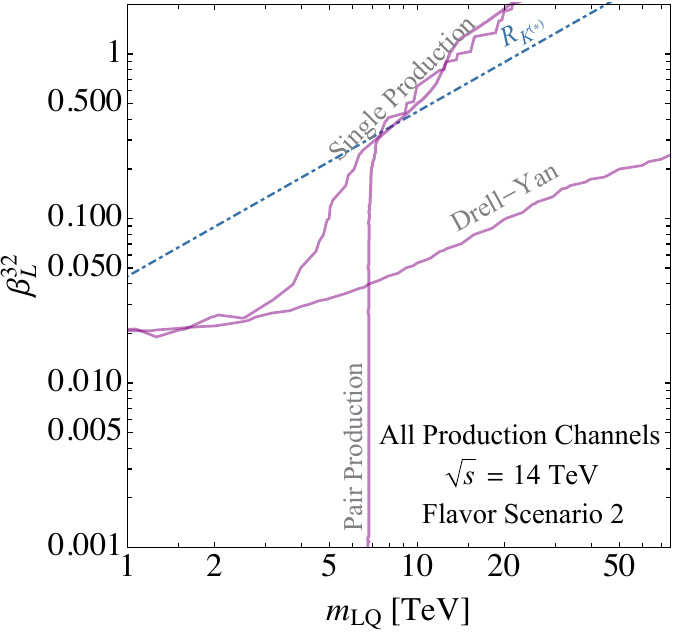} 
}
\caption{The $5\sigma$ discovery reach of 3 (14) TeV $\mc$ with 1 (20) ab$^{-1}$ of data. The reach is calculated using the flavor scenario described in Eq.~\ref{eq:betas}. The straight-line boundary of the pair-production channel corresponds to pure EW production, and is therefore independent of $\beta_L$. Figure taken from Ref.~\cite{Asadi:2021gah}.}
\label{fig:RKreach}
\end{figure}

Combining the results of all production channels, the reach of a 3 and 14 TeV $\mc$ in the mass $m_\text{LQ}$ and coupling $\beta_L^{32}$ for a $U_1$ model is shown in Fig.~\ref{fig:RKreach}. In Ref.~\cite{Asadi:2021gah} four different flavor scenarios, i.e. texture of yukawa couplings, were considered. The plots here are reproduced with flavor scenario 2 ($\beta_L^{22}=\beta_L^{32}$) of Ref.~\cite{Asadi:2021gah}.

Note that the pair-production channel is dominant and ultimately independent of $\beta_L$ at sufficiently small couplings as the EW production takes over. Here it was observed that with the cuts and the rudimentary analyses proposed in Ref.~\cite{Asadi:2021gah}, the Drell-Yan-like channel has the best sensitivity for most of the parameter space for these choices of $\sqrt{s}$. In particular, it was found that the line for the best fit to $R_{K^{(*)}}$ anomalies, see Eq.~\ref{eq:RKline}, can be probed even at a 3~TeV $\mc$. If the anomalies are supported by the upcoming LHCb or Belle~II experiments, these results provide an irrefutable case for building a high energy $\mc$.

Since the construction of a future $\mc$ has not begun, this analysis has not attempted to simulate systematics or detector effects. An attempt at emulating the systematics in searches for $U_1$ at a future $\mc$ can be found in~\cite{Huang:2021biu} . Inclusion of systematics and different statistical analysis led Ref.~\cite{Huang:2021biu} to a slightly lower reach than shown in Fig.~\ref{fig:RKreach}. Yet, both analyses agree that a $\mc$ with a few to 10 TeV center of mass energy, and with predicted attainable luminosities \cite{Delahaye:2019omf}, can cover the entire parameter space of $U_1$ that explains the flavor anomalies. Once the research and design of the collider is underway, further studies will be needed to refine the reach plot provided in this proof-of-concept study.

\section{Lepton Flavour Violation \label{sec:LFV}}

The SM exhibits a distinctive pattern of fermion masses and mixing angles, for which we currently have no deep explanation.
Delicate symmetries also lead to a strong suppression of flavor-changing processes in the quark and lepton sectors, which may be reintroduced by new particles or interactions.
The non-observation of such processes thus leads to some of the most stringent constraints on BSM physics, while a positive signal could give us insight into the observed structure of the SM.
A number of precision experiments searching for lepton flavor violating (LFV) processes such as $\mu \to 3e$, $\tau \to 3\mu$ or $\mu$-to-$e$ conversion within atomic nuclei will explore these processes with orders of magnitude more precision in the coming decades~\cite{Baldini:2018uhj}.
As we will see, a high-energy $\mc$ has the unique capability to explore the same physics --- either via measuring effective interactions or by directly producing new states with flavor-violating interactions --- at the TeV scale.

\subsection{Effective LFV Contact Interactions}\label{cilfv}

In this section, we study $\mc$ bounds on $\mu\mu \ell_i \ell_j$-type contact interactions, and demonstrate the complementarity with precision experiments looking for lepton-flavor violating decays, as first studied in ref.~\cite{AlAli:2021let}. We will focus on $\tau 3\mu$ and $\mu 3e$ operators, since constraints on them can be compared directly with the sensitivity from $\tau \to 3\mu$ and $\mu \to 3 e$ decays. We parametrize the four-fermion operators relevant for the $\tau \to 3\mu$ decay via
\begin{equation}
\mathcal{L} \supset 
V_{LL}^{\tau 3\mu}\big( \bar{\mu} \gamma^{\mu} P_L \mu \big) \big(\bar{\tau} \gamma_{\mu} P_L \mu\big)
+ V_{LR}^{\tau 3\mu} \big( \bar{\mu} \gamma^{\mu} P_L \mu\big) \big(\bar{\tau} \gamma_{\mu} P_R \mu\big) + \big( L \leftrightarrow R \big) + \textrm{h.c.}\,,
\end{equation}
with an equivalent set for the $\mu \to 3e$ decay. 
We will assume all the $\tau 3\mu$ coefficients are equal:
In what follows, we will assume all the $V_{ij}^{\tau 3\mu}$ coefficients are equal to $c^{\tau 3\mu} / \Lambda^2$, 
where $c^{\tau 3\mu}$ is a dimensionless coefficient and $\Lambda$ is to be interpreted as the scale of new physics, and similarly for $\mu 3 e$ coefficients.

\begin{figure}[t!]
\centering
\includegraphics[width=13cm]{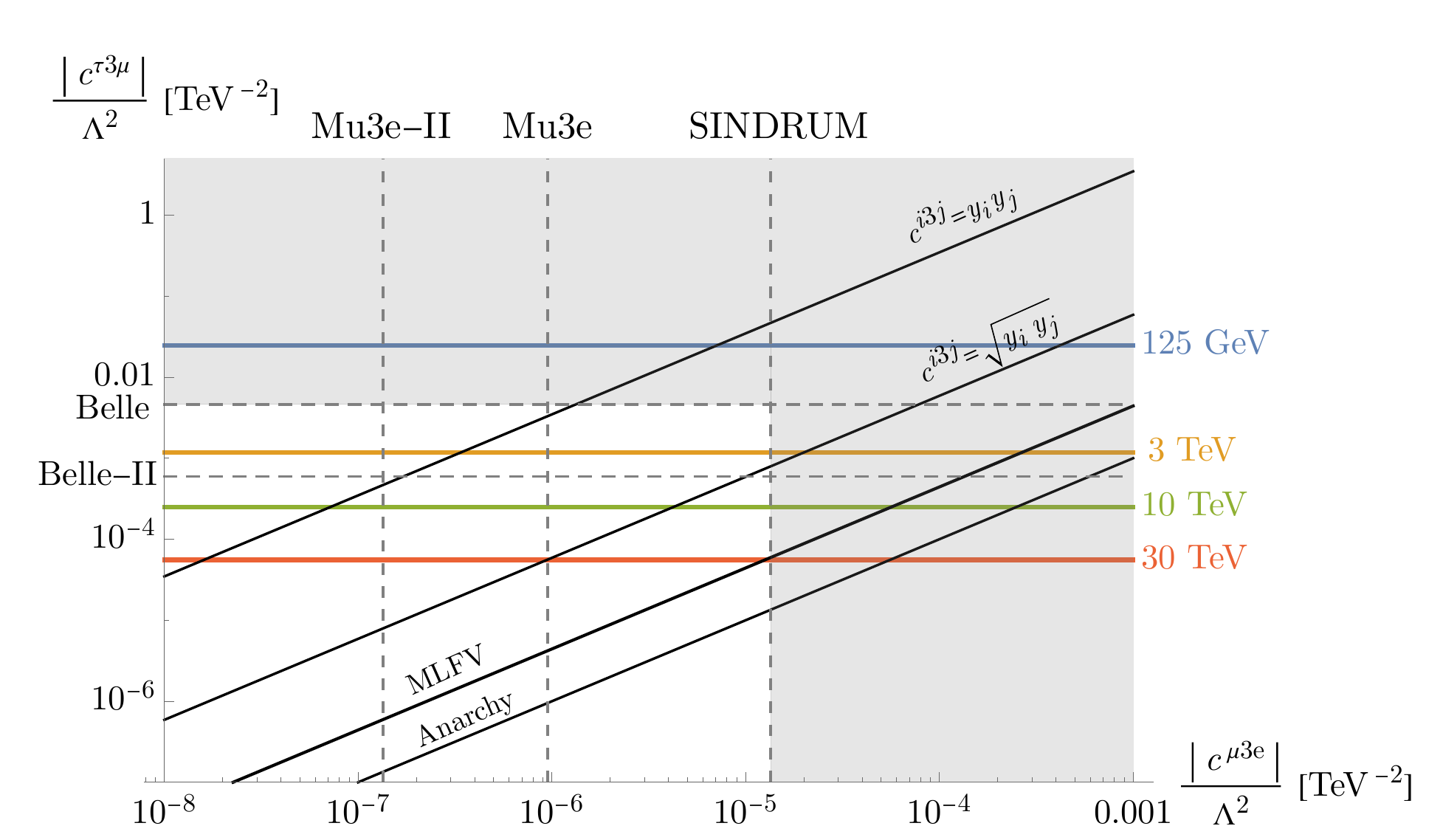}
\caption{Summary of $\mc$ and low-energy constraints on flavor-violating 3-body lepton decays. The colored horizontal lines show the sensitivity to the $\tau 3\mu$ operator at various energies, all assuming $1\text{ ab}^{-1}$ of data. The dashed horizontal (vertical) lines show the current or expected sensitivity from $\tau \to 3\mu$ ($\mu \to 3e$) decays for comparison. The diagonal black lines show the expected relationship between different Wilson coefficients with various ansatz for the scaling of the flavor-violating operators (e.g., ``Anarchy'' assumes that all Wilson coefficients are $\mathcal{O}(1)$).
}\label{fig:lfv-operator-summary}
\end{figure}

At a $\mc$, the $\tau 3 \mu$ coefficients are probed via $\mu^+\mu^- \to \mu \tau$. 
Our analysis closely follows an analogous study at an $e^+ e^-$ collider in ref.~\cite{Murakami:2014tna}. 
As discussed in ref.~\cite{AlAli:2021let}, the SM backgrounds from $\tau^+\tau^-$ and $W^+W^-$ production can be substantially mitigated by a simple set of cuts, whereas the signal can be largely retained up to $\sim 10\%$ effects due to initial state radiation.
The resulting bounds, assuming integrated luminosities of $1\,\textrm{ab}^{-1}$ at $0.125$, $3$, $10$ and $30\,\textrm{TeV}$ are shown in Fig.~\ref{fig:lfv-operator-summary}, alongside current and future sensitivities of $\tau \to 3\mu$ and $\mu\to 3 e$ experiments.
A $3\,\textrm{TeV}$ machine would set a direct bound at the same level as the future Belle~II sensitivity.

Given an ansatz regarding the flavor structure, the constraints on the $\tau 3\mu$ operators can be compared to the constraints on the analogous $\mu 3e$ operator in the $\mu \to 3 e$ decay. 
The diagonal lines in Fig.~\ref{fig:lfv-operator-summary} show the expected relationship between the two Wilson coefficients for several different ansatz, including flavor anarchy (where all coefficients $\sim 1$), Minimal Leptonic Flavor Violation~\cite{Cirigliano:2005ck}, or scalings with different powers of the involved Yukawa couplings.
While muon decays set the strongest limits assuming anarchical coefficients, a $\mc$ could set competitive constraints for other ansatz: in the most extreme case, where the Wilson coefficients scale like the product of the Yukawas, a $3\,\textrm{TeV}$ machine would have sensitivity comparable to the final Mu3e sensitivity.

In addition to the $\tau 3\mu$ operators considered here, similar sensitivity should be attainable for the $\mu^+ \mu^- \to \mu^{\pm} e^{\mp}$ process, as well as to the processes such as $\mu^+ \mu^- \to \tau^{\pm} e^{\mp}$ that violate lepton flavor by two units.
Overall, we see that a $\mc$ would be capable of directly probing flavor-violating interactions that are quite complementary to future precision constraints.

\subsection{Direct Probes: Lepton-Flavor Violation in the MSSM}

An exciting possibility is that the flavor-changing processes that might be observed in low-energy experiments arise from loops of new particles near the TeV scale.
As a motivated example, consider the MSSM. The scalar superpartners of the SM leptons can have soft supersymmetry-breaking contributions to their mass matrix that are off-diagonal in the SM lepton eigenbasis.
As a result, the slepton interactions with the leptons will be flavor-violating and lead to processes such as muon-to-electron conversion and rare muon decays at one loop. 
In well-motivated constructions, the mixing between the selectron and smuon states can be quite large, as the low-energy processes are protected by a ``Super-GIM'' mechanism~\cite{Arkani-Hamed:1996bxi}, allowing the new states to be near the TeV scale while consistent with current bounds.

A 3 TeV $\mc$ would dramatically extend the reach for electroweak-charged superpartners beyond a TeV, raising the possibility of directly producing the new states responsible for lepton flavor-violation. 
Moreover, the unique environment of a $\mc$ makes it possible to not only produce these new states, but measure their LFV interactions. 
This would provide detailed insight into both the mechanism of supersymmetry breaking and the origin of the flavor structure of the SM.
A detailed investigation of these prospects is carried out in ref.~\cite{Homiller:2022xxx};\footnote{These prospects were also reviewed in ref.~\cite{AlAli:2021let}.}
here we briefly review their results for the 3 TeV case. 

To understand the complementarity of low-energy cLFV probes and the $\mc$ reach, we consider the scenario in which only the right-handed selectron and smuon, along with one light neutralino (which we will assume to be a pure bino with mass $M_1$) are in the spectrum.
If the slepton masses $m_{\tilde{\ell}} > M_1$, the sleptons decay directly to a lepton and bino, and the LFV interactions can be measured directly via the pair-production process: $\mu^+ \mu^- \to \tilde{e}^+_{1,2} \tilde{e}^-_{1,2} \to \mu^{\pm} e^{\mp} \chi_1^0 \chi_1^0$, where the binos appear as missing momentum.
In this simplified scenario, both the low-energy LFV processes and the pair-production process at a $\mc$ depend only on the slepton masses and mixing angle, as well as $M_1$. 

\begin{figure}[t!]
\centering
\includegraphics[width=7cm]{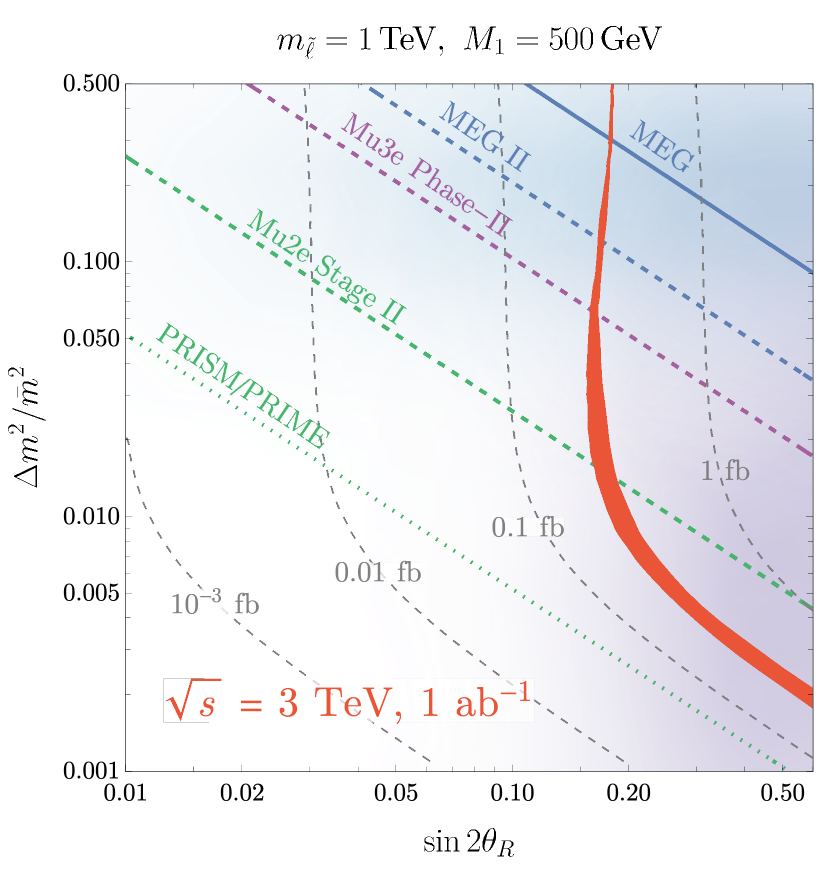}
~
\includegraphics[width=7cm]{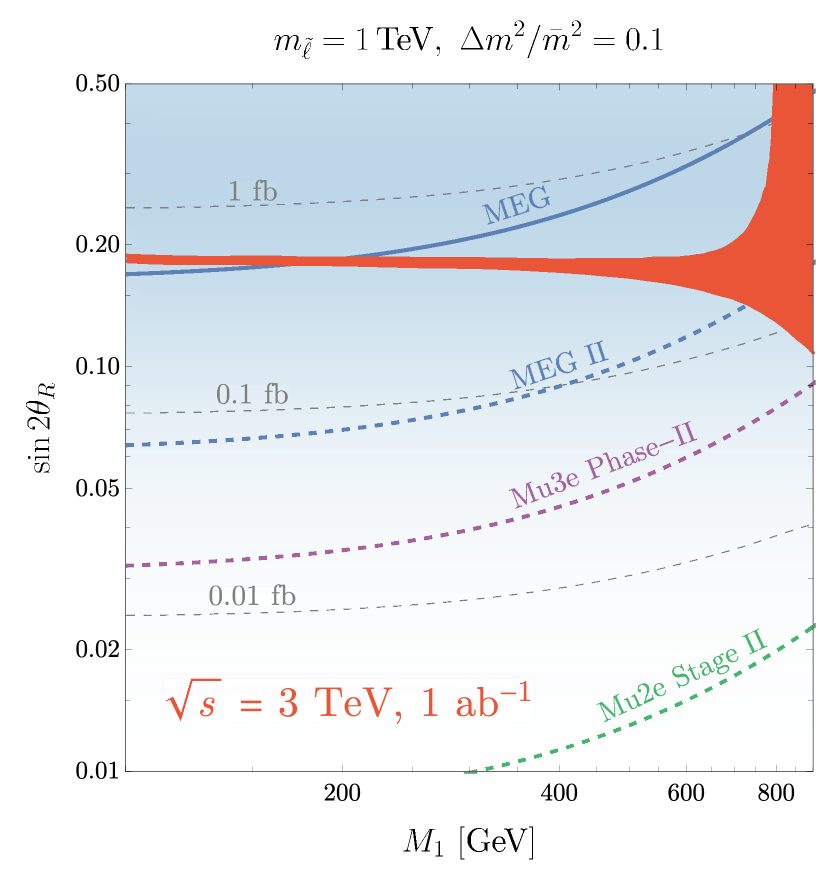}
\caption{Constraints on lepton flavor violation in the MSSM in the $\Delta m^2 / \bar{m}^2$ vs. $\sin 2\theta_R$ plane (left) and the $\sin 2\theta_R$ vs. $M_1$ plane (right) from measurements of the slepton pair production process with flavor-violating final states (red band) at a 3 TeV $\mc$, assuming $1\,\textrm{ab}^{-1}$ of luminosity.
The width of the band represents the uncertainty on the reach from the measurement of the slepton and neutralino masses in flavor-conserving channels.
The purple and blue shaded lightly shaded regions indicate parameters preferred in Gauge-Mediated Supersymmetry Breaking scenarios and flavor-dependent mediator scenarios, respectively.
Both plots assume a mean slepton mass of $1\,\textrm{TeV}$. In the left plot we fix the neutralino mass $M_1 = 500\,\textrm{GeV}$, while in the right figure $\Delta m^2 / \bar{m}^2$ is fixed to 0.1.
The current (solid) and expected (dashed, dotted) limits from low-energy lepton flavor violation experiments are indicated by the blue, purple and green lines.}
\label{fig:mssm-3tev-summary}
\end{figure}

In Fig.~\ref{fig:mssm-3tev-summary}, we show the $5\sigma$ reach for a $3\,\textrm{TeV}$ $\mc$, assuming an average slepton mass of $1\,\textrm{TeV}$.
The left panel shows the reach as a function of the mixing angle and mass-splitting, $\Delta m^2 = m_{\tilde{e},2}^2 - m_{\tilde{e},1}^2$, with $M_1 = 500\,\textrm{GeV}$. The right panel shows the constraints for fixed $\Delta m^2 / \bar{m}^2 = 0.1$ in the $M_1$ vs. $\sin 2\theta_R$ plane. Large mixing angles are motivated in models involving gauge-mediated supersymmetry breaking (GMSB), indicated by the purple region, while larger mass splittings are motivated in scenarios where the messengers carry flavor-dependent charges, such as $L_{\mu} - L_{\tau}$, indicated by the blue regions (see ref.~\cite{Homiller:2022xxx} for more details).
The complementary constraints from low-energy experiments searching for $\mu\to e\gamma$, $\mu\to 3e$ decays or $\mu$-to-$e$ transitions are shown in blue, purple and green, respectively.
We see that the $\mc$ reach can extend to small mass splittings in the GMSB scenario, and can cover a substantial part of the most well-motivated parameter space.

\subsection{Gauge $L_\mu-L_\tau$ Interactions}

It is not straightforward to test the $L^{}_{\mu}-L^{}_{\tau}$ model at laboratories due to the preferred couplings to the second and third family leptons, unless we have a facility to directly collide muons. Here we summarize the findings of \cite{Huang:2021nkl} regarding searches of a gauged $L^{}_{\mu}{-}L^{}_{\tau}$ interaction at a MuC.
The discussion focuses on a collider with an energy of $\sqrt{s} = 3~{\rm TeV}$ and an integrated luminosity of $L=1~{\rm ab}^{-1}$. In particular, the parameter space which explains the $(g-2)^{}_{\mu}$ as well as $B$-physics anomalies is found to be fully explored by such a facility given a reasonable integrated luminosity.
The relevant interaction with the new boson $Z'$ reads
\begin{eqnarray} \label{eq:L}
	\mathcal{L} \supset  g^{\prime} \left( \overline{\ell^{}_{\rm L}} Q^{\prime} \gamma^{\mu} \ell^{}_{\rm L} + \overline{E^{}_{\rm R}}Q^{\prime} \gamma^{\mu} E^{}_{\rm R} \right) Z^{\prime}_{\mu}\;,
\end{eqnarray}
where $g'$ stands for the coupling constant of gauged $L^{}_{\mu}{-}L^{}_{\tau}$ symmetry, $\ell \equiv (\nu, E)^{\rm T}$ is the lepton doublet with $\nu$ and $E$ being the neutrino and the charged lepton, respectively, and $Q^{\prime} = {\rm Diag}(0,1,-1)$ represents the charge matrix in the basis of $(e,\mu,\tau)$.
The $Z'$ will inevitably mix with the SM gauge bosons,  i.e., $\gamma$ and $Z$. It is found that the mixing with $\gamma$ is strongly suppressed by the $Z'$ mass, while the mixing with $Z$ can be relevant if their masses are of the similar order.
For simplicity, we assume a negligible mixing in the following, which actually represents a conservative estimate of the sensitivity.

In such a setup, the relevant processes for the analysis
include the final-state signatures of dimuon ($+$ photon), ditau ($+$ photon) as well as monophoton.
Even though the process with  initial photon radiation is of higher order compared to the trivial two-body scatterings, its impact is comparable and in some circumstances even larger than the two-body ones, due to the radiative return of resonant $Z'$ production~\cite{Chakrabarty:2014pja,Karliner:2015tga}.

The two-body scattering is very clean, as the final back-to-back dimuon or ditau carries all the energy delivered by the initial colliding muons. The only background of our concern should be the intrinsic SM processes, such as $\mu^+  \mu^- \to \gamma/Z \to l^+  l^-$ as well as $t$-channel exchanges.
Here one also benefits from the interference between the $Z'$ and SM-mediated diagrams. For instance, the cross section for $\mu^+ \mu^- \to \tau^+ \tau^-$ is approximately $e^2 g'^2/(4\pi s)$ for $s \gg M^{2}_{Z'}$ and $-e^2 g'^2/(4\pi M^2_{Z'})$ for $s \ll M^{2}_{Z'}$, which actually dominates over the $Z'$-only cross section $\propto g'^4$ when $g'$ is small.
The SM cross section approximately takes $\sim e^4 /(8\pi s) \sim 10^4~{\rm ab}~(3~{\rm TeV}/\sqrt{s})^2$.
Hence one can readily estimate the excellent sensitivity to the gauge coupling even before the event generation: 
\small
\begin{eqnarray} 
	\hspace{-0.3cm}
	g' < 3.4 \times 10^{-2} \left(\frac{\sqrt{s}}{3~{\rm TeV}} \right)^{\frac{1}{2}}  \left(\frac{1~{\rm ab^{-1}}}{L} \right)^{\frac{1}{4}}  {\rm max}\left(1, \frac{M^{}_{Z'}}{\sqrt{s}} \right).
\end{eqnarray}
\normalsize

\begin{figure*}[t!]
	\begin{center}
		\hspace{-1cm}
\includegraphics[width=0.6\textwidth]{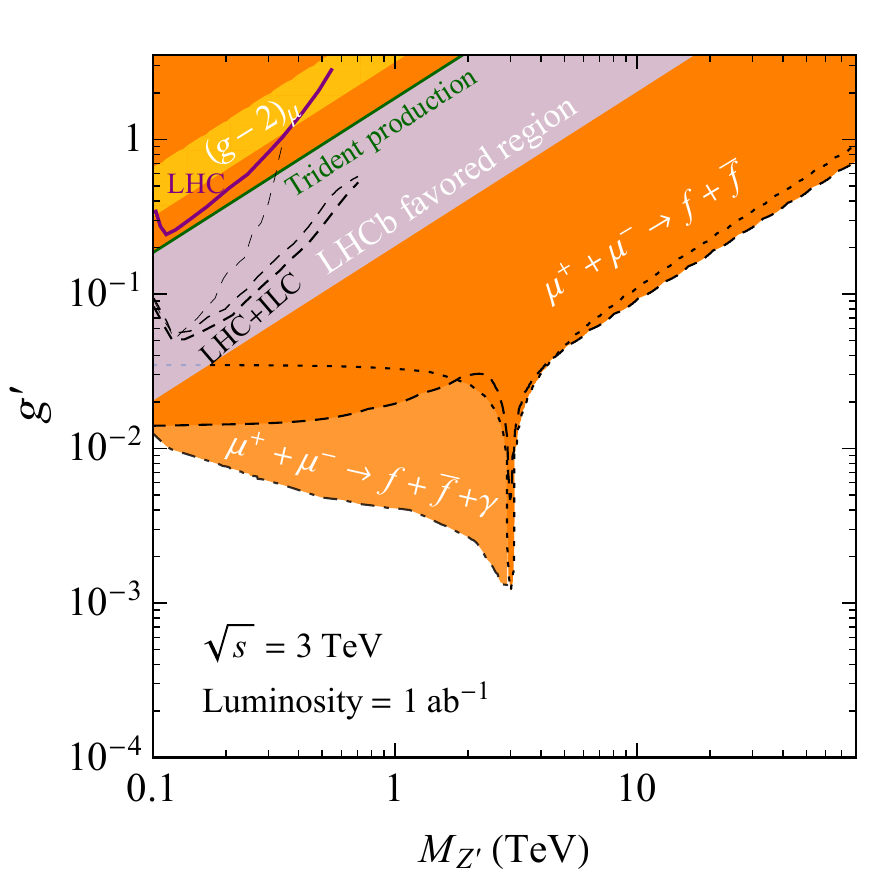}
	\end{center}
	\vspace{-0.3cm}
	\caption{The sensitivity of the $\mc$ with the COM energy $\sqrt{s}=3~{\rm TeV}$ and luminosity $L=1~{\rm ab^{-1}}$, given as orange regions. Other limits and projections are also shown for comparison. Our concerned parameter regions explaining the $(g-2)^{}_{\mu}$ and $B$ anomalies are given as yellow and blue bands, respectively. Figure from Ref.~\cite{Huang:2021nkl}.}
	\label{fig:contour}
\end{figure*}

To obtain the final sensitivity to the parameter space, one has to make a few assumptions about the particle identification and detection prospects.
For the two-body scatterings, it has been assumed an efficiency for dimuon identification of $100\%$ and that for ditau of $70\%$, which is rather conservative.
The  search of resonance for the radiative return process severely relies on the energy resolution of photon or equivalently dilepton.
For photon, it has been considered the energy resolution of the current CMS detector with ${\rm PbWO}^{}_{4}$ crystals~\cite{Chatrchyan:2008aa}, and for dimuon one can take $\Delta m^{}_{\mu^+\mu^-} \simeq 5\times 10^{-5}~{\rm GeV^{-1}}\cdot s$~\cite{Freitas:2004hq}.
Moreover, a systematic uncertainty of $0.1\%$ level has been assumed.

The projected sensitivity is presented in Fig.~\ref{fig:contour}. The limits using $\mu^+  \mu^- \to \ell^+ \ell^-$ (dashed and dotted curves for $\ell=\mu$ and $\tau$, respectively) are given as the darker orange region, while the radiative return process yields the lighter orange region.
Other limits and projections are also shown for comparison, such as $e^{+}   e^{-} \to \mu^{+}  \mu^{-}   Z',~Z' \to \mu^{+}   \mu^{-} $ from the BaBar experiment \cite{TheBABAR:2016rlg}, 
the LHC searches~\cite{Ma:2001md,Heeck:2011wj},
the trident production in neutrino scattering experiments~\cite{Altmannshofer:2014pba}.
The parameter spaces which can explain the $g$-2 and $B$ anomalies are shown as the yellow and blue bands, respectively. It is obvious that the parameter space of our concern with $M^{}_{Z'} > 100~{\rm GeV}$ is entirely covered by the $\mc$ setup $\sqrt{s} = 3~{\rm TeV}$ and $L=1~{\rm ab}^{-1}$.

\section{Muon Yukawa Couplings \label{sec:muon-higgs}}
\subsection{Modified muon-Higgs Coupling}

The Higgs couplings to the second generation of SM fermions still remain to be measured precisely. Recently, the Higgs-charm coupling was observed to be $|\kappa_c|=\sqrt{2}|y_c|/m_c<8.5$ at 95\% confidence level by ATLAS~\cite{ATLAS:2022ers}. In comparison, the Higgs-muon coupling can be measured more precisely due to the cleaner background of $H\to\mu^+\mu^-$.  First evidence suggests its value to be of the order of magnitude predicted by the SM~\cite{ATLAS:2020fzp,CMS:2020xwi}, but $\mathcal{O}(100\%)$ deviations from the SM value are still possible.
During the upcoming high-luminosity phase of the LHC, the muon Yukawa coupling can be pinned down to tens of percent, albeit in a model-dependent way~\cite{ATL-PHYS-PUB-2014-016}. 

A high-energy $\mc$ with multi-TeV center-of-mass energy and high luminosity would allow to measure the Higgs-muon coupling in a model-independent way, directly probing the mass generation mechanism of the muon. Considering its general applicability, the proposal presented in \cite{Han:2021lnp} can be extended to study related new physics effects involving final states of charged leptons and jets. Here we summarize their key findings.

\subsubsection*{The EFT parameterization}

In the Higgs Effective Field Theory (HEFT), the physical Higgs singlet together with the triplet Goldstone bosons is introduced in a non-linear parameterization as
\begin{equation}
U=e^{i\phi^a\tau_a/v},~\textrm{with}~
\phi^{a} \tau_{a}=\sqrt{2}\left(\begin{array}{cc}
\frac{\phi^{0}}{\sqrt{2}} & \phi^{+} \\
\phi^{-} & -\frac{\phi^{0}}{\sqrt{2}}
\end{array}\right).
\end{equation}
The HEFT Lagrangian can describe a generic Yukawa sector as follows,
\begin{equation}
\begin{aligned} 
\mathcal{L}_{U H}\supset
-\frac{v}{2 \sqrt{2}}\left[\sum_{n \geq 0} y_{n}\left(\frac{H}{v}\right)^{n}\left(\bar{\nu}_{L}, \bar{\mu}_{L}\right) U\left(1-\tau_{3}\right)\left(\begin{array}{c}\nu_{R} \\ \mu_{R}\end{array}\right)+\text { h.c. }\right]. 
\end{aligned}
\end{equation}
With these definitions, the muon mass and the prefactor of the Yukawa coupling are given by $m_\mu=y_0 v/\sqrt{2}$ and $\kappa_\mu=y_1 v/(\sqrt{2}m_\mu)$, respectively.

The case $y_1=y_0=y_\mu$ corresponds to the SM reference value, $\kappa_\mu=1$. In a generic new-physics scenario, the relation between the coefficients $y_0$ and $y_1$ is unknown; it depends on the specific underlying dynamics. 
In the effective-theory description, new operators in the $H/v$ expansion will appear as contact terms which directly couple the muon to Higgs or Goldstone bosons.  By means of the Goldstone-Boson Equivalence Theorem (GBET), one can associate a modification of the muon-Higgs coupling $y_\mu$ with new contributions to multiple vector-boson production which generically can become large in the high-energy limit.

Alternatively, a new-physics contribution to the Yukawa interaction can be parameterized in terms of the Standard-Model Effective Field Theory (SMEFT) formalism.  A generic Yukawa part of the Lagrangian takes the form
\begin{equation}
\mathcal{L}_{\varphi}\supset\left[-\bar{\mu}_{L} y_{\mu} \varphi \mu_{R}+\sum_{n=1}^{N} \frac{C_{\mu \varphi}^{(n)}}{\Lambda^{2 n}}\left(\varphi^{\dagger} \varphi\right)^{n} \bar{\mu}_{L} \varphi \mu_{R}+\text { h.c. }\right],
\end{equation}
where
\begin{equation}
\varphi=\frac{1}{\sqrt{2}}\left(\begin{array}{c}
\sqrt{2} \phi^{+} \\
v+H+i \phi^{0}
\end{array}\right).
\end{equation}
Higher-dimensional effective operators in the SMEFT Lagrangian ($n\geq1$) result in modifications to the muon mass and the corresponding Yukawa coupling,
\begin{equation}\label{eq:kappaSMEFT}
m_{\mu} =\frac{v}{\sqrt{2}}\left[y_{\mu}-\sum_{n=1}^{N} \frac{C_{\mu \varphi}^{(n)}}{\Lambda^{2 n}} \frac{v^{2 n}}{2^{n}}\right],
\qquad
\kappa_\mu=1-\frac{v}{\sqrt{2} m_{\mu}} \sum_{n=1}^{N} \frac{C_{\mu \varphi}^{(n)}}{\Lambda^{2 n}} \frac{n v^{2 n}}{2^{n-1}},
\end{equation}
respectively.
In this approach, the SM reference value $\kappa_\mu=1$ is reproduced if only a dimension-4 operator ($n=0$) is present.  Starting from dimension-6 operators, we receive new contributions to the muon-Higgs coupling.  These are associated with contact terms involving Higgs or Goldstone bosons.  They lead to an enhanced production of multi-boson final states in the high energy limit, in complete analogy with the HEFT formalism.  Assuming a modification of the Yukawa coupling, one can translate an experimental bound on $\Delta\kappa_\mu$ to a new-physics scale $\Lambda$ via (assuming $C_{\mu\varphi}^{(1)}\sim\mathcal{O}(1)$)
\begin{equation}\label{eq:k2L}
\Lambda\sim\sqrt{\frac{v^3}{\sqrt{2}m_\mu\Delta\kappa_\mu}}.
\end{equation}

\subsubsection*{Multiple boson production}

In the context of the above model-independent $\kappa_\mu$ parameterizations, the authors in~\cite{Han:2021lnp} have extensively studied multi-boson production at high-energy $\mc$.  They demonstrate that at high collision energies, a modification of the Yukawa coupling can induce a significant enhancement of the multi-boson production rate that grows with energy.  The effect becomes more striking for a final-state multiplicity of three or more bosons.  It provides a unique opportunity to test the muon Yukawa coupling which is independent from the measurement via the Higgs decay to muons.  Focusing on the examples of $ZHH$ and $WWH$ production, one can explore various relevant kinematic distributions in order to compute the achievable precision on the coupling and thus on the corresponding operator coefficients. 

\begin{figure}
\begin{center}
\includegraphics[width=0.45\textwidth]{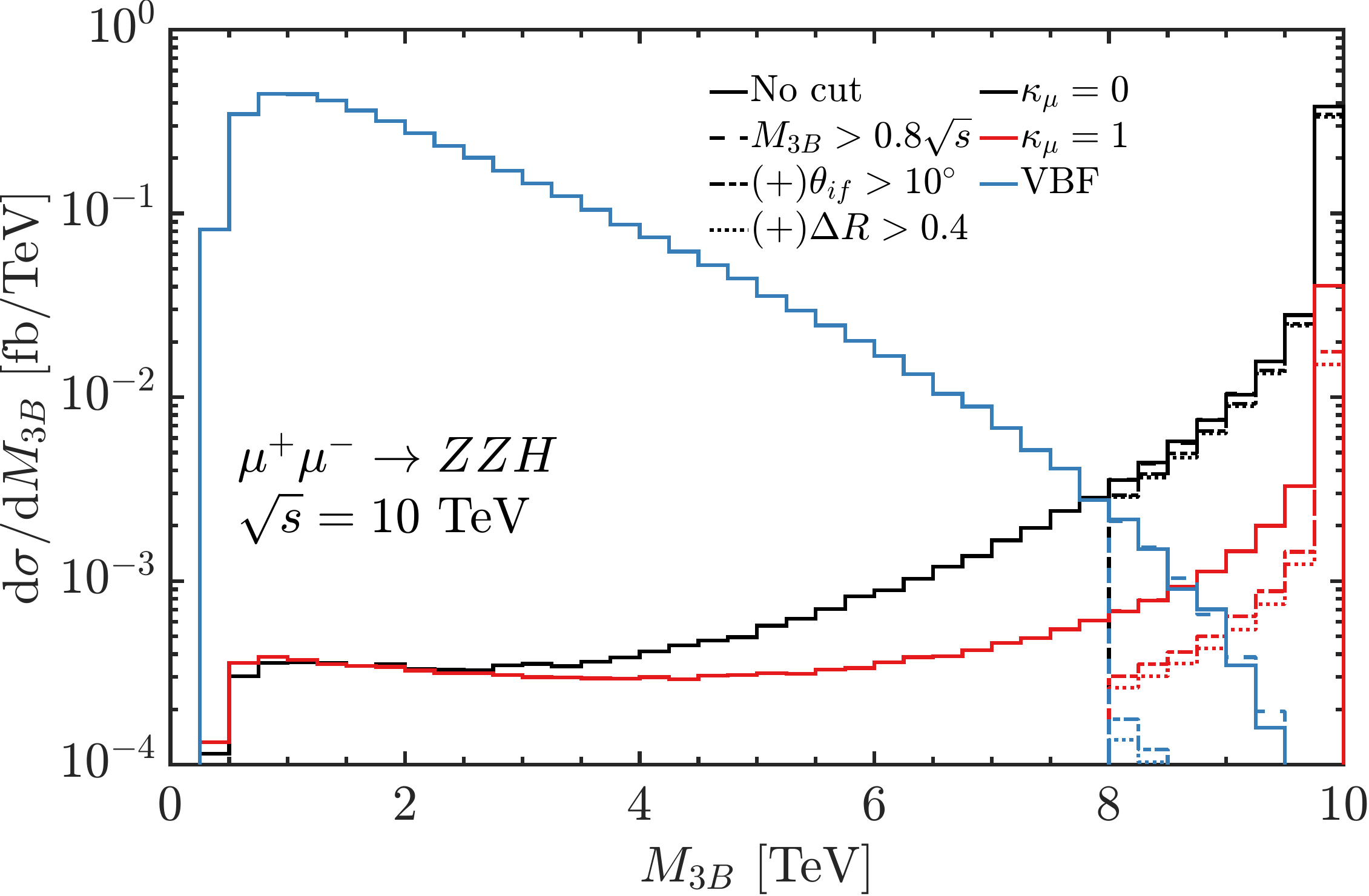}
\includegraphics[width=0.45\textwidth]{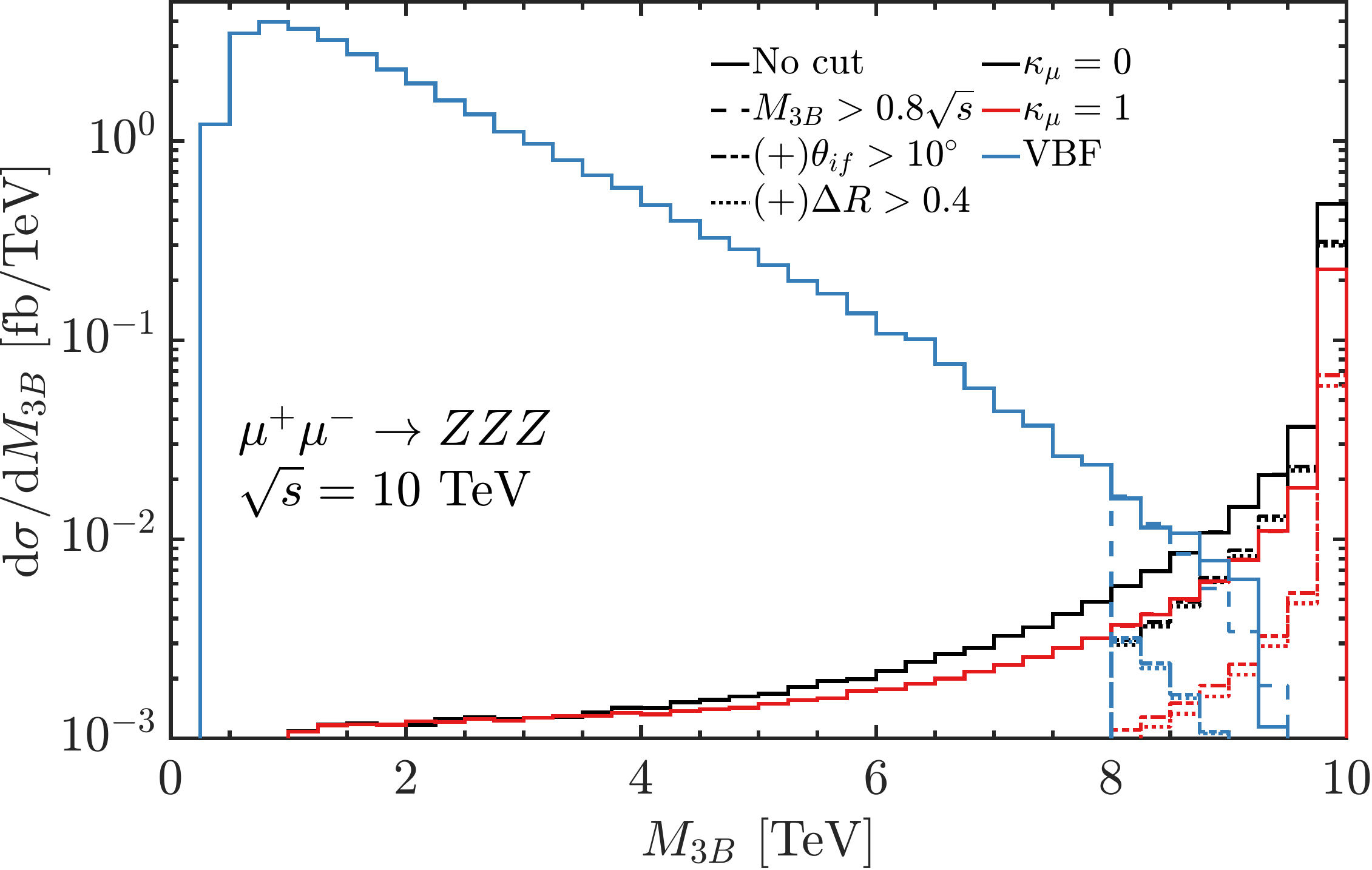}
\includegraphics[width=0.45\textwidth]{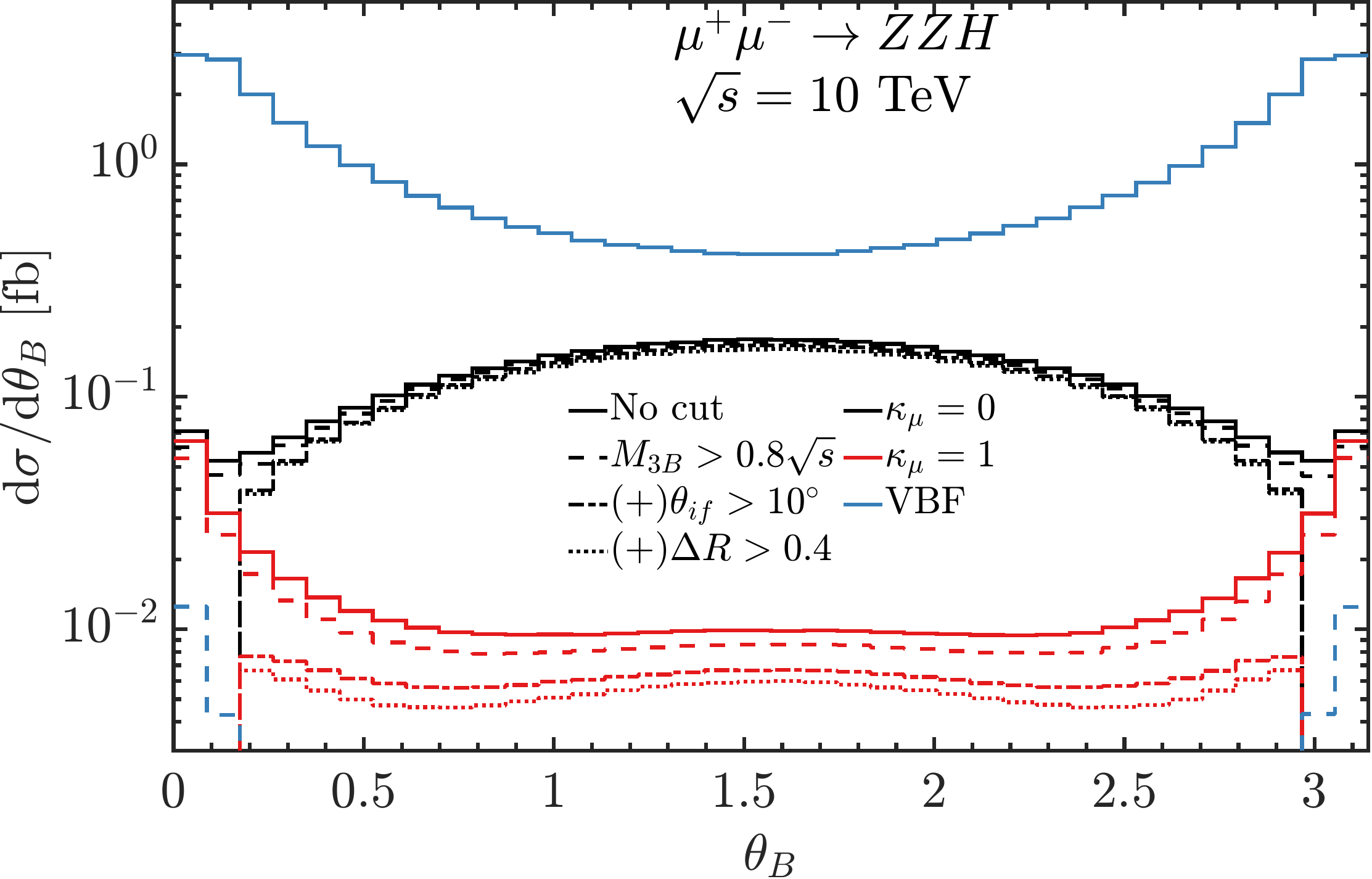}
\includegraphics[width=0.45\textwidth]{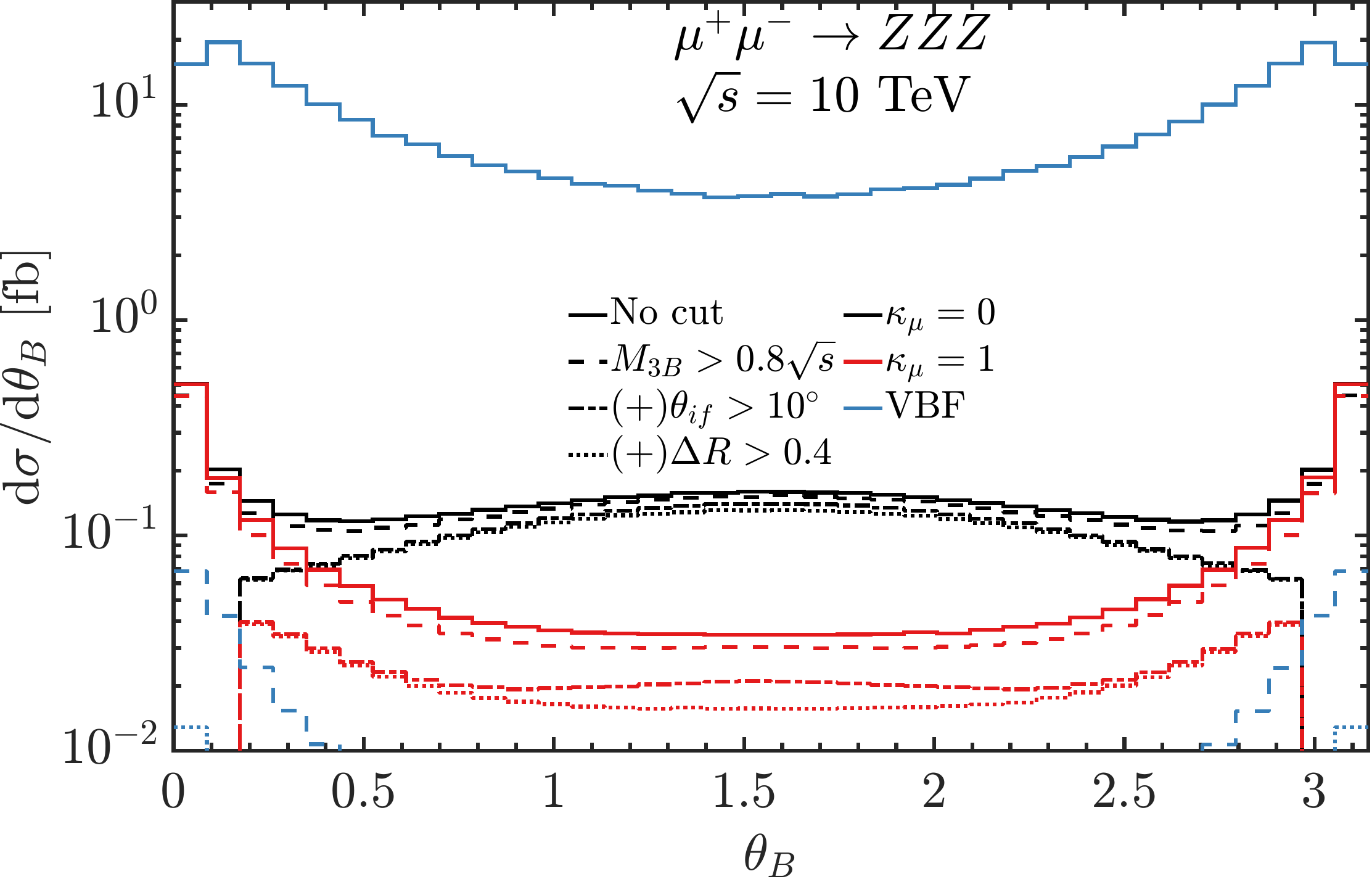}
\includegraphics[width=0.45\textwidth]{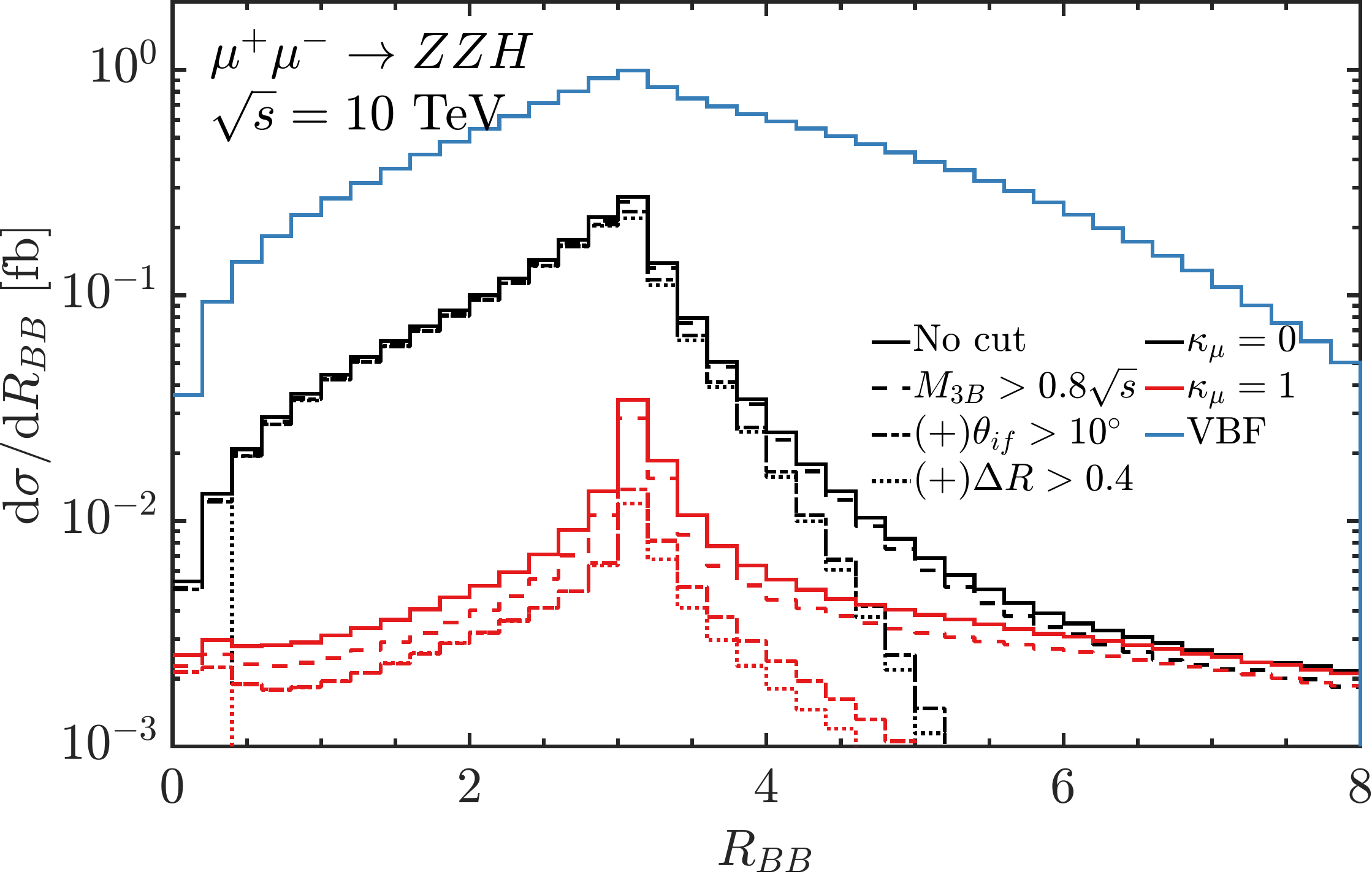}
\includegraphics[width=0.45\textwidth]{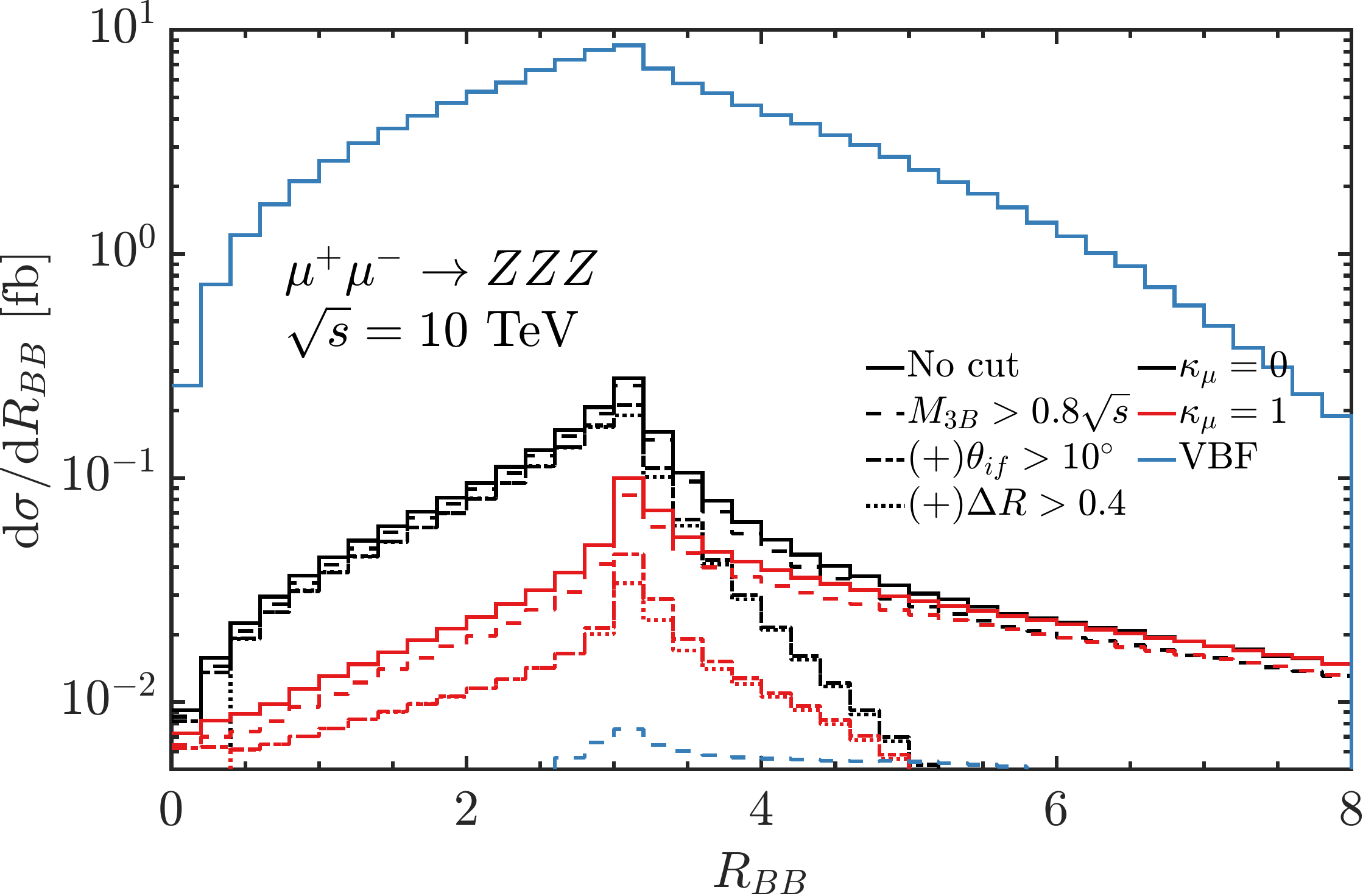}
\end{center}
\caption{The kinematic distributions $\theta_B,R_{BB},M_{3B}(B=Z,H)$ of $ZZH$ (left) and $ZZZ$ (right) production at a $\sqrt{s}=10$ TeV $\mc$.}
\label{fig:dist}
\end{figure}

This report extends the explicit coverage of multi-boson final states by presenting distributions of $ZZH$ and $ZZZ$ production, adopting a reference value of $10\;\TeV$ for the muon-collider c.m.\ energy.   The inclusive boson angle $\theta_B$, diboson distance $R_{BB}$ and triboson invariant mass $M_{3B}$ distributions are shown in Fig.~\ref{fig:dist}, respectively.  A few features stand out. First, one can verify that for the annihilation process, the invariant mass $M_{3B}$ sharply peaks at the collision energy $\sqrt{s}$, with a small spread as a consequence of the initial-state radiation (ISR).  The vector-boson fusion contribution to the same three-boson final state mainly accumulates around the threshold. One can take advantage of this characteristic feature to filter the vector-boson funsion (VBF) background, by imposing an invariant mass cut such as $M>0.8\sqrt{s}$, explicitly shown as the dashed lines in Fig.~\ref{fig:dist}.  Another feature that clearly discriminates the extreme cases of the SM $\kappa_\mu=1$ vs.\ $\kappa_\mu=0$ (\emph{i.e.}, the BSM scenario with an order-one modification of the muon Yukawa coupling) is that $\kappa_\mu=0$ enhances the annihilation to bosons mostly in the central region, while the SM produces a large fraction of the bosons in the forward region. With a reasonable acceptance cut $10^\circ<\theta_B<170^\circ$ to require bosons to be detectable, one can further reduce the irreducible SM background.  Finally, a basic separation cut $R_{BB}>0.4$ has been imposed in order to resolve the final-state boson within a generic detector setup. 

\begin{figure}
\begin{center}
\includegraphics[width=0.45\textwidth]{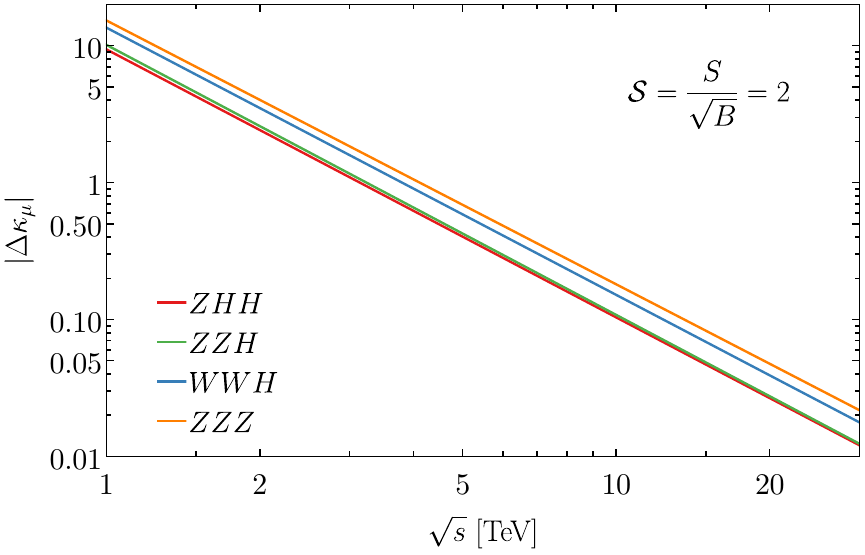}
\includegraphics[width=0.45\textwidth]{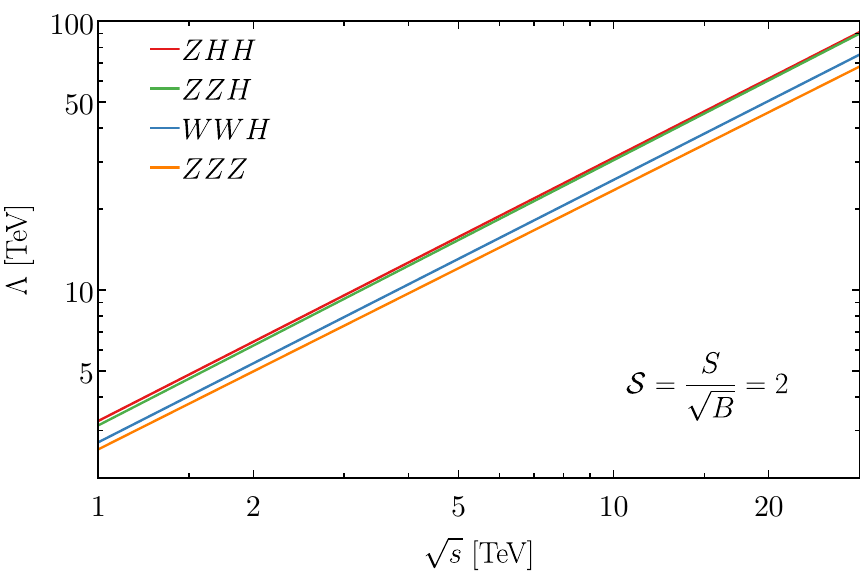}
\end{center}
\caption{(a) A high-energy MC's statistical sensitivity contour $\mathcal{S}=2$ to probe the muon-Higgs coupling $\kappa_\mu$ based on the measurement of three-boson production. (b) The probe of new physics scale with the assumption in Eq.~(\ref{eq:k2L}).}
\label{fig:sens}
\end{figure}

Assuming some deviation of multi-boson production from the SM background, one can estimate the sensitivity that follows from analyzing the tri-boson channels as $\mathcal{S}={S}/{\sqrt{B}}$,
where
\begin{equation}
S=N_{\kappa_\mu}-N_{\kappa_\mu=1}, ~B=N_{\kappa_\mu=1}+N_{\rm VBF}.
\end{equation}
Regarding the energy dependence of this sensitivity, the integrated luminosity is taken as quadratically scaling with energy, $\mathcal{L}=10~{\rm ab}^{-1}(\sqrt{s}/10~\TeV)^2$~\cite{Delahaye:2019omf}. Fig.~\ref{fig:sens}(a) shows the sensitivity contours that correspond to $\mathcal{S}=2$.  One can conclude that at a 3 TeV $\mc$, the muon Yukawa coupling can be probed by this method at the order of 100\%.  With an increased collision energy of 10 (30) TeV, this result can be improved to 10\% (1\%), respectively.  Based on the translation in Eq.(\ref{eq:k2L}), the precision of this muon-Higgs coupling measurement can be translated into a Yukawa-sector new-physics scale of 10 (30) TeV to be probed at a 3 (10) TeV $\mc$, respectively.

\subsection{Heavy Higgses through the Radiative Return Process}

A unique feature of $\mc$ is the possibility of generating $s$ channel-resonant of Higgs boson~\cite{Han:2012rb,Eichten:2013ckl,Alexahin:2013ojp,Greco:2016izi,deBlas:2022aow}. 
However, when identifying the heavier additional (pseudoscalar) scalars, the lack of {\it a priori} knowledge of mass makes finding new particles very difficult. A wide range of new physics scenarios from supersymmetry (SUSY) to neutrino mass generative models, motivates an extended sector of basic scalars.
Due to the weak couplings and sizable SM backgrounds, the LHC will have limited coverage for such search. 
At a future lepton collider is clean, and it would be straightforward to identify a heavy Higgs signal once produced on resonance~\cite{Eichten:2013ckl}.

The exact value of center-of-mass energy required for optimal detection of heavy Higgs depends on its unknown mass, particularly for the $s$-channel resonant production at a $\mc$. If we consider the associated production of a Higgs boson with other particles, the situation may improve.
A compelling process is the ``radiative return'' (RR) process, 
\begin{equation}
\mu^{+} \mu^{-} \rightarrow  \gamma H, \gamma A ,
\label{eq:rr}
\end{equation}
where $H$ ($A$) is a neutral CP-even (CP-odd) Higgs state. 
When the center-of-mass energy of the $\mc$ is above the heavy Higgs mass, the photon emission from the initial state enables an opportunity for the heavy Higgs boson to ``back'' to the resonance.
In this case, we do not need to know the exact value of the (unknown) heavy scalar mass.
In this section we illustrate the main points in the context of two-Higgs-doublet models (2HDM), summarizing the findings of~\cite{Haber:1978jt}.

The relevant heavy Higgs boson couplings can be parametrized as
\begin{equation}
\label{eq:int}
\mathcal{L}_{int} = -\kappa_\mu \frac {m_\mu} {v} H \bar \mu \mu + i \kappa_\mu \frac {m_\mu} {v} A \bar \mu \gamma_5 \mu
    + \kappa_Z \frac {m_Z^2} {v} H Z^\mu Z_\mu 
    +\frac {g \sqrt{ (1-\kappa_Z^2)}} {2\cos\theta_W} (H\partial^\mu A-A\partial^\mu H) Z_\mu.
\end{equation}

The two parameters $\kappa_\mu$ and $\kappa_Z$ characterize the coupling strength relative to the SM Higgs boson couplings. The coupling $\kappa_\mu$ controls the heavy Higgs resonant production and the radiative return cross-sections. $\kappa_Z$ controls the cross-sections for $ZH$ associated production and heavy Higgs pair $HA$ production.
Using $\kappa_\mu$ as the common rescale parameter for the Yukawa couplings for both the CP-even $H$ and the CP-odd $A$. Although, in principle, these couplings could be different. For the $HAZ$ coupling one can use the generic 2HDM relation: $\kappa_Z$ is proportional to $\cos(\beta-\alpha)$ and the $HAZ$ coupling is proportional to 
$\sin(\beta-\alpha)$. In the decoupling limit of 2HDM at large $m_A$, $\kappa_{Z}\equiv \cos(\beta-\alpha)\sim m_{Z}^{2}/m_{A}^{2}$ is highly suppressed and $\kappa_\mu \approx \tan\beta\ (-\cot\beta)$ in Type-II and lepton-specific (Type-I and flipped) 2HDM.
The choices of parameters and their 2HDM correspondences are shown in Table.~\ref{tab:parameters}. 

When kinematically allowed, the photon emission from the initial state enables an opportunity for the heavy Higgs boson ``back'' to resonance. The signature is quite striking: a monochromatic photon. The ``recoil mass'' would be a sharp resonant peak at $m_{H/A}$, standing out of the continuous background. The reconstruction of the heavy Higgs boson from its decay product provides an extra handle. 

The characteristic of this RR signal is a photon with the energy given by
\begin{equation}
E_\gamma= \frac {\hat s - m_{H/A}^2} {2\sqrt{\hat s}},
\end{equation}
from which a recoil mass peaked at the heavy Higgs mass $m_{H/A}$ can be reconstructed. The energy of this photon is broadened by detector photon energy resolution, beam energy spread, additional (soft) ISR/FSR, and heavy Higgs width. 
The beam energy spread and additional soft ISR/FSR are GeV level~\cite{Delahaye:2013jla}.
When the Higgs boson is significantly below the beam energy, the recoil mass construction receives considerable smearing dominated by the photon energy resolution. 

\begin{table}[t]
\centering
\begin{tabular}{|c|c|c|c|}
  \hline
  Coupling & $\kappa\equiv g/g_{\rm SM}$ & Type-II \& lepton-specific& Type-I \& flipped\\ \hline
  $g_{H\mu^+\mu^-}$ & $\kappa_\mu$ &$\sin\alpha/\cos\beta$ & $\cos\alpha/\sin\beta$ \\ \hline
  $g_{A\mu^+\mu^-}$ & $\kappa_\mu$ &$\tan\beta$ & $-\cot\beta$ \\ \hline
  $g_{HZZ}$ & $\kappa_Z$ & $\cos(\beta-\alpha)$ & $\cos(\beta-\alpha)$ \\ \hline
  $g_{HAZ}$ & $1-\kappa^2_Z$ & $\sin(\beta-\alpha)$ & $\sin(\beta-\alpha)$ \\
  \hline
\end{tabular}
\caption[]{Parametrization and their 2HDM models correspondence.}
\label{tab:parameters}
\end{table}

\begin{figure}[t]
\centering
\includegraphics[bb=0.0in -0.1in 7in 6.0in,width=300pt]{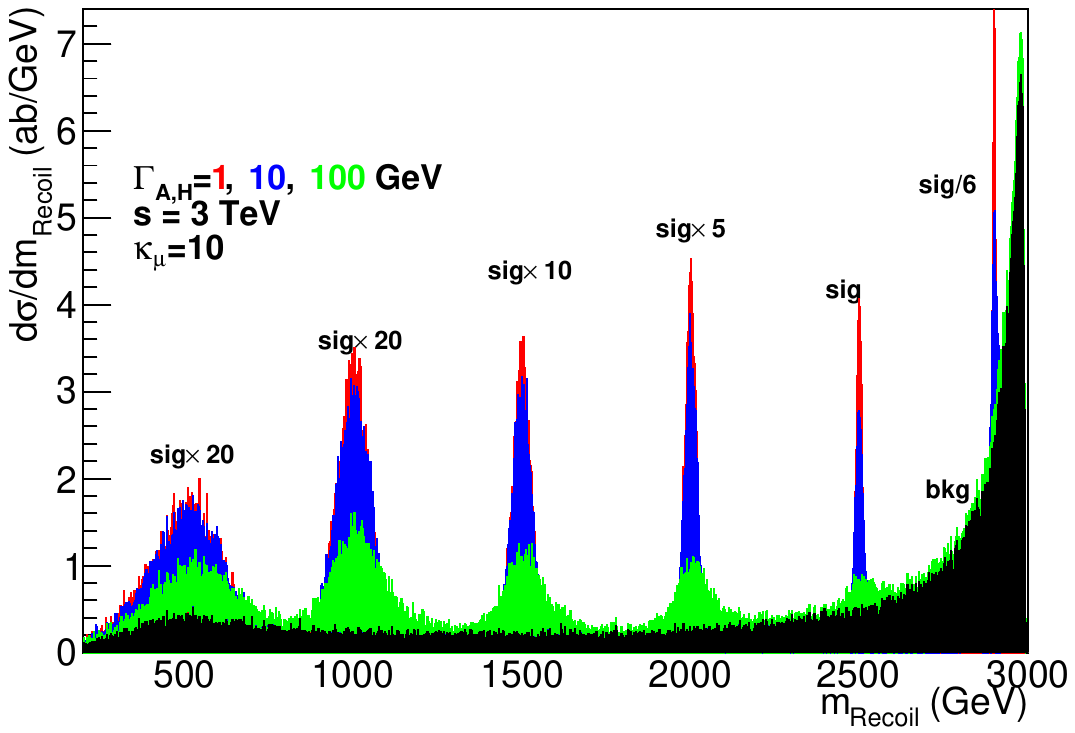}
\caption{Recoil mass distribution for heavy Higgs mass of 0.5, 1, 1.5, 2, 2.5, 2.9 TeV with a total width 1 (red), 10 (blue), and 100 (green) GeV at a 3 TeV $\mc$. ISR and FSR are included in this calculation. Background (black shaded region) includes all events with a photon of $p_T>10~$GeV. Note that signal and background have different re-scale factors for clarity. This figure is obtained from~\cite{Chakrabarty:2014pja} and more detailed discussion can be found there.}
\label{fig:sigbkg}
\end{figure}

Besides the mass, the other most important parameter is its total width, which effectively smears the monochromatic photons. In Type-II 2HDM, $\kappa_\mu=\tan\beta$ in the decoupling limit. The total width is minimized when $\tan\beta=\sqrt{m_t/m_b}$ but typically O(GeV) to O(100~GeV). 

The inclusive cross-section for the mono-photon background is substantial compared to the radiative return signal. The background is mainly from the M\"roller scattering with ISR/FSR  $\mu^+\mu^- \to \mu^+\mu^- \gamma$, 
and the $W$ exchange with ISR $\mu^+\mu^- \to \nu\nu \gamma$. 
The signal background ratio is typically the order $10^{-3}$ for a 3 TeV $\mc$. Consequently, to discover through RR, we rely on some exclusive processes.

A Type-II 2HDM has been adopted for concrete illustration with the $b\bar b$ final state as a benchmark with the decaying branching fraction be $80\%$. Also, it has been assumed an $80\%$ $b$-tagging efficiency and require at least one $b$-jet tagged. 

Madgraph5~\cite{Alwall:2011uj} has been used for parton level signal and background simulations and then Pythia~\cite{Sjostrand:2006za} for ISR and FSR. Detector smearing and beam energy spread has been also implemented.
Fig.~\ref{fig:sigbkg} shows the recoil mass distribution at a 3 TeV $\mc$. Both cross-sections of the signal and the background at fixed beam energy increase as the recoil mass increase from the photon emission.
One can clearly see the pronounced mass peaks look and the
RR process is an essential discovery production mechanism. 

\begin{figure}[t]
\centering
\includegraphics[scale=0.38]{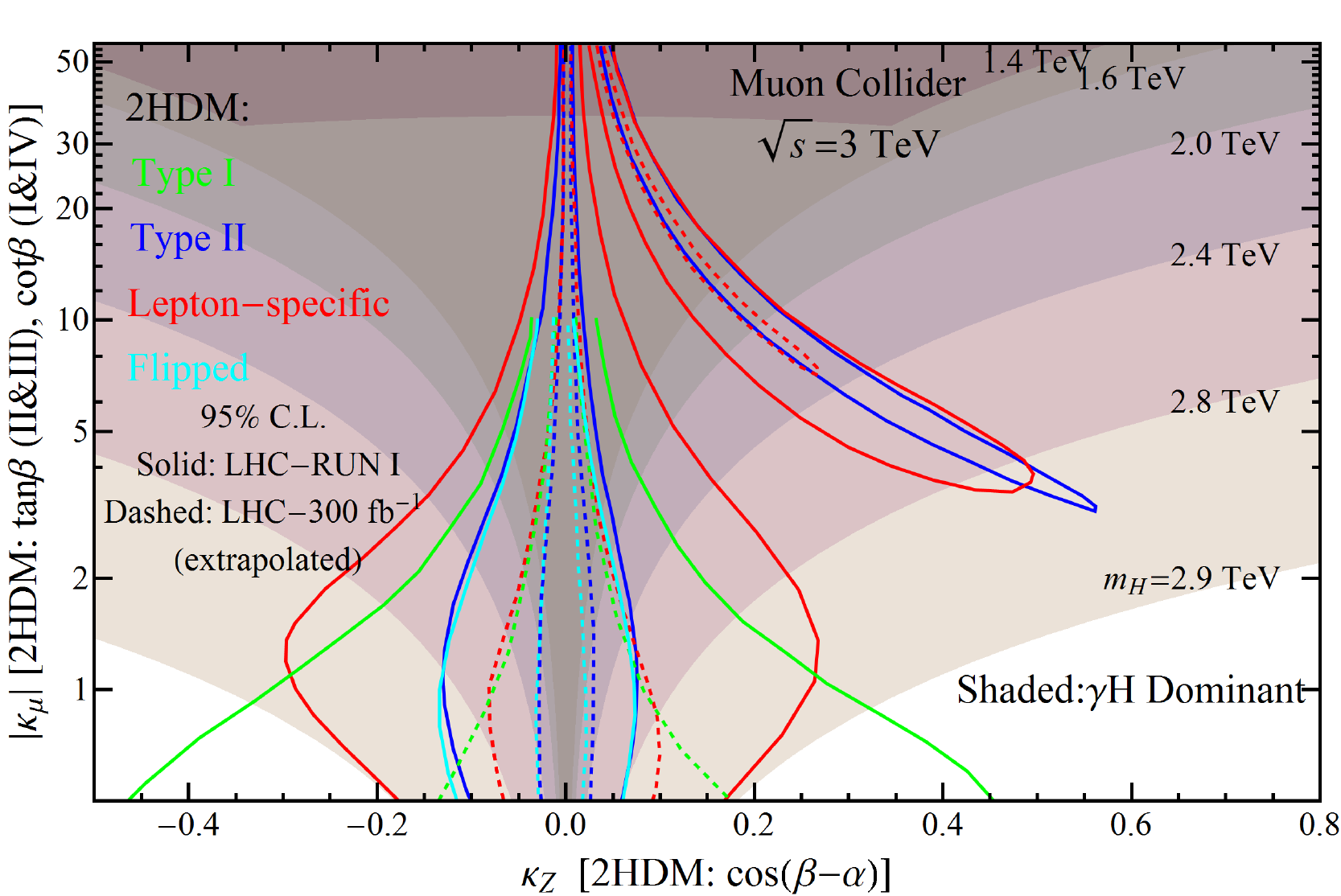}
\caption[5pt]{Comparison of sensitivities between different production mechanisms in the parameter plane $\kappa_\mu$\nobreakdash-$\kappa_Z$ for different masses of the heavy Higgs boson at the $3~\TeV$ $\mc$. The shaded regions show a higher direct signal rate from the RR process than the $ZH$ associated production and $HA$ pair production channels. 
One can also see the allowed parameter regions (extracted from Ref.~\cite{Barger:2013ofa}). This figure is obtained from~\cite{Chakrabarty:2014pja} and more detailed discussion can be found there.}
\label{fig:kappas}
\end{figure}
It is informative to put the reach of the two theory parameters side-by-side via the RR and pair production, as in Fig.~\ref{fig:kappas}. The shaded regions represent when the RR process dominants over the $ZH$ associated production and $HA$ pair production. 
The RR production mode covers a large region of $\kappa_\mu$ ($\tan\beta$ in Type II 2HDM).
 The closer the Higgs mass to the $\mc$ energy threshold, the more critical the RR channel is than the $ZH$ channel.
Well below the threshold, these two processes scale the same way as $1/s$.
The RR process is only dependent on $\kappa_\mu$, while both $ZH$ associated production and $HA$ pair production mainly depend on $\kappa_Z$. 
The nearly flat region in the figure for $1.4$ TeV heavy Higgs represents the good sensitivity from heavy Higgs pair production. The RR process is the leading channel for a heavy Higgs boson near the energy threshold and the decoupling regime of general Higgs extensions. %

The currently observed SM-like Higgs boson tightly constrains the $\kappa_Z$ region. The allowed parameter regions for 2HDM with current LHC data (solid) and projection are also shown in the figure for comparison. This illustrates that the RR processes are favored in all allowed 2HDM models.

In summary, this sections showed the signature and sensitivity of heavy Higgs boson signals from three production modes at a high-energy $\mc$. More detailed discussions can be found in Ref.~\cite{Chakrabarty:2014pja}. Compared to the $s$-channel resonance at $\sqrt s = m_h$, these different production mechanisms do not rely on {\it a priori} knowledge of the heavy Higgs mass. It has been found that radiative return is of particular interest, avoiding the scan process. A monochromatic photon characterizes this signal ($\gamma H$).
The coupling-mass parameter space $\kappa_{\mu}$\nobreakdash-$m$ (SUSY equivalent of $\tan\beta-M_{A}$) covered by such search through RR process at a high energy $\mc$ can substantially extend over the LHC projections.
Compared with other modes of $ZH$ and $HA$ production at a lepton collider, the RR process is advantageous, especially for the decoupling regions in all 2HDM-like models. The RR process could undoubtedly provide us an attractive option compared to the traditional scanning procedure for heavy Higgs boson at a high energy $\mc$, enabling heavy Higgs discovery opportunities.

\section{Dark Sectors \label{sec:dark}}

Dark particles can couple to SM states by means of effective higher dimensional operators, which are dominated by those of dimension five. These  operators appear for instance in dark photon (DP) coupling to SM fermions via magnetic dipole interactions, as predicted by portal dark-sector models \cite{Dobrescu:2004wz}, or axion-like particles (ALP) to di-photon couplings, as a consequence of the $U(1)$ Peccei-Quinn anomaly. In this section we review the findings of~\cite{Casarsa:2021rud} regarding the reach of searches for operators like these at a MuC.

On dimensional grounds, the leading production cross section of a dark particle in association with a photon at high energy tends to a constant proportional to $1/\Lambda^2$, with $\Lambda$ the effective scale associated to  the dimension five operators. This behavior must be compared  to that of the cross section for dark particles production  by  renormalizable couplings to SM particles, the cross section of which is expected to decrease as $\sigma \sim 1/s$ at high center of mass (CM) energy $\sqrt{s}$. In addition, the corresponding cross section for the SM background, characterized by a photon plus a neutrino pair, scales as $1/s$ at high energy which leads to the enhanced ratio of signal over background at high energy for dark-particle productions in association to a photon. 
These features make a $\mc$ with both  high energy and high luminosity a very promising machine for the study of the dark sector~\cite{Long:2020wfp}. 

In~\cite{Casarsa:2021rud}, the authors focus on the annihilation of a muon pair into a light dark particle $X$ and a photon 
$\gamma$
\begin{equation}
\mu^+ \, \mu^- \to \gamma \, X\, ,
\end{equation}
the dark-sector particle behaving as missing energy inside the detector.
As the dark particle is invisible and assumed to be light, the events shows up as  a mono-chromatic single photons with almost half of the center-of-mass energy.

Experimental searches for the same mono-photon signature have been performed at the LEP~\cite{OPAL:1998aqw,L3:2003yon,DELPHI:2003dlq}, the Tevatron~\cite{CDF:2008njt,D0:2008ayi} and the LHC~\cite{ATLAS:2016zxj,CMS:2018ffd} though only providing  rather weak bounds on their couplings to SM particles. A $\mc$ with CM energy of 3 and 10 TeV offers a large potential to increase the sensitivity to this signal with respect to the aforementioned colliders.

\subsection*{Dark particles}
Two possible candidates for the invisible state in the single photon signature are considered: a massless, spin 1 particle (the DP) and a light pseudo-scalar particle (an ALP).

The DP $A^\prime _\mu$ with field strength $F^{\prime \mu \nu}$ can couple to the muons via the magnetic-dipole interaction
\begin{equation}
{\cal L}^{\rm \tiny dipole}_{\rm \tiny DP}= \frac{1}{2\, \Lambda} \, ( \bar \mu \, \sigma_{\mu\nu}\,  \mu  ) \, F^{\prime \mu \nu} \, ,  \label{L-DP}
\end{equation}
where $ \sigma_{\mu\nu}$ is defined to be $i [\gamma^\mu,\, \gamma^\nu]/2$. The scale $\Lambda$ modulates the strength of the interaction. In a UV completion of the theory this effective scale can be generated at one-loop by the exchange of heavy particles in the portal sector \cite{Dobrescu:2004wz,Gabrielli:2016cut}.
The coupling in \eqref{L-DP} is the only one in the case of a massless dark photon.
On the other hand, in the case of a massive dark photon, in addition to the  Pauli dipole term, an ordinary coupling to the vectorial muon current is also possible
$
{\cal L}^{\rm \tiny tree}_{\rm \tiny DP}=\varepsilon e ( \bar \mu \, \gamma^{\mu}\,  \mu  ) A^\prime _\mu$,
arising from a tree-level contribution of kinetic mixing of dark-photon with ordinary photon \cite{Holdom:1985ag}, that in the massive case cannot be rotated away. This Pauli operator has not been constrained by current massive DP searches because they have been performed at low-energies, where its effect is strongly suppressed. Therefore, we assume here, the interaction in \eqref{L-DP} be the dominant mechanism also in production of a massive dark-photon at $\mc$.

The ALP $a$ couples to the muons by means of the portal operator
${\cal L}^{\rm \tiny muon}_{\rm \tiny ALP}= ( \bar \mu \, \gamma_5   \gamma^\mu \, \mu  ) \, \partial_\mu a /\Lambda$ 
and to photons by means of
\begin{equation}
{\cal L}^{\rm \tiny photon}_{\rm \tiny ALP}=\frac{1}{\Lambda} \,  a \, F^{\mu\nu} \widetilde F_{\alpha \beta} \, , \label{L-ALP2}
\end{equation}
where $\widetilde F_{\alpha \beta}=1/2\epsilon_{\alpha\beta\mu\nu}F^{\mu\nu}$ is the dual field strength of the photon, with $\epsilon_{\alpha\beta\mu\nu}$ the Levi-Civita antisymmetric tensor satisfying  $\epsilon_{0123}=1$.  The scale $ \Lambda$ controls the strength of the interactions. However, in the high-energy regime  the interaction with the muon axial current  is chirally suppressed by terms proportional to the muon mass over energy \cite{Darme:2020sjf}  and so only the interaction in \eqref{L-ALP2} is retained which significantly contributes to the cross section. Being \eqref{L-DP} and \eqref{L-ALP2} effective interactions, the $\Lambda$ scale is assumed to be larger or at most of the same order as the CM energy.

\subsection*{Constraints}

 \begin{figure*}[t!]
\begin{center}
  \includegraphics[width=4.5in]{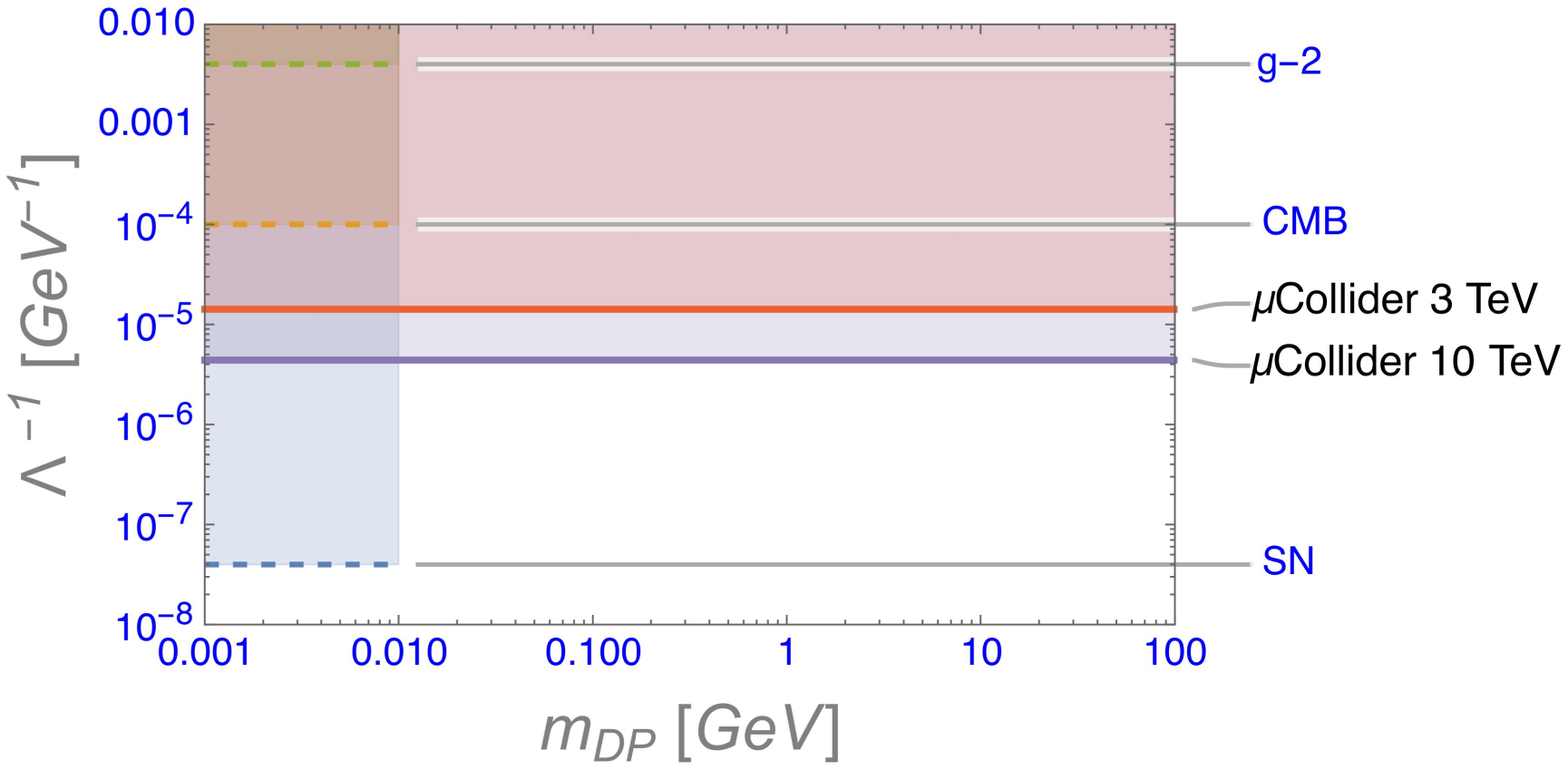}
  \vspace{-1.5cm}
\caption{\small 
  \label{fig:bounds1} Limits on $1/\Lambda$ scale for the dark-photon as a function of the dark-photon mass
  $m_{DP}$: for SN the scale of the coupling to muons has been set  at  $10^{7.4}$ GeV  \cite{Bollig:2020xdr} by the effect of dark radiation on  Supernovae dynamics.  For CMB see~\cite{DEramo:2018vss}. For $g-2$ see~\cite{Pospelov:2008zw,Escudero:2019gzq}. Comparable bounds hold  for the ALP to muons because of the similar structure of the interaction vertex. For masses up to 100 GeV the $\mu$Collider limits are for all practical purposes mass independent.  
}
\end{center}
\end{figure*}
 
The SM process $\mu^+ \mu^- \to \gamma \nu \bar \nu$ gives rise to the same signature as the signal and it provides the main source of background.  The SM cross section grows with the CM energy but the number of events with a high-energy photon decreases~\cite{Bento:1985in,Berends:1987zz}. However, background events at the end of the photon energy spectrum  around
$ E_\gamma=\sqrt{s} (1 - m_Z^2/s)/2$
are enhanced by the radiative return of the $Z$-boson pole. This feature reduces the sensitivity to the signal that---it being a two-body process---is centered in the same range of energies (for  $s\gg m_Z^2$). Therefore a suitable statistical analysis is necessary in order to distinguish the signal from this background. 

\begin{table}[h]
\renewcommand{\arraystretch}{1.3}
\begin{center}
\caption{Explorable values of the effective energy scale $\Lambda$ for DP and ALP (95\% CL) for the two benchmark scenarios of the future $\mc$ under consideration.}
\label{tab:lambda}
\begin{tabular}{p{2cm}ccccc}
&\multicolumn{2}{c}{\bf DP} & & \multicolumn{2}{c}{\bf ALP} \\
\hline\hline
{\bf Energy} & 3 TeV & 10 TeV & & 3 TeV & 10 TeV  \\
\hline
{\bf Exclusion} & 141 TeV & 459 TeV & & 112 TeV & 375 TeV \\
{\bf Discovery} & 92 TeV & 303 TeV & & 71 TeV & 238 TeV\\
\hline\hline
\end{tabular}
\end{center}
\end{table}

In the analysis of~\cite{Casarsa:2021rud}, the authors consider two benchmark collider scenarios, namely with CM energy of 3 TeV and 10 TeV with total integrated luminosity of 1 ab$^{-1}$ and  10 ab$^{-1}$ respectively. Then, they study the generation of events with a single, monochromatic photon plus missing energy in the final states.
The events for the signal and the background are generated by means of \textsc{MadGraph5}~\cite{Alwall:2011uj}.  A 10-GeV cut on the photon generated transverse momentum is imposed to remove most of the soft radiation. The output of \textsc{MadGraph5} is automatically fed into \textsc{Pythia}~\cite{Sjostrand:2006za} and the events thus generated are processed by the detector simulation.
The full-simulated events are reconstructed with a particle-flow algorithm~\cite{Marshall:2015rfa}, which is integrated in the ILCSoft reconstruction software.
A suitable choice of cuts on the photon energy and polar angle, to suppress the large  background induced by the radiative return effect, has been implemented to increase signal over background sensitivity \cite{Casarsa:2021rud}. Results for the limits (95\% CL) and discovery ($5\sigma$) for the largest $\Lambda$ reachable are reported in Table I.

\begin{figure*}[t!]
\begin{center}
\includegraphics[width=4.5in]{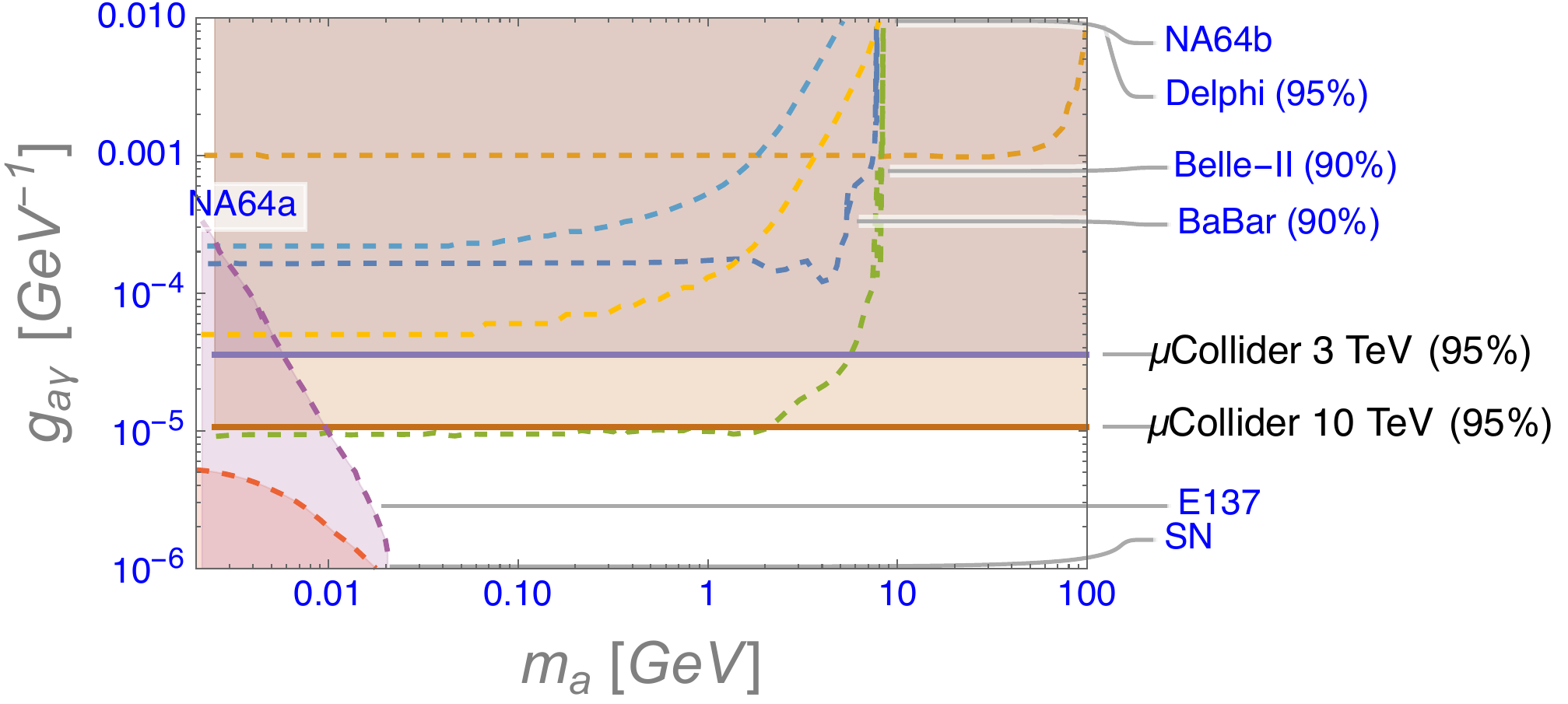}
\caption{
\small
  \label{fig:bounds2} Limits on $g_{a\gamma}=4/ \Lambda$ as a function of the ALP mass $m_a$: NA64a~\cite{NA64:2020qwq}, Delphi~\cite{DELPHI:2008uka}  and BaBar~\cite{BaBar:2017tiz} are actual limits. Belle-II~\cite{Belle-II:2018jsg,Dolan:2017osp}, NA64b~\cite{NA64:2020qwq} and $\mu$Collider~\cite{Casarsa:2021rud} are future estimates. The limit indicated by E137 is  the one from \cite{Bjorken:1988as} as modified for a small ($10^{-4}$) visible branching fraction~\cite{Darme:2020sjf}. For masses up to 100 GeV the $\mu$Collider limits are for all practical purposes mass independent. }
\end{center}
\end{figure*}

Finally, in Figs.\ref{fig:bounds1} and \ref{fig:bounds2} the bounds for $1/\Lambda$ and the $g_{a\gamma}$ couplings respectively, are compared with current and future limits from low-energy, cosmological, astrophysical and collider physics, where the following notation is adopted for the coupling $g_{a\gamma}\equiv 4/ \Lambda$ associated to the dipole operator in \eqref{L-DP}, in order to compare it with the common notation used in the various experiments.

When and if a signal is found, it will be important to know which dark sector particle is responsible for it.  In \cite{Casarsa:2021rud} the authors show that a $\mc$ operating at 3 or 10 TeV has  the potential to distinguish the spin-0 ALP from the spin-1 DP scenario. For a common energy scale $\Lambda=300$ TeV---about 200 events (which  can be accumulated in order five years) are required to separate the two spin scenarios at the 95\% CL.

%% file: Section_Conclusions/section.tex
\section{Key findings}\label{sec:KF}

We conclude by collecting the highlights of the muon collider potential in each of the areas of interest presented in the report. 

\begin{center}
{\bf Higgs Physics}
\end{center}
\vspace{-0.25cm}

Higgs physics at high-energy muon colliders mainly benefits  from the energy growth of the rates in vector-boson-fusion processes allowing opening up the possibility of vast programme of measurements covering not only single and double Higgs but also and multi vector bosons, and top quarks. With 1 ab$^{-1}$ of collected luminosity, precision measurements of the single Higgs couplings at 3 TeV would significantly improve in many cases the percent-level knowledge gained from the HL-LHC, and hence the sensitivity to a large class of BSM scenarios  predicting modifications of the Higgs interactions. Reaching permille level precision in the couplings to $W$, $Z$ and the bottom Yukawa would be possible at the higher energy 10~TeV stage with the target 10~ab$^{-1}$ luminosity. 
Assessing the reach of a high-energy muon collider in precision for the $htt$ interaction is still an open question. In particular, the top Yukawa coupling could be accessible via measurements not only of $\mu^+ \mu^-\to t\bar{t}h$ but, with higher rates, also via $VV\to t\bar{t}$ or, at 10 TeV, $VV\to t\bar{t} h$, yet a detailed analysis is not available.

High-energy muon colliders open the way to direct measurements of the Higgs trilinear self-coupling, $\lambda_3$, and at above 10 TeV, even the potential observation of multi-Higgs production, which is sensitive to the quartic self-coupling.  We find that the precision in the determination of $\lambda_3$ of the 3~TeV muon collider would substantially benefit from an increase in the total luminosity by a factor$\sim 2$ with respect to the proposed benchmark of 0.9 ab$^{-1}$, suppressing a second mode in the likelihood for $\lambda_3$ and allowing a determination at the $15\%$ level. Percent level uncertainties will be achieved at the higher energy stages.   

In this report, we have also briefly presented the physics prospects of a low energy muon collider option operating at the Higgs pole, $\sqrt{s}=125$ GeV. With 5 fb$^{-1}$, such a collider would provide the model independent determination of the Higgs boson width at the few percent level, which is not possible at higher energies, and the determination of the muon Yukawa coupling at the one percent level.  On the other hand, higher luminosities $\sim 20$ fb$^{-1}$ are required to achieve a precision in the determination of other Higgs boson couplings comparable to that of a generic Higgs factory. We note that a measurement of $\Gamma_H$ is important on its own as it helps to resolve a specific flat direction in the global Higgs boson coupling determination. However, an assessment of the usefulness of such constraints in specific BSM scenarios is still lacking.  We conclude stressing that while a model-independent determination of the Higgs boson width is not possible at a high-energy muon collider, this absence could be solved when combined with the information that will be available
from future $e^+ e^-$ Higgs factories.


\begin{center}
{\bf Effective Field Theories}
\end{center}
\vspace{-0.25cm}

The overall reach in terms of constraining new interactions at high-energy muon colliders is not limited to the determination of the in Higgs boson final states.  
A global assessment of the physics potential for indirect constraints has been performed here within the framework of the SMEFT at dimension six, including a considerable number of new  interactions.
One of the main advantages of operating at multi-TeV centre-of-mass energies is the augmented sensitivity to operators whose relative contributions to SM electroweak processes grow with the energy $\sim E^2_{\rm cm}/\Lambda^2$. This is the case of, for instance, four-fermion contributions in 2 to 2 fermion processes, which could be generated at low energies by a variety of heavy new particles. Such enhancements due to virtual effects of new resonances allow to set stringent bounds on their properties, even if experimental precision is limited, e.g. testing $\Lambda\sim 100$ TeV for percent-level precision measurements at $\sqrt{s}=10$ TeV.
Although the set of projections for measurements interpreted in the EFT framework at a high-energy muon collider is still limited, preventing a full exploration of the EFT parameter space, the results discussed here, which combine information from Higgs, difermion and diboson measurements, clearly indicate the potential for massive gains in terms of sensitivity to new interactions with respect to the HL-LHC. In particular, sensitivities to new physics interaction scales up to $\Lambda/\sqrt{c}\sim 30$ (100) TeV would be possible at $\sqrt{s}=$3 (10) TeV.

For a clear illustration of the sensitivity gain, we interpreted the EFT results in terms of concrete new physics scenarios, among which composite Higgs models. In this case a $\sqrt{s}=$3~TeV muon collider can test values of the typical mass of the composite sector, $m_\star$, in the range of $\sim 15$-$30$~TeV, depending on the value of the typical coupling $g_\star \in [1,4\pi]$. The sensitivity is comparable to the one of the combination of the FCC-ee and FCC-hh colliders. The sensitivity will reach $m_*\sim 50$-$90$~TeV at the 10~TeV muon collider.

\begin{center}
{\bf BSM - New Scalars}
\end{center}
\vspace{-0.25cm}

On top of investigating the couplings of the SM particles, a very intriguing possibility which can be explored at the 3~TeV muon collider is multiple new scalar boson production and their interactions.  The 3~TeV muon collider generically has sensitivity to discover new Higgs bosons up to half of the center of mass energy when they can be produced in pairs via gauge interactions, e.g., pair production of charged Higgs bosons. For singly produced Higgs bosons,  the reach in mass depends on the strength of the coupling that mediates the single production. In the simple examples of extended Higgs sectors featuring new singlet scalars coupled to the SM only via mixing with the Higgs boson,  the 3~TeV muon collider is sensitive to new Higgs bosons up to around 2~TeV. This mass reach significantly extends that of the HL-LHC and complements the sensitivity from indirect  probes such as Higgs boson couplings measurement. The 3~TeV muon collider, as it simultaneously operates as a Higgs factory at the intensity frontier and as an exploration machine at the energy frontier, can provide multiple probes of new physics in the scalar sector. 

Focusing on interpretations of general searches of extra Higgs bosons  allows to quantify the reach of the 3~TeV muon collider. In many cases, very promising results are expected some of which leading to significant progress about fundamental open issues of the SM. For instance, the measurements of the Higgs boson couplings and the direct search for new bosons can put very stringent bounds on models that modify the strength of the electroweak phase transition and essentially rule out new scalar states as possible agents of modification of the Higgs boson potential. In this particular class of models,  a 3~TeV muon collider could have an exciting interplay with gravity waves observations expected from the electroweak phase transition. The scenario where space-born gravity waves observatories will come online during the late 2030s marries nicely with the timeline of the 3~TeV muon collider as initial stage of a high energy exploration based on muon beams.

In addition, the thorough exploration of trans-TeV masses for new scalars is a significant step in the understanding of role of the Higgs boson in shaping fundamental interactions. For instance the role of the Higgs as symmetry breaking agent can be further clarified by finding, or excluding, a new scalar in the TeV mass range. A discovery enabled by the 3~TeV muon collider would open up a vista on a whole new scalar sector. Such a finding would call for a deeper understanding of the origin of spin-0 particles and their possible point-like nature. Not finding a new scalar in the TeV mass range would, on the other hand, stress even further the already peculiar role played by the Higgs boson in the SM, motivating the determination of each and every of its properties even more.  

\begin{center}
{\bf BSM - Dark Matter}
\end{center}
\vspace{-0.25cm}

A high-energy muon collider has a great potential to probe dark matter particles, in particular weakly charged ones. The interesting mass range for this type of dark matter covers a rather large span from fractions of TeV up to fractions of PeV. The lighter dark matter candidates can be embedded in more ambitious BSM scenarios such as perturbative supersymmetric extensions of the SM in principle valid up to very high scales. The heavier candidates, roughly above O(10) TeV, are typical of BSM constructions that feature non-perturbative regimes at some short distance above the weak scale. A 3~TeV muon collider has a potential to probe, and potentially discover, dark matter candidates around the TeV scale employing three different search modes: $i)$ the direct search for signatures such as the stub-track of the higgsino dark matter candidate; $ii)$ the direct and very general search for dark matter production in association with SM states, e.g. electroweak vector bosons;  $iii)$ the indirect search for precision effects beyond the SM from loops of weakly charged dark matter. 
Several probes from indirect precision effects are expected to have sensitivity to the thermal higgsino dark matter at the 3~TeV muon collider.

A long list of weakly charged dark matter candidates can be probed at higher energy muon colliders, with few candidates already in the reach of the 3~TeV machine. By increasing the energy and luminosity  of the first stage of the muon collider,  one can establish a systematic path to cover the entire list of weakly charged dark matter candidates. 

Very importantly for the livelihood of the field, the timeline for the realization of a high energy muon collider can interleave nicely with both direct and indirect searches of astrophysical dark matter.  These experiments are expected to probe new ground in data-taking expected in the 2030s. After these new runs, there might be first claims for the observation of TeV scale dark matter, thus calling for action already during the next decade. A high energy muon collider would have a unique opportunity to clarify the veracity of these claims in a timely and accurate manner.

\begin{center}
{\bf Muon-Specific Opportunities}
\end{center}
\vspace{-0.25cm}
Muon colliders have a clear advantage over any other collider when it comes to searches for new physics that interacts more with muons than with first-generation particles.
Already in the Standard Model, the Yukawa interactions of the Higgs boson are an example of such a flavor non-universal physics.
Hints of the existence of 
new physics coupled to muons can be found in experimental anomalies like the muon $g$-2 and the $B$-physics anomalies. 
We find that a muon collider running at an energy of a few TeV is guaranteed to discover or disprove the physics responsible for these anomalies.

$B$-physics anomalies point to the existence of flavored new physics that couples more strongly to muons than to electrons. Various constraints in quark- and lepton-flavor physics suggest that flavor violation in these new interactions should be suppressed by small mixing angles, similarly to what happens in the Standard Model. If this is the case, the $(bs)(\mu\mu)$ operator is accompanied by larger flavor-conserving interactions, and the new physics scale must be in the few TeV range. Although not necessarily in reach of the HL-LHC, a 3 TeV muon collider could fully test this new physics either with high-energy probes of the effective interactions, or by directly discovering their mediators. 
Even in the `nightmare scenario' where only the very $(bs)(\mu\mu)$ interactions responsible for the experimental anomalies are present, with a scale of a few tens of TeV, a muon collider with a slightly higher energy of about 7 TeV could test the full parameter space, thus providing a full-fledged no-lose theorem, a goal that no other collider could achieve.


A muon collider program in the several TeV range 
is also highly motivated due to the no-lose theorem for discovering the new physics responsible for the  $g$-2 anomaly.
A 3 TeV muon collider can test all beyond the Standard Model scenarios where the muon $g$-2 is generated by a semi-leptonic interaction between muons and charm quarks. At the same time, it would be able to discover electroweak singlets with masses above a GeV, thus probing all models where the $g$-2 is generated at one loop by the exchange of these bosons (lighter singlets will be in reach of upcoming low-energy experiments).
If new states with electroweak quantum numbers contribute to $g$-2 their mass can be larger, but a muon collider in the 10 TeV range will discover all theoretically reasonable solutions. 
Larger particle masses near the perturbativity limit of $\mathcal{O}(100 \ \mathrm{TeV})$ are only possible if the new physics generates a calculable new hierarchy problem for the Higgs boson and the Muon mass, or involves highly tuned flavour structures. 
However, even that case can be probed at a 30 TeV muon collider, which is guaranteed to detect a signal in the Higgs plus photon channel due to the same interaction responsible for the $g$-2, verifying the anomaly in a fully general and model independent way.

Finally, colliding muon beams allows very powerful searches for lepton flavor violating interactions at high energies. A muon collider with center-of-mass energy between 3 and 10 TeV can probe new physics scales between 10-1000 TeV. This is in some cases comparable with the reach of searches for lepton flavor violation processes at flavor factories -- some of the measurements that are sensitive to the highest new physics scales in high energy physics.
Further opportunities arise in Higgs physics, where modified muon Yukawa couplings can be tested, or in searches for portal interactions between muons and a dark sector, which all add to the rich physics potential of a TeV-scale muon collider.